\documentclass[amsmath,amssym,aps,reprint,prx,twocolumn,superscriptaddress,floatfix]{revtex4-2}

\usepackage[bookmarks=true, colorlinks=true, linkcolor=blue, urlcolor=blue, citecolor=blue, bookmarks=true, hyperindex=true]{hyperref}
\usepackage{lipsum}
\usepackage{braket}
\usepackage{mathtools}
\usepackage{enumerate}
\usepackage{xfrac}
\usepackage{soul}

\newcommand*{\scale}[2][4]{\scalebox{#1}{$#2$}}

\begin{document}
\title{Higher spin Richardson-Gaudin model with time-dependent coupling: Exact dynamics}

\author{Suvendu Barik}
\email[]{s.k.barik@uva.nl}
\affiliation{Institute for Theoretical Physics, Universiteit van Amsterdam, Science Park 904, 1098XH Amsterdam, The Netherlands}

\author{Lieuwe Bakker}
\email[]{l.bakker@friam.nl}
\affiliation{Institute for Theoretical Physics, Universiteit van Amsterdam, Science Park 904, 1098XH Amsterdam, The Netherlands}

\author{Vladimir Gritsev}
\affiliation{Institute for Theoretical Physics, Universiteit van Amsterdam, Science Park 904, 1098XH Amsterdam, The Netherlands}
\affiliation{Russian Quantum Center, Skolkovo, Moscow 143025, Russia}

\author{Ji\v{r}\'{i} Min\'{a}\v{r}}
\affiliation{Institute for Theoretical Physics, Universiteit van Amsterdam, Science Park 904, 1098XH Amsterdam, The Netherlands}
\affiliation{QuSoft, Science Park 123, 1098 XG Amsterdam, the Netherlands}

\author{Emil A. Yuzbashyan}
\affiliation{Department of Physics and Astronomy, Center for Materials Theory, Rutgers University, Piscataway, New Jersey 08854 USA}

\date{\today}

\begin{abstract}
\noindent
We determine the exact asymptotic many-body wavefunction of a spin-$s$ Richardson-Gaudin  model with a coupling inversely proportional to time, for time evolution starting from the ground state at $t = 0^+$ and for arbitrary $s$. Contrary to common belief, the resulting wavefunction cannot be derived from the spin-$1/2$ case by merging spins, but instead requires independent treatment for each spin size. The steady state is non-thermal and, in contrast to the spin-$1/2$ case, does not conform to a natural Generalized Gibbs Ensemble. We show that mean-field theory is exact for any product of a finite number of spin operators on different sites. We discuss how these findings can be probed in cavity QED and trapped ion experiments.
\end{abstract}
\maketitle

\section{Introduction}
Explicit time-dependent Hamiltonians are ubiquitous across all domains of quantum physics. The time dependence may take various forms, such as a quantum quench---a step-like change in parameters~\cite{fan_quenched_2024}---or a continuous, often monotonic variation, as encountered in Landau-Zener (LZ) problems~\cite{landau_theory_1932,zener_non-adiabatic_1997}, atom-atom and atom-ion collisions~\cite{delos_theory_1981}, and nuclear dynamics~\cite{nakatsukasa_time-dependent_2016}. Alternatively, it can be periodic, as in Floquet protocols~\cite{Bai_2021_AdvPhys, Claeys_2019_PRL, Mizuta_2023_Quantum}, which are used, for example, in studies of time crystals~\cite{wilczek_quantum_2012,yao_time_2018,khemani_brief_2019}, the generation of gauge potentials~\cite{Goldman_2014_PRX}, and spin physics~\cite{Geier_2021_Science} in cold atom experiments, as well as in studies of integrable models~\cite{gritsev_integrable_2017, fan_emergent_2020}. The study of such systems remains central to quantum simulation, sensing, and computation, and is closely related to quantum optimal control~\cite{DAlessandro_2021_Book}. Here, the goal is to evolve an easily prepared initial state of the system into a desired target state. Designing efficient protocols for such state preparation is an active area of research~\cite{Yuan_2019_Quantum, Cerezo_2021_NatRevPhys}, with the two standard paradigms being discrete (gate-based)\cite{Nielsen_2010_Book, Georgescu_2014_RMP, Fauseweh_2024_NatComm, Zhang_2022_PRL} and continuous time evolution \cite{Lidar_2018_RMP, monroe_programmable_2021}.

In most relevant situations, time-dependent many-body Hamiltonians are not amenable to exact treatment. This necessitates further simplifications, such as mean-field analysis~\cite{senese_out--equilibrium_2024} or numerical perturbation theory~\cite{puzzuoli_algorithms_2023}, which often limit these models to weak couplings, small Hilbert spaces, or short-time evolutions. As a result, they remain notoriously difficult to solve analytically. In this context, \emph{exactly solvable models} remain an important recourse, serving as reference points for numerical and approximate methods, and offering clear analytical guidance in the design of quantum state preparation protocols. A well-studied class of such models is periodically driven systems, where, by exploiting their underlying Lie-algebraic structure, exact solutions have been constructed in both closed~\cite{gritsev_integrable_2017,fan_emergent_2020,oka_floquet_2019} and open~\cite{scopa_lindblad-floquet_2018,scopa_exact_2019,bakker_lie-algebraic_2020} settings. More generally, it has been conjectured~\cite{Patra_2015} that a \emph{necessary} condition for the exact solvability of a time-dependent quantum Hamiltonian is its quantum integrability, in the sense of possessing nontrivial parameter-dependent commuting partners. A corresponding \emph{sufficient} condition was later formulated~\cite{sinitsyn_integrable_2018} as the existence of a non-Abelian gauge field with zero curvature in the space of system parameters. This framework has unified existing solvable models under the broader umbrella of \emph{integrable time-dependent models}~\cite{yuzbashyan_integrable_2018,PhysRevLett.123.123605,Chernyak_2020,Chernyak_2021,PhysRevLett.129.033201,zabalo_nonlocality_2022,barik_knizhnik-zamolodchikov_2024,suzuki2025competing}.

An important class of such solvable models comprises Gaudin magnets and their generalizations, such as the Richardson-Gaudin (RG), inhomogeneous Dicke (also known as Tavis-Cummings), and various central spin Hamiltonians. The celebrated BCS theory maps onto the spin-$1/2$ RG model when expressed in terms of Anderson pseudospins---bilinear combinations of fermionic creation and annihilation operators~\cite{anderson_random-phase_1958,coleman_introduction_2015}. In the time-independent case, this model has been solved using a classical version of the Bethe ansatz~\cite{richardson1964exact,richardson1965exact,Gaudin_book,Sklyanin_1989}. Introducing time dependence in the Hamiltonian parameters generally breaks integrability. However, for certain special choices---in particular, when the interaction strength is inversely proportional to time---the model remains integrable~\cite{sinitsyn_integrable_2018,yuzbashyan_integrable_2018,zabalo_nonlocality_2022}. Closely related integrable  models include generalized bosonic and fermionic Tavis-Cummings Hamiltonians with linear time dependence~\cite{PhysRevLett.123.123605,PhysRevLett.129.033201,suzuki2025competing}.

Time-dependent RG models also give rise to a variety of exactly solvable multilevel LZ problems that have attracted considerable attention over the past decades~\cite{child_curve-crossing_1971,bandrauk_long-range_1972,osherov_threshold_1996,ostrovsky_nonstationary_2003,sinitsyn_exact_2014,barik_knizhnik-zamolodchikov_2024}, with applications ranging from ion-atom collisions~\cite{nikitin_theory_1970} and transitions between vibrational modes~\cite{mies_effects_1964} to Rydberg transitions~\cite{nascimento_2009,baumgartner_2010,feynman_quantum_2015} and molecular collisions~\cite{tantawi_two-state_2000}.

The dominant focus so far has been on spin-$1/2$ RG models. The prevailing assumption is that the physics of higher-spin models can be obtained from the spin-$1/2$ case by appropriately combining spins and projecting onto the relevant subspace. For example, an effective spin-1 Hilbert space can be constructed by combining two spin-$1/2$ particles and projecting onto the triplet subspace, as relevant for the spin-1 BCS problem (see also Sec.\ref{sec:Discrepancy}). This approach is closely analogous to the construction of the paradigmatic AKLT model \cite{Affleck_1987_PRL}, or to numerical studies of the Haldane gap in alternating-bond Heisenberg spin chains~\cite{hung_numerical_2005}. Gaudin employed this strategy to derive the exact solution of higher-spin Gaudin magnets from the spin-$1/2$ case~\cite{Gaudin_book}: ``It is sufficient to bring the $\varepsilon$ together in packets to obtain operators $\ldots$ the size of each spin being arbitrary.'' What is meant here is that the solution to the higher-spin Gaudin model is obtained from the spin-$1/2$ problem by  merging several spin-$1/2$ degrees of freedom, as described above. While Gaudin’s assertion is entirely correct, we note that he considered only time-independent Hamiltonians.

In this work, we show that, in striking contrast to the time-independent case, for time-dependent RG Hamiltonians one \emph{cannot} obtain higher-spin solutions from the spin-$1/2$ ones by following a similar procedure. Consequently, it becomes necessary to analyze spin-$s$ RG models on a case-by-case basis for different spin magnitudes $s$, without expecting solutions for lower spins to recover their higher-spin counterparts. Moreover, we find that the late-time asymptotic wavefunctions of higher-spin RG models do not conform to the natural emergent generalized Gibbs ensemble (GGE), unlike in the spin-$1/2$ case~\cite{zabalo_nonlocality_2022}.

We begin by presenting a simple example of two spins, illustrating the fundamental difference between the time-dependent and time-independent cases, followed by a brief introduction to the Richardson-Gaudin  Hamiltonian.

The first major result of this paper is the derivation of the exact long-time dynamics of the many-body wavefunctions for spin-1 and spin-$3/2$ systems 
(Sec.~\ref{sec:Spin-1_Model} and Sec.~\ref{sec:Higher_spin}, respectively). The novelty of this analysis lies in the incorporation of the exact solution for a two-site spin-$s$ time-dependent RG Hamiltonian, recently obtained in~\cite{barik_knizhnik-zamolodchikov_2024}. By comparing the asymptotic form of the two-site wavefunction with results from a saddle-point analysis, we extract the full $t \rightarrow \infty$ asymptotic wavefunction of the corresponding $N$-body RG Hamiltonian. These results extend the limited set of known exact solutions for time-dependent many-body Hamiltonians.

The second major result is a partial solution to the general time-dependent spin-$s$ RG  Hamiltonian, presented in Sec.~\ref{sec:Higher_spin}. This solution is complete up to a corrective term. We propose a conjectural form, based on previous results and supported by numerical evidence, which may also solve the spin-2 RG Hamiltonian.

In Sec.~\ref{sec:Thermodynamic_Limit}, we compute the thermodynamic limit of local observables for the spin-1 model and, in Sec.~\ref{sec:steadystate}, investigate the corresponding steady states for the spin-1 and spin-$3/2$ models. The third major result is the finding that these steady states conform to an ensemble \emph{different from} the standard GGE. This contrasts with the spin-$1/2$ case studied in~\cite{sinitsyn_integrable_2018,zabalo_nonlocality_2022}, which does converge to a GGE distribution. We also briefly comment on the general spin-$s$ case.

Finally, in Sec.~\ref{sec:Experimental_Realizations}, we discuss possible experimental realizations of the theory developed in this work. Higher-spin representations of RG Hamiltonians arise naturally in systems with twofold degeneracies, such as in graphene~\cite{andrei_2020,zhang_2025}, superconducting ladders~\cite{dagotto_1992,takahashi_2015,jiang_2018}, and double-layer structures in the strong co-tunneling regime~\cite{zhou_2021,guo_2025}. Motivated by the high degree of control in current cold atom and ion experiments, we identify experimental platforms that feature the two main ingredients of our model: spin-1 (or more generally, spin-$s$) degrees of freedom and all-to-all spin connectivity. These conditions are naturally realized in cavity QED~\cite{mivehvar_cavity_2021} and trapped ion setups~\cite{monroe_programmable_2021, schneider_experimental_2012}, and we argue that the physics studied here can be probed using existing experiments, such as those in~\cite{davis_photon-mediated_2019, kotibhaskar_programmable_2024}, as detailed in Sec.~\ref{sec:Experimental_Realizations}.

\subsection*{Non-analytic behavior}\label{sec:analyticitybreakdown}
We begin by illustrating the non-analytic behavior with respect to merging spin degrees of freedom with a simple  example.
Consider a two-site spin-$1/2$ BCS Hamiltonian with an interaction strength  inversely proportional to time:
\begin{equation}\label{eq:N2_spin-1/2_Hamiltonian}
\hat{H}_\alpha(t) = 2\varepsilon_1 \hat{s}_1^z + 2\varepsilon_2 \hat{s}_2^z - \frac{1}{\nu t^\alpha} (\hat{s}_1^+ + \hat{s}_2^+)(\hat{s}_1^- + \hat{s}_2^-)
\end{equation}
Here, $\varepsilon_{1,2}$ are the on-site Zeeman fields, $\nu > 0$ is the ramp rate, $\hat{s}^{(x,\,y,\,z)}$ are spin-$1/2$ operators, and $\hat{s}^\pm = \hat{s}^x \pm i \hat{s}^y$. The exponent $\alpha$ is a positive real number. The $4 \times 4$ Hamiltonian in Eq.~\eqref{eq:N2_spin-1/2_Hamiltonian} is block-diagonalized according to the eigenvalues of the total magnetization  $\hat{J}^z = \hat{s}_1^z + \hat{s}_2^z$. This yields three blocks, from which we focus on the largest sub-block corresponding to $J^z = 0$. This sub-block is identified as a $2 \times 2$ hyperbolic Landau-Zener problem~\cite{barik_knizhnik-zamolodchikov_2024}. In the singlet-triplet basis
\begin{equation}
\begin{split}
\ket{\sigma_z = -1} &= \ket{1,0} = \frac{1}{\sqrt{2}} \left( \ket{\uparrow\downarrow} + \ket{\downarrow\uparrow} \right), \\
\ket{\sigma_z = +1} &= \ket{0,0} = \frac{1}{\sqrt{2}} \left( \ket{\uparrow\downarrow} - \ket{\downarrow\uparrow} \right),
\end{split}
\end{equation}
this sub-block takes the form:
\begin{equation}\label{eq:N2_spin-1/2_LZ_Hamiltonian_1}
\hat{H}^{0}_\alpha(t) = \frac{1}{\nu t^\alpha} \sigma^z - \Delta \sigma^x,
\end{equation}
where $\Delta = \varepsilon_2 - \varepsilon_1$, $\sigma^{x,\,z}$ are the Pauli matrices, and we have omitted a multiple of the $2 \times 2$ identity matrix, as it does not affect the transition probabilities.

It is also helpful to introduce a dimensionless 
time variable $\tau = |\Delta| t$. The non-stationary Schr\"odinger equation becomes
\begin{equation}
i\frac{d\ket{\Psi(\tau)}}{d\tau}=\hat{H}^0_{\alpha}(\tau)\ket{\Psi(\tau)},
\label{nstSch}
\end{equation}
where
\begin{equation}
\label{eq:N2_spin-1/2_LZ_Hamiltonian}
\hat{H}^0_\alpha(\tau) = \frac{|\Delta|^{\alpha-1}}{\nu \tau^\alpha}\sigma^z -\text{sgn}(\Delta)\sigma^x.
\end{equation}
Note that we define the sign function so that $\text{sgn}(0)=0$.

\begin{figure}[t]
\centering
\includegraphics[width=\linewidth]{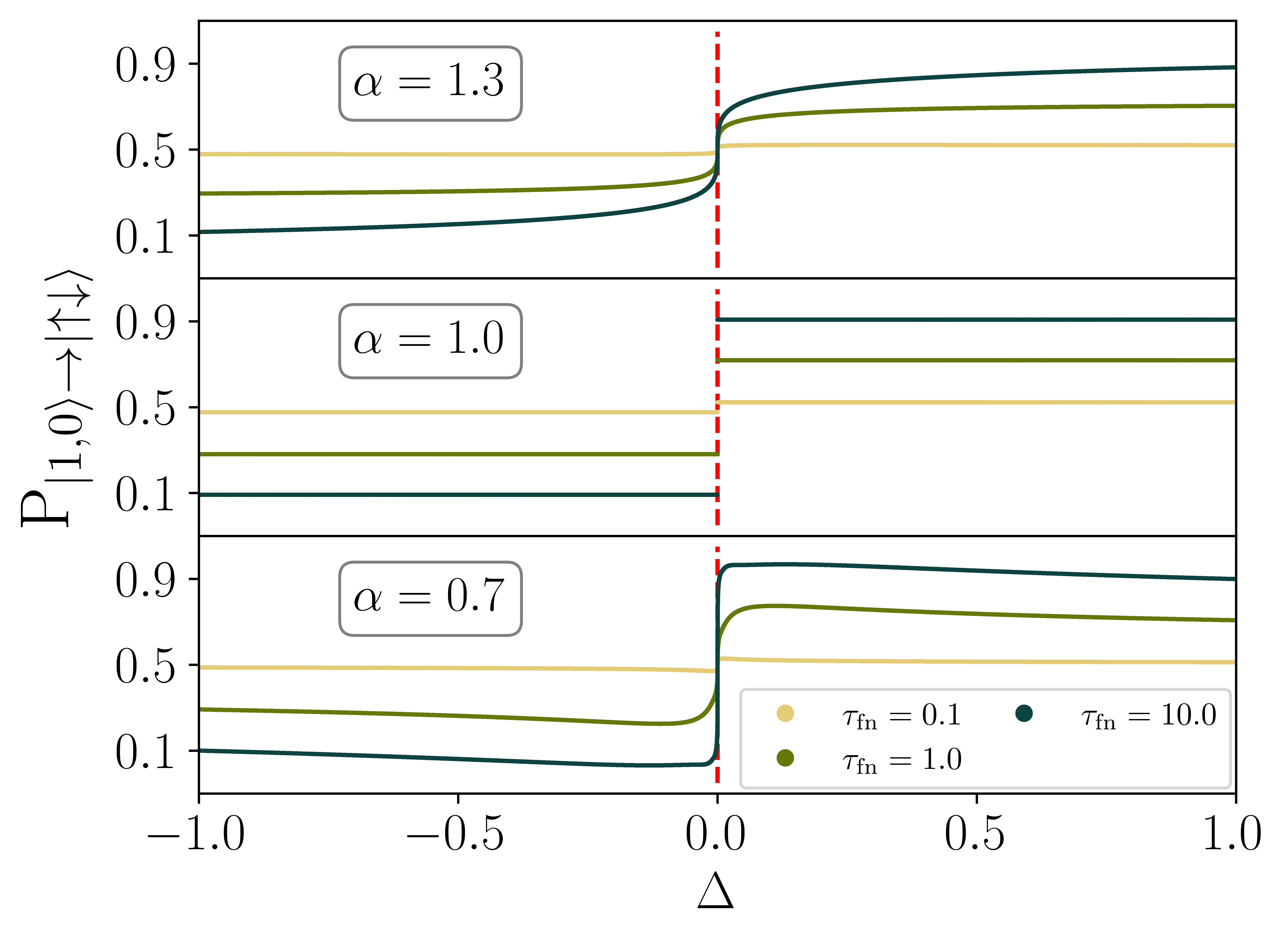}
\caption{Transition probabilities of \eqref{eq:N2_spin-1/2_LZ_Hamiltonian} as a function of $\Delta$. Initial time is set to $\tau_\mathrm{in}=10^{-5}$ and final times correspond to the colors outlined in the legend. For $\alpha=1$, the probability curves are discontinuous for any final time $\tau$ at $\Delta \rightarrow 0$ (indicated by the vertical, dashed line), unlike for $\alpha \neq 1$ as is seen in the upper and lower panels. For this figure, $\nu = 3.0$ and $\text{min}(|\Delta|)=10^{-12}$.}
    \label{fig:Transition_Probabilities}
\end{figure}

It is straightforward to see that $\alpha = 1$ exhibits special behavior in the limit $\Delta \to 0$, which corresponds to merging the two spin-$1/2$ degrees of freedom, as proposed by Gaudin and discussed above. Let us initialize the system in the triplet state $\ket{1,0}$ at $\tau = 0^+$ and consider the transition probability $P_{\ket{1,0} \rightarrow \ket{\uparrow\downarrow}}$ at $\tau \gg 1$. The curves in Fig~\ref{fig:Transition_Probabilities} display a sharp discontinuity at $\Delta = 0$ for $\alpha = 1\,$---a feature absent when $\alpha \neq 1$. We note that $\alpha = 1$ corresponds to the integrable model~~\cite{sinitsyn_integrable_2018,yuzbashyan_integrable_2018,zabalo_nonlocality_2022,barik_knizhnik-zamolodchikov_2024}. Thus, the observed non-analytic behavior in this finite-size system is a signature of integrability and persists for any $\nu > 0$. To see this explicitly, consider two scenarios:
\begin{enumerate}[(A)]
\item Evolve the system and then take the limit $\Delta \to 0$.
\item Set $\Delta = 0$ first and then evolve the system.
\end{enumerate}

In scenario (B), the Hamiltonian in Eq.~\eqref{eq:N2_spin-1/2_LZ_Hamiltonian_1} is proportional to $\sigma^z$. Since the initial state is an eigenstate of $\sigma^z$, the system remains in that state throughout the evolution from any given initial time, yielding $P_{\ket{1,0} \rightarrow \ket{\uparrow\downarrow}} = \frac{1}{2}$.

In scenario (A) we initialize the system in the triplet state at a small $\tau_\mathrm{in}$ and evolve it to a large $\tau_\mathrm{fn}$. For $\alpha = 1$, the transition probabilities in scenario (A) are known explicitly for any $\tau_\mathrm{in}$ and $\tau_\mathrm{fn}$~\cite{barik_knizhnik-zamolodchikov_2024}. These depend only on the sign of $\Delta$ [as already evident from Eqs.~\eqref{eq:N2_spin-1/2_LZ_Hamiltonian} and~\eqref{nstSch}], and $P_{\ket{1,0} \rightarrow \ket{\uparrow\downarrow}}$ is greater than $1/2$ for any $\Delta > 0$ and less than $1/2$ for any $\Delta < 0$, at any finite $\nu$. Therefore, there is a discontinuity at $\Delta = 0$ for $\alpha = 1$.

Now consider scenario (A) for $\alpha \ne 1$. At small $|\Delta|$, the first term on the right-hand side of Eq.~(\ref{eq:N2_spin-1/2_LZ_Hamiltonian}) dominates for $\alpha > 1$ and becomes negligible for $\alpha < 1$, so the system effectively evolves with a Hamiltonian proportional to $\sigma^z$ or $\sigma^x$, respectively. In either case, $P_{\ket{1,0} \rightarrow \ket{\uparrow\downarrow}}$ tends to $1/2$ in the limit $\Delta \to 0$. Hence, the transition probability is a continuous function of $\Delta$ for all $\alpha \ne 1$.

\section{The Spin-1 Model}\label{sec:Spin-1_Model}
We begin with the spin-1 time-dependent Richardson-Gaudin  Hamiltonian
\begin{equation}
\label{eq:mainBCS}
\hat{H}(t) = 2\sum_{j=1}^{N} \varepsilon_{j} \hat{s}_{j}^{z} - g(t) \sum_{j,k=1}^{N} \hat{s}_{j}^{+} \hat{s}_{k}^{-},
\end{equation}
where the time-dependent coupling, 
\begin{equation}
g(t) = \frac{1}{\nu t}
\end{equation} is taken to be identical to that in the spin-$1/2$ case studied in~\cite{zabalo_nonlocality_2022}, and $N$ is the number of spins. The operators $\hat{s}^{z}$, $\hat{s}^{+}$, and $\hat{s}^{-}$ are spin-1 operators that satisfy the standard $\mathfrak{su}(2)$ algebra, with each single-site Hilbert space spanned by the eigenstates of $\hat{s}^{z}$: ${ \ket{-1}, \ket{0}, \ket{1} }$.

Since Eq.\eqref{eq:mainBCS} conserves the $z$-component of the total spin,
\begin{equation}
\boldsymbol{\hat{J}} = \sum_{k=1}^{N} \hat{\boldsymbol{s}}_{k},
\end{equation}
where $\hat{\boldsymbol{s}}_{k} = (\hat{s}^{x}_{k}, \hat{s}^{y}_{k}, \hat{s}^{z}_{k})$ and $[\hat{H}, \hat{J}^{z}] = 0$, the Hamiltonian can be block-diagonalized into sectors labeled by different values of $J^{z}$. In the remainder of the text, we refer to the first (second) term in Eq.~\eqref{eq:mainBCS} as the Zeeman (interaction) term, respectively.

\subsection{Ground state of the model}
At the initial time $t = 0^{+}$, up to a divergent multiplicative constant, the Hamiltonian~\eqref{eq:mainBCS} becomes $\hat{H}_{\text{int}} \propto \boldsymbol{\hat{J}}^{+} \boldsymbol{\hat{J}}^{-}$. After fixing the total magnetization $J^{z}$, the lowest energy state in a given $N_+ = N + J^{z}$ sector is a symmetric superposition of all states obtained by applying $N_+$ raising operations to the pseudo-vacuum $\ket{\odot} \equiv \ket{-1}^{\otimes N}$. This state is given, up to normalization, by
\begin{equation}
\label{eq:bcss1gs}  
\begin{aligned}
\ket{\Psi^{\scale[0.55]{(N_+)}}\scale[0.85]{(0)}} \propto& \;\left(\boldsymbol{\hat{J}}^{+}\right)^{\!N_{+}} \ket{\odot} \\
=& \smashoperator{\sum_{\substack{\left|\{\alpha\}\right| + 2\left|\{\beta\}\right| \\ = N_+}}} 2^{\frac{N_1}{2}} \ket{\{\alpha\}} \ket{\{\beta\}} \ket{\oslash},
\end{aligned}
\end{equation}
where $\ket{\{\alpha\}}$ and $\ket{\{\beta\}}$ are product states composed of $\ket{0}$ at positions $\{\alpha\} \equiv (\alpha_{1}, \dots, \alpha_{N_{1}})$, and $\ket{+1}$ at positions $\{\beta\} \equiv (\beta_{1}, \dots, \beta_{N_{2}})$, while $\ket{\oslash}$ is a product of $\ket{-1}$ over all remaining sites. Here, $\alpha_1 < \ldots < \alpha_{N_1}$ and $\beta_1 < \ldots < \beta_{N_2}$, and we refer to the $\alpha$ ($\beta$) sites as singly (doubly) raised. The sum includes \textit{all possible ordered sets} $\{\alpha\}$ and $\{\beta\}$ satisfying $N_+ = \left|\{\alpha\}\right| + 2\left|\{\beta\}\right|$.

Alternatively, one may express $\ket{\Psi^{\scale[0.55]{(N_+)}}\scale[0.85]{(0)}}$ in Eq.~\eqref{eq:bcss1gs} as a projected spin-1 BCS state:
\begin{equation}
\label{eq:BCSgsspin1}
\ket{\Psi^{\scale[0.55]{(N_+)}}\scale[0.85]{(0)}} \propto P_{N_{+}} \bigotimes_{j=1}^{N} \left( \ket{-1} + \sqrt{2}\ket{0} + \ket{1} \right),
\end{equation}
where $P_{N_{+}}$ is the projector onto states with fixed $J^{z}$.  

In what follows, we refer to this lowest energy state (with fixed $N_+$) simply as the ground state.

\subsection{The formal solution of the wavefunction}\label{sec:FormalSolutionSpin-1}
\begin{figure}
\centering
\includegraphics[width=\linewidth]{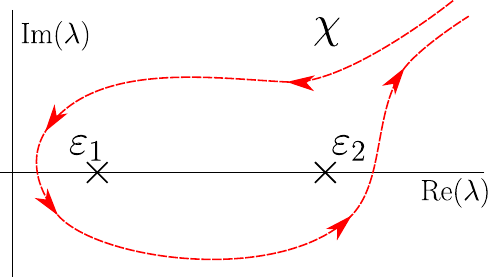}
\caption{ A sample contour $\chi$ (red dashed line) for the integral in Eq.~\eqref{eq:formalwavesolution} for a spin-1 BCS Hamiltonian with two sites.  This contour encircles the poles of the action, located at the on-site Zeeman fields $\varepsilon_{1,2}$. In general, the number of contours is determined by $N_+$. For a detailed example of a spin-$1/2$ solution with $N = 2$ involving contour integrals, see~\cite{barik_knizhnik-zamolodchikov_2024}.}
\label{fig:Contour_Saddle_Schematic}
\end{figure}
In this work, we use the solution of the non-stationary Schr\"odinger equation for the Hamiltonian~\eqref{eq:mainBCS}, derived in~\cite{yuzbashyan_integrable_2018,zabalo_nonlocality_2022} by means of the off-shell Bethe ansatz method. This method was pioneered in~\cite{babujian_off-shell_1993,babujian_off-shell_1994,babujian_generalized_1998} and further developed in~\cite{Sedrakyan_2010,fioretto_exact_2014}. The general solution   $\ket{\Psi^{\scale[0.55]{(N_+)}}\scale[0.85]{(t)}}$ for arbitrary spin-$s$ and total number of raising operations $N_{+}$ is given by an $N_{+}$\,--\,fold contour integral over variables $\lambda_{1}, \dots, \lambda_{N_{+}}$.
\begin{equation}
\label{eq:formalwavesolution}
\ket{\Psi^{\scale[0.55]{(N_+)}}\scale[0.85]{(t)}} =\oint_\chi d \boldsymbol{\lambda} \exp \left(-\frac{i \mathcal{S}(\boldsymbol{\lambda}, \varepsilon, t)}{\nu}\right) \Xi(\boldsymbol{\lambda}, \varepsilon),
\end{equation}
\\
where $\varepsilon=\left(\varepsilon_{1},\dots,\varepsilon_{N}\right)$, $\boldsymbol{\lambda}=(\lambda_{1},\dots,\lambda_{N_{+}})$, $d\boldsymbol{\lambda}=d\lambda_{1}\dots d\lambda_{N_{+}}$ and
\begin{equation}
\label{eq:contourstate}
\Xi(\boldsymbol{\lambda}, \varepsilon)=\prod_{\lambda_r \in \boldsymbol{\lambda}} \hat{L}^{+}\left(\lambda_r\right)\ket{\odot},\; \hat{L}^{+}(\lambda_r)=\sum_{j=1}^N \frac{\hat{s}_j^{+}}{\lambda_r-\varepsilon_j} .
\end{equation}
The quantity $\mathcal{S}(\boldsymbol{\lambda},\varepsilon,t)$, known as the Yang-Yang action,  reads
\begin{equation}
\label{eq:yangyangaction}
\begin{split}
\mathcal{S}(\boldsymbol{\lambda}, \varepsilon, t)=2 \nu t &\sum_{\alpha} \lambda_\alpha+2 s \sum_{j=1}^{N} \sum_{\alpha} \ln \left(\varepsilon_j-\lambda_\alpha\right)
\\&-\sum_{\alpha} \sum_{\beta \neq \alpha} \ln \left(\lambda_\beta-\lambda_\alpha\right).
\end{split}
\end{equation}
For the remainder of this section, we  set $s = 1$.  Higher spins are considered in Sec.~\ref{sec:Higher_spin}. The choice of the contour $\chi$ in \eqref{eq:formalwavesolution} should be such that the integrand is single valued and $\ket{\Psi^{\scale[0.55]{(N_+)}}\scale[0.85]{(t)}}$ satisfies the initial condition $\ket{\Psi^{\scale[0.55]{(N_+)}}\scale[0.85]{(0)}}$ in \eqref{eq:bcss1gs}. An example of such a contour for  $N=2$ is shown in Fig. \ref{fig:Contour_Saddle_Schematic}, see also \cite{barik_knizhnik-zamolodchikov_2024}.

\subsection{Saddle point analysis}
We use the formal solution~\eqref{eq:formalwavesolution} of the model to determine the long-time asymptotic wavefunction $\ket{\Psi_{\scale[0.75]{\infty}}^{\scale[0.55]{(N_+)}}\scale[0.85]{(t)}}$.

Specifically, we apply the saddle point method~\cite{fedoryuk_asymptotic_1989} to evaluate the integral~\eqref{eq:formalwavesolution} in the limit $t \rightarrow \infty^{+}$. In this regime, the integral localizes around the stationary points of the Yang-Yang action, determined by the equations $\partial \mathcal{S} / \partial \lambda_{p} = 0$, which take the form
\begin{equation}
\label{eq:betheeq}
\nu t +  \sum_{j=1}^{N} \frac{1}{\lambda_p-\varepsilon_j}=\sum_{j \neq p} \frac{1}{\lambda_p-\lambda_j},\;p=1, \dots, N_{+}.
\end{equation}
These equations are known as the Richardson equations~\cite{richardson_application_1963,richardson_exact_1964,richardson_pairing_1977}, which play a central role in computing the exact spectrum of time-independent RG Hamiltonians. In our context, they determine the instantaneous spectrum at time $t$. We also note that at $t = 0^{+}$, all $\lambda_{p}$ associated with the ground state diverge as $(\nu t)^{-1}$~\cite{yuzbashyan_strong-coupling_2003}. This behavior implies that the contour $\chi$ must be deformable toward infinity in order to enclose all $\varepsilon_{j}$ without encountering any essential singularities.

As $t\to\infty$, each $\lambda_{p}$ tends toward one of the $\varepsilon_{p}$. Moreover,   the $\frac{1}{\lambda_p-\varepsilon_p}$ term in~\eqref{eq:betheeq}  must compensate the divergent $\nu t$ term.  This suggests the following ansatz for the solution:
\begin{equation}\label{eq:betheeq_ansatz}
\lambda_{p} = \varepsilon_{p} + \frac{z_{p}}{t},
\end{equation}
which we substitute into Eq.~\eqref{eq:betheeq} and solve for $z_{p}$.

In this limit, the raising operator $L^+(\lambda_p)\to \frac{\hat{s}_p^+}{\lambda_p-\varepsilon_p}$. 
In the spin-1 problem,  two raising operations are allowed per site---namely, $\ket{-1} \rightarrow \ket{0}$ and $\ket{0} \rightarrow \ket{1}$. Therefore, at most two $\lambda_p$ can converge to the same $\varepsilon_p$. By contrast, in the spin-$1/2$ case, only a single raising operation ($\ket{\downarrow} \rightarrow \ket{\uparrow}$) is permitted, and hence each $\lambda_p$ approaches a unique $\varepsilon_p$.
Thus, for the spin-1 problem, we first identify the pairings $(\lambda_{q}^{+}, \lambda_{q}^{-})$ that converge to the same $\varepsilon_{q}$, while the remaining $\lambda_{p}^{\circ}$ each converge to  distinct $\varepsilon_{p}$.

\begin{figure}
\centering
\includegraphics[width=\linewidth]{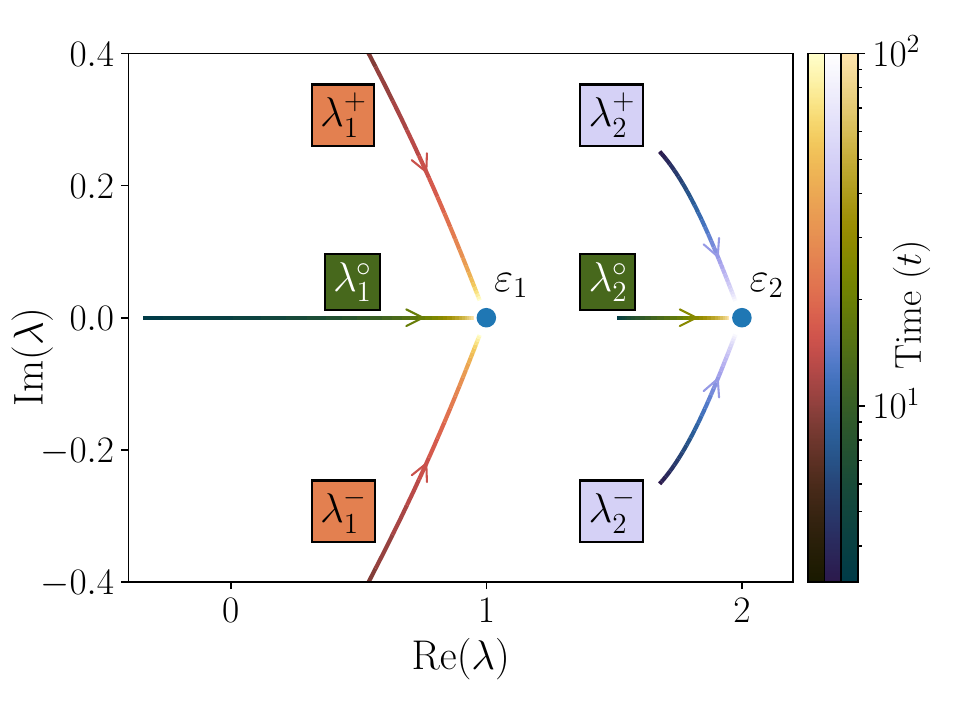}
\caption{Simulation of the saddle point coordinates $\lambda_{p,q}^{0,\pm}$, as determined by Eq.~\eqref{eq:betheeq}, approaching the poles at $\varepsilon_{1,2}$ at large time for the Hamiltonian~\eqref{eq:mainBCS} with $N = 2$, $N_+ = 2$, $\nu = 0.2$, and $\varepsilon_{1,2} = \{1, 2\}$. The color gradient along each curve represents time, while the different color schemes highlight distinct saddle points---three in this case: $(\lambda_1^\circ, \lambda_2^\circ)$, $(\lambda_1^+, \lambda_1^-)$, and $(\lambda_2^+, \lambda_2^-)$. The late-time asymptotic wavefunction is given by the sum over contributions from all three saddle points.}
\label{fig:lambda_Saddle_Schematic}
\end{figure}

Using Eqs.~\eqref{eq:betheeq} and~\eqref{eq:betheeq_ansatz}, we determine the large-$t$ behavior of the saddle points as follows:
\begin{equation}
\label{eq:sollambda}
\lambda_{p}^{\circ} = \varepsilon_{p} - \frac{1}{\nu t},\;\; \lambda^{\pm}_{q} = \varepsilon_{q} - \frac{(1 \pm i)}{2\nu t},
\end{equation}
where the index $p$ corresponds to singly raised basis states and $q$ to doubly raised ones.

A simulation of these saddle points evolving in time according to Eq.~\eqref{eq:betheeq} for a simple $N = 2$, $N_+ = 2$ (i.e., $J^z = 0$) case is shown in Fig.~\ref{fig:lambda_Saddle_Schematic}. The basis states that form the initial condition (the $t = 0^+$ ground state in this $J^z$ sector) are $\ket{1,-1}$, $\ket{-1,1}$, and $\ket{0,0}$. The corresponding saddle points are $(\lambda_1^+, \lambda_1^-)$, $(\lambda_2^+, \lambda_2^-)$, and $(\lambda_1^\circ, \lambda_2^\circ)$, respectively. Note that $N_+$ sets the dimension of saddle point coordinates.

Using the general form of the saddle points in Eq.~\eqref{eq:sollambda}, we evaluate  the state~\eqref{eq:contourstate} for $t \rightarrow \infty^{+}$ as
\begin{equation}
\Xi(\boldsymbol{\lambda}, \varepsilon) = t^{N_{1} + 2N_{2}} (-\nu)^{N_{1}} (2\nu^{2})^{N_{2}} \ket{\{\alpha\}} \ket{\{\beta\}} \ket{\oslash}.
\end{equation}
Similarly, the determinant of the Hessian matrix $\mathcal{S}'' = [\partial^{2}\mathcal{S}/\partial\lambda_{p}\partial\lambda_{q}]\big|_{t \rightarrow \infty}$ evaluates to
\begin{equation}
\det \mathcal{S}''(\boldsymbol{\lambda}, \varepsilon, t) \approx t^{2(N_{1} + 2N_{2})} (-2\nu^{2})^{N_{1}} (4\nu^{2})^{2N_{2}},
\end{equation}
where we have neglected terms that vanish in the infinite-time limit. Thus, we conclude that
\[
\frac{\Xi(\boldsymbol{\lambda}, \varepsilon)}{\sqrt{\det \mathcal{S}''}} \propto \ket{\{\alpha\}} \ket{\{\beta\}} \ket{\oslash}.
\]
This expression is written up to global prefactors, which we omit in the solution.

Now we are left with evaluating the Yang-Yang action at the stationary points. By substituting Eq.~\eqref{eq:sollambda} into Eq.~\eqref{eq:yangyangaction} and neglecting terms of order $t^{-1}$, we find
\begin{equation}\label{eq:YYaction_spin1_unsimplified}
\begin{split}
&\mathcal{S}_{\{\alpha\}\{\beta\}} = 2\smashoperator[l]{\sum_{i\in \{\alpha\}}}\sum_{\substack{j=1 \\j\neq i}}^{N}l_{ji}(\varepsilon)+ 4\smashoperator[l]{\sum_{i\in \{\beta\}}}\sum_{\substack{j=1 \\j\neq i}}^{N}l_{ji}(\varepsilon) \\
&-\smashoperator{\sum_{\substack{i,j\in \{\alpha\}\\j \neq i}}}l_{ji}(\varepsilon) - 2\smashoperator{\sum_{\substack{i\in \{\alpha\}\\j\in \{\beta\}}}}l_{ji}(\varepsilon) - 2\smashoperator{\sum_{\substack{i\in \{\beta\}\\j\in \{\alpha\}}}}l_{ji}(\varepsilon) \\
&-4\smashoperator{\sum_{\substack{i,j\in \{\beta\}\\j \neq i}}}l_{ji}(\varepsilon) +2\nu t\smashoperator{\sum_{i\in\{\alpha\}}}\varepsilon_{i}+4\nu t\smashoperator{\sum_{i\in\{\beta\}}}\varepsilon_{i}\\
&-N_1[1+\ln(\nu t)] -N_2\ln(4),
\end{split}
\end{equation}
where we used $l_{ji}(\varepsilon)\equiv\ln\left(\varepsilon_{j}-\varepsilon_{i}\right)$ as a shorthand and dropped a global prefactor. 

We further simplify the first two terms on the right-hand side of Eq.~\eqref{eq:YYaction_spin1_unsimplified} as follows:
\begin{equation}\label{eq:YYaction_spin1_simplification_example}
\begin{split}
2\smashoperator[l]{\sum_{i\in \{\alpha\}}}\sum_{\substack{j=1\\j\neq i}}^{N}l_{ji}(\varepsilon) &\rightarrow 2\smashoperator{\sum_{k\in \{\alpha\}}}\Biggl(
-ik\pi + \sum_{\substack{j=1\\j\neq k}}^{N}l_{jk}|\varepsilon| \Biggr), \\
4\smashoperator[l]{\sum_{i \in \{\beta\}}}\sum_{\substack{j=1\\j\neq i}}^{N}l_{ji}(\varepsilon) &\rightarrow 4\smashoperator{\sum_{k \in \{\beta\}}}\Biggl(-ik\pi +\sum_{\substack{j=1\\j\neq i}}^{N}l_{jk}|\varepsilon| \Biggr), 
\end{split}    
\end{equation}
where $l_{ji}|\varepsilon|\equiv\ln|\varepsilon_{j}-\varepsilon_{i}|$. Here, we choose the branch of the logarithm such that $\ln(-1) = \ln(e^{-i\pi}) = -i\pi$, with the $i k \pi$ terms in the above expression arising from counting the number of $\varepsilon_{j} < \varepsilon_{k}$. This branch cut choice is made for the same reason as in~\cite{zabalo_nonlocality_2022}---namely, that in the $\nu \rightarrow 0^+$ limit (which we will refer to as the `slow-ramp' or `adiabatic' limit), the system must remain in the ground state as $t \rightarrow \infty^+$. Each term contributes to $\ln(-1)$ a total of $k - 1$ times, and replacing $k - 1 \rightarrow k$  only modifies a global prefactor.

After simplifying the remaining terms in $\mathcal{S}_{\{\alpha\}\{\beta\}}$ using the same approach as in Eq.~\eqref{eq:YYaction_spin1_simplification_example}, and again neglecting global prefactors that affect only the norm and overall phase of the wavefunction, we finally arrive at
\begin{equation}
\label{eq:yyactionsaddle}
\begin{aligned}
&\mathcal{S}_{\{\alpha\}\{\beta\}} = 2\nu t\smashoperator{\sum_{i\in\{\alpha\}}}\varepsilon_{i}+4\nu t\smashoperator{\sum_{i\in\{\beta\}}}\varepsilon_{i} -2i\pi\smashoperator{\sum_{k\in\{\alpha\}}} k \\
&-4i\pi\smashoperator{\sum_{k\in\{\beta\}}} k+2\smashoperator[l]{\sum_{i\in \{\alpha\}}}\sum_{\substack{j=1\\j\neq i}}^{N}l_{ji}|\varepsilon|+4\smashoperator[l]{\sum_{i\in \{\beta\}}}\sum_{\substack{j=1\\j\neq i}}^{N}l_{ji}|\varepsilon|
\\
&-2\smashoperator{\sum_{\substack{i<j\\i,j\in\{\alpha\}}}}l_{ji}|\varepsilon|-4\smashoperator{\sum_{\substack{i\in\{\alpha\}\\j\in\{\beta\}}}}l_{ji}|\varepsilon|-8\smashoperator{\sum_{\substack{i<j\\i,j\in\{\beta\}}}}l_{ji}|\varepsilon|\\
&-N_1\left[1+\ln(\nu t/2)-i\pi/2\right].
\end{aligned}
\end{equation}
A compact way to write this expression is 
\begin{equation}
\begin{split}
\mathcal{S}_{\{\alpha\}\{\beta\}} = &2\nu t\smashoperator{\sum_{i\in\{\alpha\}}}\varepsilon_{i}+4\nu t\smashoperator{\sum_{i\in\{\beta\}}}\varepsilon_{i} -2i\pi\smashoperator{\sum_{k\in\{\alpha\}}} k \\
&-4i\pi\smashoperator{\sum_{k\in\{\beta\}}} k -2\smashoperator{\sum_{\substack{i,j=1,\;j>i}}^{N}}\hat{s}_{i}^{z}\hat{s}_{j}^{z}\ln|\varepsilon_j-\varepsilon_i|\\
&-N_1\left[1+\ln(\nu t/2)-i\pi/2\right],
\end{split}
\end{equation}
which is equivalent to \eqref{eq:yyactionsaddle} up to a prefactor. This prefactor  depends only on $N$ and $J^{z}$ and is identical for all $\mathcal{S}_{\{\alpha\}\{\beta\}}$ and thus is safely absorbed in the overall normalization of the wavefunction. 

The asymptotic wavefunction is a sum over all  stationary points,
\begin{equation}
\ket{\Psi_{\scale[0.75]{\infty}}^{\scale[0.55]{(N_+)}}\scale[0.85]{(t)}} = \sum_{\{\alpha\},\{\beta\}}e^{-\frac{i\mathcal{S}_{\{\alpha\}\{\beta\}}}{\nu}}\ket{\{\alpha\}}\ket{\{\beta\}}\ket{\oslash}.   
\end{equation}
Using \eqref{eq:yyactionsaddle}, we obtain 
\begin{equation}
\label{eq:s1asympsol}
\ket{\Psi_{\scale[0.75]{\infty}}^{\scale[0.55]{(N_+)}}\scale[0.85]{(t)}} = \smashoperator{\sum_{\substack{N_1+2N_2\\=N_{+}}}}e^{-\gamma^\times_{N_1}}\smashoperator{\sum_{\substack{\text{\tiny$\left|\{\alpha\}\right|$=$N_1$}\\\text{\tiny$\left|\{\beta\}\right|$=$N_2$}}}}e^{i\Lambda_{\text{\tiny$\{\alpha\}\{\beta\}$}}}\zeta^{{\text{\tiny$\{\alpha\}$}}}_{_{\text{\tiny$\{\beta\}$}}}\ket{\{\alpha\}}\ket{\{\beta\}}\ket{\oslash},
\end{equation}
\begin{figure}[t]
\centering
\includegraphics[width=\linewidth]{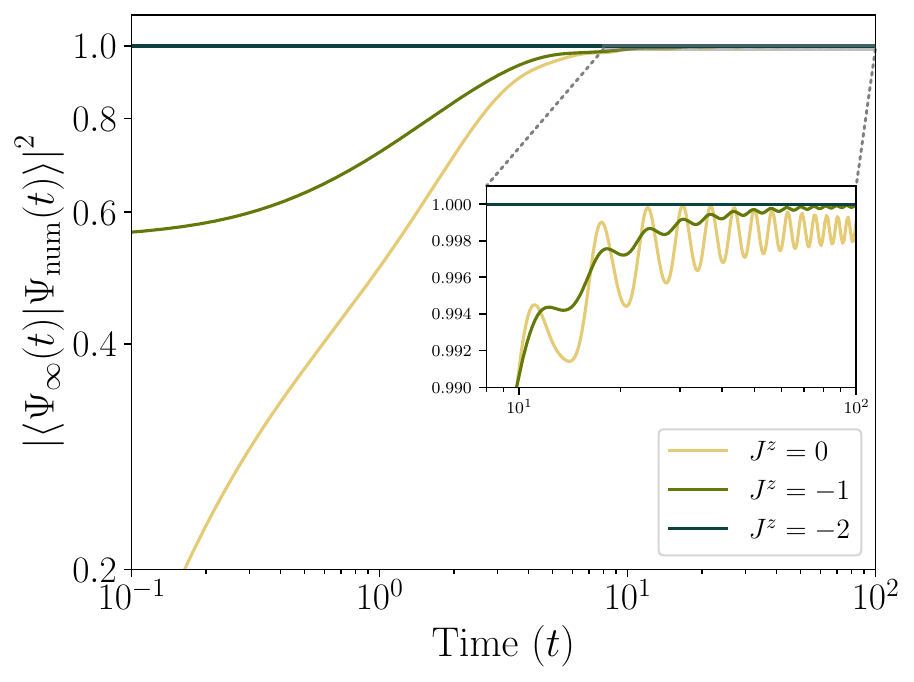}
\caption{Overlap between the asymptotic wavefunction $\ket{\Psi_\infty(t)}$ from Eq.~\eqref{eq:s1asympsol} and the numerically simulated wavefunction $\ket{\Psi_{\text{num}}(t)}$, for various magnetization sectors $J^z$ with $N = 2$, $\nu = N$, and $\varepsilon_i = i/N$. As $t \rightarrow \infty$, the overlap must converge to unity; the fact that it does not indicates that the asymptotic wavefunction, as written, is incorrect. The inset shows that the $J^z = -2, -1$ sectors appear to match correctly, in contrast to the $J^z = 0$ sector. All simulations use an initial cutoff time $t_{\text{init}} = 10^{-5}$  in the interaction term  to avoid numerical singularities. Varying  $t_{\text{init}}$ does not affect the qualitative features of the numerical results.}
\label{fig:overlap_incorrect}
\end{figure}
where,
\begin{subequations}
\begin{equation}
\nu \Lambda_{\{\alpha\}\{\beta\}}=
2\smashoperator{\sum_{\substack{i<j\\i,j\in\{\alpha\}}}}l_{ji}|\varepsilon|+4\smashoperator{\sum_{\substack{i\in\{\alpha\}\\j\in\{\beta\}}}}l_{ji}|\varepsilon|+8\smashoperator{\sum_{\substack{i<j\\i,j\in\{\beta\}}}}l_{ji}|\varepsilon|,
\end{equation}
\begin{equation}
\zeta^{{\text{\tiny$\{\alpha\}$}}}_{_{\text{\tiny$\{\beta\}$}}} = \smashoperator{\prod_{j\in\{\alpha\}}}e^{-2it\varepsilon_{j}-\frac{2\pi j}{\nu}-i2\theta_{j}}\smashoperator{\prod_{k\in\{\beta\}}}e^{-4it\varepsilon_{k}-\frac{4\pi k}{\nu}-2i\theta_{k}},
\end{equation}
\begin{equation}
\label{eq:theta_k}
\theta_{k} = \frac{1}{\nu}\sum_{j\neq k}\ln|\varepsilon_{j}-\varepsilon_{k}|,
\end{equation}
\begin{equation}
\label{eq:gammafunc}
\gamma_{N_1}^{\times} = -N_1\left[\frac{\pi+2i(1+\ln\nu)}{2\nu} +\frac{i}{\nu}\ln\left(\frac{t}{2}\right)\right].    
\end{equation}
\end{subequations}
An important new feature compared to the spin-$1/2$ solution is the appearance of $\gamma^\times_{N_1}$, which introduces additional weights and phases to the state $\ket{\{\alpha\}} \ket{\{\beta\}} \ket{\oslash}$ that depend solely on $N_1$ and $\nu$~\footnote{Linear dependence on $N_2$ can be eliminated in favor of $N_1$, since $N_1 = N_{+} - 2N_2 = N + J^z - 2N_2$, and $N + J^z$ is a conserved quantity.}. These weights play a crucial role in ensuring the correct $\nu \rightarrow 0^+$ (slow-ramp) limit. They also lead to qualitative differences from the spin-$1/2$ system, particularly in the steady-state behavior discussed in detail in Sec.~\ref{sec:steadystate}.

\subsection{Correction to the asymptotic wavefunction}
\begin{center}
\begin{figure*}
\centering
\includegraphics[width = 0.8\linewidth]{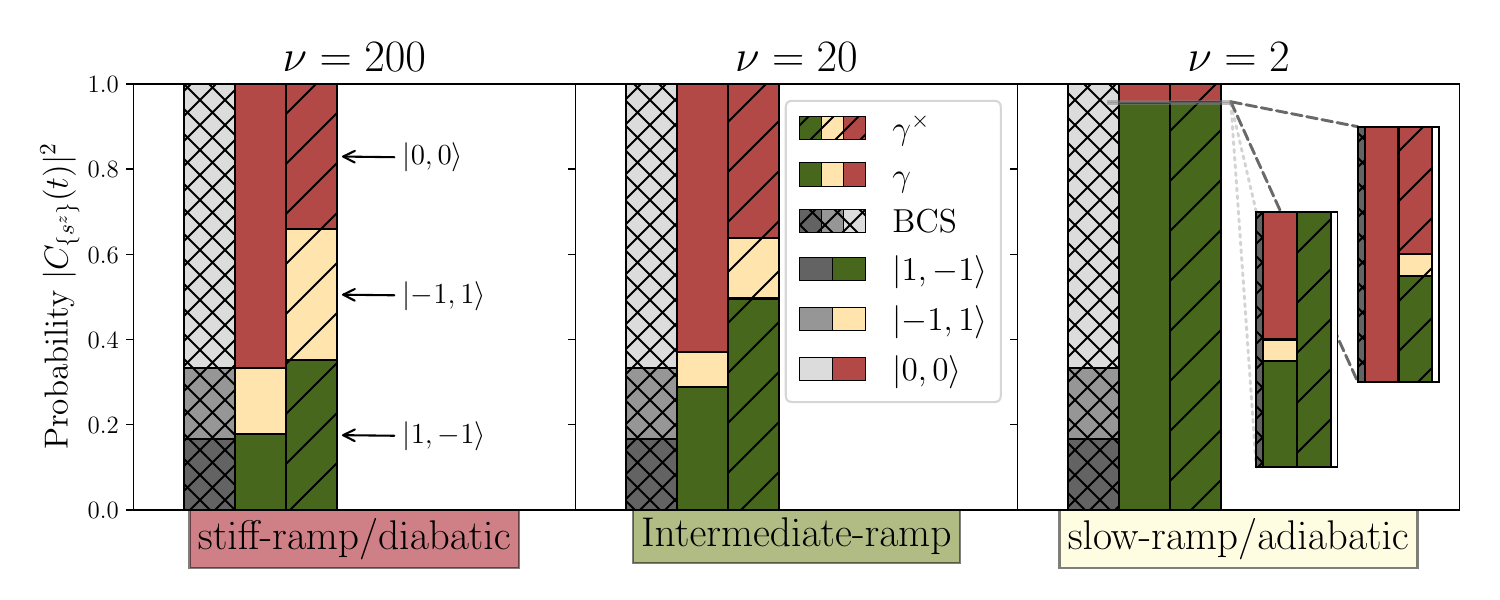}
\caption{Probabilities $|C_{\{s^z\}}(t)|^2$ of the basis states $\ket{1,-1}$, $\ket{0,0}$, and $\ket{-1,1}$ for a two-site spin-$1$ system with fixed total magnetization $J^{z} = 0$, calculated from Eq.~\eqref{eq:s1asympsol} using the incorrect weight $\gamma^{\times}$ from Eq.~\eqref{eq:gammafunc}, and the corrected weight $\gamma$ from Eq.~\eqref{eq:correctedgammafunc}, evaluated at $t = 10^4$. The calculation is performed for different values of $\nu \equiv 2/\eta$ with $\eta \in \{0.01, 0.1, 1\}$, illustrating the crossover from the stiff-ramp (diabatic) limit to the slow-ramp (adiabatic) limit. These results are compared with the probabilities derived from the $t=0^+$ ground state in Eq.~\eqref{eq:bcss1gs}.  A clear discrepancy emerges in the diabatic regime, where the corrected wavefunction reproduces the  ground state probabilities as it should, unlike the one using $\gamma^\times$.}
\label{fig:probablities-s1-N2-Jz0}
\end{figure*}
\end{center}

\begin{figure}
\centering
\includegraphics[width=\linewidth]{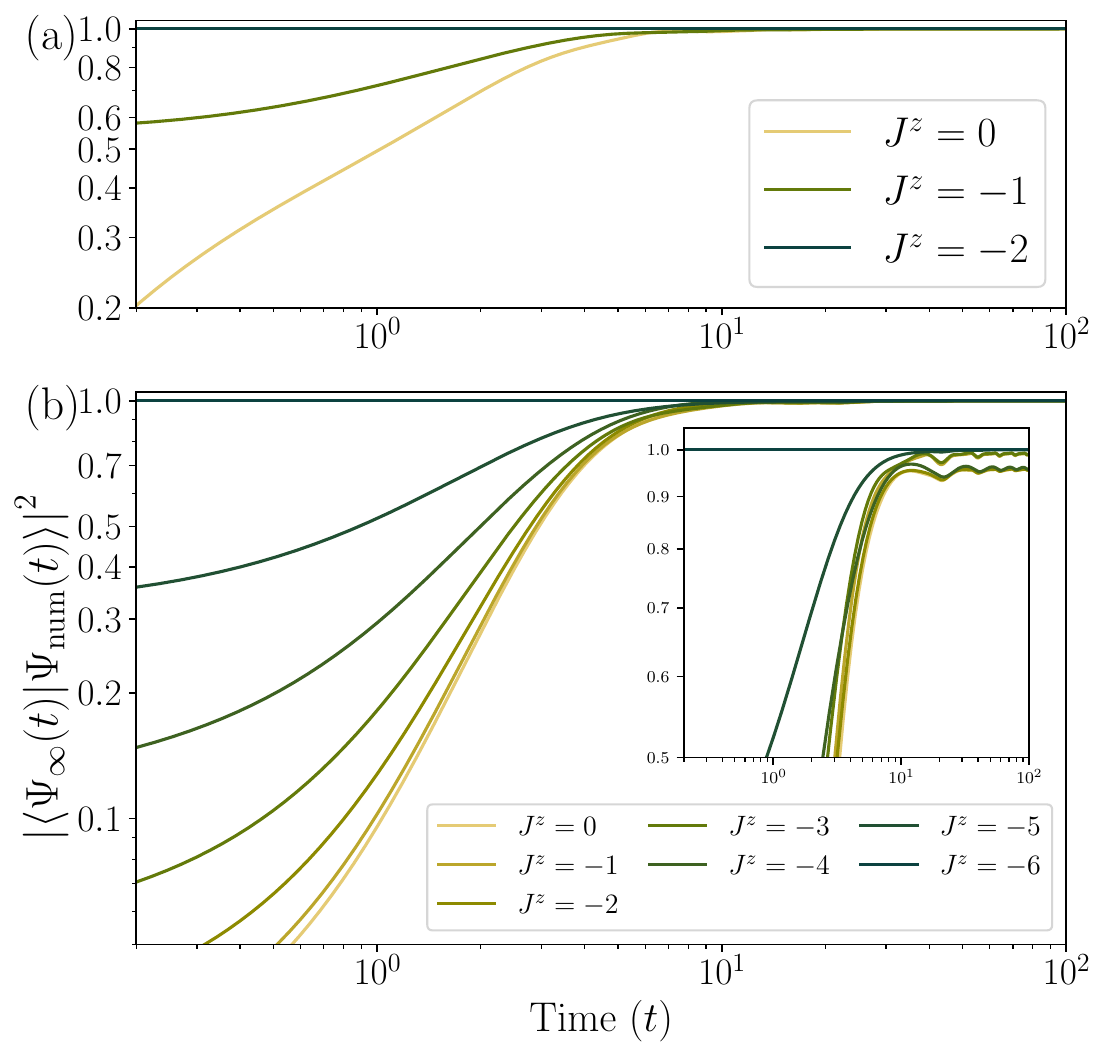}
\caption{Overlap between the asymptotic wavefunction $\ket{\Psi_\infty(t)}$ and the result of direct numerical simulation $\ket{\Psi_{\text{num}}(t)}$ for various magnetization sectors $J^z$, using the same system parameters as in Fig.~\ref{fig:overlap_incorrect}. (a) Same setup as Fig.~\ref{fig:overlap_incorrect} with $N=2$, but with the corrected asymptotic wavefunction incorporating $\gamma_{N_1}$ in Eq.~(\ref{eq:correctedgammafunc}). As $t \rightarrow \infty$, the overlap approaches unity. (b) Same as (a) but for $N=6$. The inset shows the result obtained using $\gamma^{\times}_{N_1}$, Eq.~\eqref{eq:gammafunc}, which fails to produce the correct asymptotic behavior.
}
\label{fig:overlap_correct}
\end{figure}

The saddle point calculation performed in the previous subsection does not yield the correct long-time wavefunction if we use Eq.~\eqref{eq:gammafunc} for $\gamma^\times_{N_1}$ in Eq.~\eqref{eq:s1asympsol}. Specifically, calculating the absolute square of the overlap with the numerically simulated wavefunction $\ket{\Psi_{\rm num}(t)}$, i.e., $\left| \braket{\Psi_\infty (t) | \Psi_\text{num}(t)} \right|^{2}$ for different values of $J^{z}$, does not converge to unity at large $t$, as would be expected for a valid asymptotic solution. This discrepancy is illustrated in Fig.~\ref{fig:overlap_incorrect} for $N = 2$, and becomes more pronounced for larger system sizes, as shown in the insets of Fig.~\ref{fig:overlap_correct} and in Figs.~\ref{fig_App:overlap_N=10},~\ref{fig_App:overlap_N=10spin=3/2}, and~\ref{fig_App:N8Spin2}. The subtle mismatch originates from the form of Eq.~\eqref{eq:gammafunc}, as confirmed by comparing each term in the first summation of Eq.~\eqref{eq:s1asympsol} with the corresponding projection of the numerically simulated wavefunction.

We provide simple constraints on the correct form of $\gamma_{N_1}$ by analyzing why Eq.~\eqref{eq:gammafunc} fails. Consider the $\nu \rightarrow \infty^+$ limit with finite $N$, corresponding to the stiff-ramp or diabatic regime. In this limit, the coupling term in Eq.~\eqref{eq:mainBCS} becomes negligible, and the dynamics are governed solely by the Zeeman terms. Time evolution under such a Hamiltonian results in each basis state   acquiring a phase that is linear in time $t$ and depends on the on-site energies $\varepsilon_i$. Meanwhile, real-valued prefactor such as $2^{N_1/2}$ in Eq.~\eqref{eq:bcss1gs} remain unaffected. However, the form of $\gamma^\times_{N_1}$ given in Eq.~\eqref{eq:gammafunc} fails to recover these prefactors, leading to an incorrect asymptotic wavefunction.

The two-site spin-1 ($N = 2,\, s = 1$) RG  model can be solved analytically for any $t$. Using this result, we obtain the correct expression for $\gamma_{N_1}$ as follows: first, we block-diagonalize the Hamiltonian into spin sectors labeled by $J^{z} \in \{-2, -1, 0, 1, 2\}$, resulting in a set of subproblems governed by coupled differential equations. These equations are solved exactly, yielding the time-dependent wavefunctions $\ket{\psi_{J^z}^{\scale[0.7]{(N=2,\,s=1)}}(t)}$, which are provided in Appendix~\ref{sec:Appendix_N=2DiffEqsSol}. Below, we give the large-time asymptotics of $\ket{\psi_{0}^{\scale[0.55]{(2,1)}}(t)}$, which is directly relevant to the present discussion: 
\begin{equation}
\label{eq:correct-s1-N2-Jz0}
\begin{split}
&\lim_{t\rightarrow\infty}\ket{\psi_{0}^{\scale[0.55]{(2,1)}}(t)} = \mathcal{N}t^{\frac{2i}{\nu}}e^{2i(\varepsilon_1+\varepsilon_2)t}e^{\frac{6\pi}{\nu}}\times\\&\bigg[e^{-\frac{4\pi}{\nu}}e^{-4i\varepsilon_{1}t}\ket{1,-1}+\kappa(t)e^{-\frac{6\pi}{\nu}}e^{-2i(\varepsilon_1+\varepsilon_{2})t}\ket{0,0}\\&+e^{-\frac{8\pi}{\nu}}e^{-4i\varepsilon_{2}t}\ket{-1,1}\bigg], 
\end{split}
\end{equation}
where
\begin{subequations}
\begin{equation}
\mathcal{N} =  \frac{\Gamma\left(\frac{1}{2}+\frac{2i}{\nu}\right)\Gamma\left(1+\frac{3i}{\nu}\right)}{\Gamma\left(1+\frac{i}{\nu}\right)\sqrt{6\pi}},
\end{equation}
\begin{equation}\label{eq:spin-1_BCS_asymptotic_wavefunction_correction}
\kappa(t) = \exp\left[\frac{2i}{\nu}\ln\left(\frac{t}{2}\right)\right] \frac{2\pi}{\Gamma\left(\frac{1}{2}+\frac{i}{\nu}\right)^2}.
\end{equation}
\end{subequations}
After comparing \eqref{eq:spin-1_BCS_asymptotic_wavefunction_correction} with \eqref{eq:gammafunc} and using $N_1=2$, we find that only the time-independent component of $
\gamma_{N_1}$ requires modification. Finally, we extract the appropriate expression of $\gamma_{N_1}$:
\begin{equation}
\label{eq:correctedgammafunc}
\gamma_{N_1} = -N_1\left\{\ln\left[\frac{\sqrt{2\pi}}{\Gamma\left(\frac{1}{2}+\frac{i}{\nu}\right)}\right]+\frac{i}{\nu}\ln\left(\frac{t}{2}\right)\right\}.
\end{equation}
The results for the overlap $|\braket{\Psi_\infty(t) | \Psi_{\rm num}(t)}|^2$ computed using the corrected expression converge to unity at long times. This is verified numerically in Fig.~\ref{fig:overlap_correct} for $N \in \{2, 6\}$, and in Appendix~\ref{sec_App:Additional_Figures} for $N = 12$. To further illustrate the effect of our modification, we compare the basis state probabilities of the corrected two-site wavefunction in Eq.~\eqref{eq:correct-s1-N2-Jz0} with those from Eq.~\eqref{eq:s1asympsol}, which employs the incorrect weight from Eq.~\eqref{eq:gammafunc}, as shown in Fig.~\ref{fig:probablities-s1-N2-Jz0}. While the wavefunction in Eq.~\eqref{eq:s1asympsol} always satisfies the slow-ramp (adiabatic) limit, the proposed modification resolves the discrepancy in the stiff-ramp (diabatic) regime, thereby validating our approach.

In the next section, we generalize the results presented here to higher-spin models.  

\section{Higher spin models}\label{sec:Higher_spin}
Similar to the spin-$1$ model, it is also possible to evaluate the asymptotic wavefunction of the spin-$s$ time-dependent RG  model. In this case, we consider general spin-$s$ operators $\hat{s}^{\pm,z}$ in the Hamiltonian~\eqref{eq:mainBCS}. The $t = 0^+$ ground state in a given $J^z$ sector is constructed analogously to Eq.~\eqref{eq:bcss1gs}. The calculations follow those of the spin-$1$ case, with the key difference being that each site in the pseudo-vacuum $\ket{\odot} \equiv \ket{-s}^{\otimes N}$ can be raised at most $2s$ times, transitioning from $\ket{-s}$ to $\ket{+s}$.

\subsection{Ground state}
At $t=0^+$, the   ground state of the spin-$s$ model for the fixed total magnetization $J^{z}$, with $N_+=sN+J^{z}$, is  
\begin{equation}
\label{eq:bcsssgs}  
\begin{aligned}
&\ket{\Psi^{\scale[0.55]{(s,N_+)}}\scale[0.85]{(0)}} \propto \;\left(\boldsymbol{\hat{J}}^{+}\right)^{\!N_{+}}\ket{\odot} \\
&= N_+ !\smashoperator{\sum_{\{\boldsymbol{\alpha}\}}}B_{\boldsymbol{\alpha}}\left(\bigotimes_{\{\alpha^{(q)}\}\in\boldsymbol{\alpha}}\ket{\{\alpha^{(q)}\}}\right)\!\otimes\ket{\oslash}.
\end{aligned}
\end{equation}
Here, $\{\boldsymbol{\alpha}\}$ are all $\boldsymbol{\alpha} \equiv \{\{\alpha^{(1)}\},\dots,\{\alpha^{(2s)}\}\}$ such that $\sum_{q=1}^{2s} q \left|\{\alpha^{(q)}\}\right| = N_+$. Every ordered set $\{\alpha_q\}$ contains the site indices where the vacuum is raised $q$ times. Thus, for instance, for the spin-1 case in the Sec. \ref{sec:Spin-1_Model}, we identify the sets $\{\alpha^{(1)}\} = \{\alpha\}$ and $\{\alpha^{(2)}\} = \{\beta\}$. 
Then, $\left(\bigotimes_{\{\alpha^{(q)}\}\in\boldsymbol{\alpha}}\ket{\{\alpha^{(q)}\}}\right)\otimes\ket{\oslash}$ is the corresponding product state, with $\ket{\oslash}$ being the product of $\ket{-s}$ for all remaining, unraised, sites.  The expression of $B_{\boldsymbol{\alpha}}$ is given by 
\begin{equation}\label{eq:BCS_GS_Partition-Weights}
\hphantom{=}\smashoperator[l]{\prod_{\{\alpha^{(q)}\}\in \boldsymbol{\alpha}}}\left[\!\frac{1}{{q!}}\!\prod_{m=-s}^{-s+q-1}\sqrt{s(s+1)-m(m+1)}\right]^{\!|\{\alpha^{(q)}\}|}.
\end{equation}
The square root terms originate from the matrix elements of the raising operators in the spin-$s$ representations of $\mathfrak{su}(2)$. The $q!$ terms in the denominator account for the over-counting in the $N_+!$ factor, and together these contributions are identified as multinomial coefficients,
\begin{equation}
\frac{N_+!}{\prod_{\{\alpha^{(q)}\}\in \boldsymbol{\alpha}}\,q!^{|\{\alpha^{(q)}\}|}} = \begin{pmatrix}
    N_+\\ \underbrace{1,\dots,1}_{\!|\{\alpha^{(1)}\}|},\underbrace{2,\dots,2}_{\!|\{\alpha^{(2)}\}|},\dots,2s
\end{pmatrix}.
\end{equation}
Note that Eq. 
\eqref{eq:BCS_GS_Partition-Weights} simplifies to
\begin{equation}
B_{\alpha}=\smashoperator[l]{\prod_{\{\alpha^{(q)}\}\in \boldsymbol{\alpha}}}\!\begin{pmatrix}2s\\q\end{pmatrix}^{\!\!\frac{|\{\alpha^{(q)}\}|}{2}},
\end{equation}
which allows us to write $\ket{\Psi_{\scale[0.75]{\infty}}^{\scale[0.55]{(s,N_+)}}\scale[0.85]{(0)}}$ in \eqref{eq:bcsssgs} as a projected spin-$s$ BCS state of the form
\begin{equation}
\label{eq:BCSgsspins}
\ket{\Psi^{\scale[0.55]{(s,N_+)}}\scale[0.85]{(0)}}   \propto P_{N_{+}}\bigotimes_{j=1}^{N}\left(\sum_{q=0}^{2s}\sqrt{\begin{pmatrix}
    2s \\ q   
\end{pmatrix}}\ket{q-s}\right),
\end{equation}
where $P_{N_{+}}$ is a projector onto the states of a fixed total magnetization $J^z=N_+-Ns$.

\subsection{General spin-\texorpdfstring{$s$}{s} asymptotics}
In analogy to Eq.~(\ref{eq:s1asympsol}), we present below the expression for the general asymptotic wavefunction $\ket{\Psi_{\scale[0.75]{\infty}}^{\scale[0.55]{(s,N_+)}}\scale[0.85]{(t)}}$ of the $N$-site spin-$s$ system at fixed total magnetization $J^z$:
\begin{equation}\label{eq:genasympsol}\begin{aligned}\ket{\Psi_{\scale[0.75]{\infty}}^{\scale[0.55]{(s,N_+)}}\scale[0.85]{(t)}} =&\smashoperator{\sum_{\substack{(\sum_{j=1}^{2s}jN_j=N_{+})}}}\!e^{-\gamma_{N_1,\dots,N_{2s-1}}} \times\!\!\\
&\smashoperator{\sum_{\substack{\text{\tiny$\left|\{\alpha^{(j)}\}\right|$=$N_j$}}}}e^{i\Lambda}\zeta\left(\smashoperator[r]{\bigotimes_{\{\alpha^{(j)}\}\in\boldsymbol{\alpha}}}\ket{\{\alpha^{(j)}\}}\!\!\right)\!\otimes\ket{\oslash},
\end{aligned}
\end{equation}
where
\begin{subequations}
\begin{equation}
\nu \Lambda =
2\sum_{q=1}^{2s}q^2\smashoperator{\sum_{\substack{i<j\\i,j\in\{\alpha^{(q)}\}}}}l_{ji}|\varepsilon|+2\smashoperator{\sum_{\substack{p,q=1\\p<q}}^{2s}}pq\smashoperator{\sum_{\substack{i\in\{\alpha^{(p)}\}\\j\in\{\alpha^{(q)}\}}}}l_{ji}|\varepsilon|,
\end{equation}
\begin{equation}
\zeta =\prod_{q=1}^{2s}\smashoperator[r]{\prod_{j\in\{\alpha^{(q)}\}}}e^{-2qit\varepsilon_{j}-\frac{2sq\pi j}{\nu}-i2q\theta_{j}},
\end{equation}
\begin{equation}
\theta_{k} = \frac{s}{\nu}\sum_{j\neq k}\ln|\varepsilon_{j}-\varepsilon_{k}|,   
\end{equation}
\begin{equation}
\label{eq:genasympsolgamma}
\begin{split}
\gamma_{N_1,\dots,N_{2s-1}} &= \sum_{j=1}^{2s-1}\omega_{j}+\frac{i}{\nu}\ln(t)\times\\
&\Bigg[2\smashoperator{\sum_{k=2}^{2s-1}}\,\begin{pmatrix}k\\2\end{pmatrix}N_{k}-(2s-1)\sum_{k=1}^{2s-1}kN_k\Bigg].
\end{split}
\end{equation}
\end{subequations}

These solutions remain incomplete because the general form of $\omega_j$ in Eq.~(\ref{eq:genasympsolgamma}) has not yet been fully determined. These $\omega_j$ constitute the time-independent components of the relative weights $e^{-\gamma}$ in Eq.~\eqref{eq:genasympsol} for each partition within a given magnetization sector. The correct form of $\omega_j$ must be consistent with both the slow-ramp (adiabatic) and stiff-ramp (diabatic) limits of the model, constraining it to depend on $\nu$ and $N_j$, as previously discussed for the spin-1 wavefunction. 

Although the saddle point calculation provides candidate expressions for these terms, further modifications are required. These can be obtained, for example, by independently solving the two-site spin-$s$ problem. At present, exact solutions exist only for $s = 1/2$, $1$, and $3/2$, as well as for the $J^z = \pm(-2s+1)$ and $J^z = \pm(-2s+2)$ sectors for arbitrary spin~\cite{barik_knizhnik-zamolodchikov_2024}. Notably, these known solutions also allow for an exact treatment of the $s = 3/2$ case, enabling us to determine the correct form of $\omega_j$ (or equivalently, $\gamma_{N_1, N_2}$).

\subsection{Spin 3/2 asymptotics}
We immediately obtain the asymptotic wavefunction $\ket{\Psi_\infty^{\scale[0.55]{(3/2, N_+)}}(t)}$ by substituting $s = 3/2$ into Eq.~\eqref{eq:genasympsol}. The expression for $\gamma_{N_1, N_2}^{\times}$, obtained from the saddle point calculation detailed above, is given by:
\begin{equation}
\label{eq:gammafunc_spin3/2}
\begin{aligned}
\gamma^{\times}_{N_1,N_2}= &-\frac{1}{\nu}(N_1+N_2) \bigg\{2 i \log (\nu  t)+\pi\\
&-i \left[\log \left(\frac{27}{4}\right)-2\right]\bigg\},
\end{aligned}
\end{equation}
while the correct expression for $\gamma_{N_1,N_2}$  reads:
\begin{equation}
\label{eq:correctedgammafunc32}
\gamma_{N_1,N_2} =-(N_1+N_2)\ln\left[\frac{\sqrt{3\pi}\Gamma\left(1+\frac{2i}{\nu}\right)t^{\frac{2i}{\nu}}}{\Gamma\left(\frac{1}{2}+\frac{i}{\nu}\right)\Gamma\left(1+\frac{3i}{\nu}\right)}\right].
\end{equation}
Similarly to the spin-1 case, the correct solution~\eqref{eq:correctedgammafunc32} is obtained by first solving the $J^z = \pm 1$ sectors of the corresponding $N = 2,\,s = 3/2$ problem. An asymptotic solution to the differential equations governing these sectors is provided in Appendix~\ref{sec_App:s32N2asym}. Computing the overlap between the asymptotic wavefunction $\ket{\Psi_\infty^{\scale[0.55]{(3/2, N_+)}}(t)}$, with $\gamma_{N_1,N_2}$ given by Eq.~\eqref{eq:correctedgammafunc32}, and the numerically obtained wavefunction shows convergence to unity, as illustrated in Fig.~\ref{fig_App:overlap_N=10spin=3/2} in Appendix~\ref{sec_App:Additional_Figures}.

The asymptotic solutions for spin-$1$ and spin-$3/2$ presented in this work are exact results. As discussed earlier, obtaining exact expressions for $\gamma_{N_1, \dots, N_{2s}}$ in the general spin-$s$ asymptotic wavefunction using the same approach requires solving increasingly large blocks of coupled differential equations for the two-site spin-$s$ problem. For $s \geq 2$, this becomes significantly more challenging. Nevertheless, in Appendix~\ref{sec_App:highspinrep}, we propose a plausible form of $\gamma_{N_1, N_2, N_3}$ for the spin-$2$ case, motivated by available evidence and consistency with known limits.

\section{Thermodynamic limit}\label{sec:Thermodynamic_Limit}
In this section, we investigate local observables in large systems for the spin-$1$ case. We focus on operators of the form $\hat{o}_1 \dots \hat{o}_n$, where each $\hat{o}_j$ is one of the spin components $\hat{s}_j^x$, $\hat{s}_j^y$, $\hat{s}_j^z$, their exponents, or any of their linear combinations, and $n$ is a positive integer that remains finite in the thermodynamic limit. Such operators are referred to as \emph{$n$-local}. We show that the expectation values of $n$-local operators computed from the exact spin-$1$ wavefunctions~\eqref{eq:s1asympsol} precisely match the corresponding mean-field predictions in the limit $N \rightarrow \infty$.

However, mean-field theory breaks down for non-local observables. By \emph{non-local}, we refer to operators whose support---that is, the number of sites they act on --- scales linearly with the system size $N$.

We have obtained analogous results for the spin-$3/2$ case, but omit the full derivations here for brevity.

For the observable $\hat{s}_j^z$ in the asymptotic wavefunction~\eqref{eq:genasympsol} of a general spin-$s$, we compute the variance and higher moments, all of which are found to be time-independent. This analysis ultimately shows that the probability distribution of $\hat{s}_j^z$ in our asymptotic state approaches a delta function in the large-spin limit, $s \gg 1/2$.

\subsection{Generalized projected BCS state}
Before analyzing the thermodynamic limit of the late-time quantum dynamics of the spin-$1$ system, it is useful to recast the asymptotic wavefunction in a form that resembles a projected BCS state, except that the coefficients $\hat{U}_k$, $\hat{V}_k$, and $\hat{W}_k$ are operators rather than $c$-numbers. Similar to Eq.~\eqref{eq:BCSgsspin1}, and based on inspection of Eq.~\eqref{eq:yangyangaction}, we find that---up to a normalization constant---the late-time wavefunction~\eqref{eq:s1asympsol} can be written as
\begin{equation}
\label{eq:ProjectedBCSs1asym}
\ket{\Psi_{\scale[0.75]{\infty}}^{\scale[0.55]{(N_+)}}\scale[0.85]{(t)}} = P_{N_{+}}\bigotimes_{k=1}^{N}\left(\hat{U}_{k}\ket{-1}+\hat{V}_{k}\ket{0}+\hat{W}_{k}\ket{1}\right),
\end{equation}
with
\begin{subequations}
\label{eq:UVW}
\begin{align}
\hat{U}_{j} &= \exp\left(-\frac{i}{\nu}\hat{\varphi}_j\right),\\\hat{V}_{j} &= \exp\left(-\gamma_1-2it\varepsilon_j-\frac{2\pi j}{\nu}\right),
\\\hat{W}_{j} &= \exp\left(\frac{i}{\nu}\hat{\varphi}_{j}-4it\varepsilon_j-\frac{4\pi j}{\nu}\right),\\ \label{eq:varphiopr}
\hat{\varphi}_{j} &= \sum_{i\neq j}\hat{s}_{i}^{z}\ln\left|\varepsilon_i-\varepsilon_j\right|.
\end{align}    
\end{subequations}
We rewrite equation \eqref{eq:ProjectedBCSs1asym} as an integral
\begin{equation}\label{eq:projectedBCS_deltafunction}
\begin{aligned}
\frac{1}{2\pi}\smashoperator{\int_{0}^{2\pi}}&d\phi \, e^{i\phi N_{+}} \times \\
&\bigotimes_{k=1}^{N}\left(\hat{U}_{k}\ket{-1}+e^{-i\phi}\hat{V}_{k}\ket{0}+e^{-2i\phi}\hat{W}_{k}\ket{1}\right).
\end{aligned}
\end{equation}
The phase $\phi$ integral is set up to project onto states with a fixed total magnetization $J^z = N_+ - N$. There is a slight subtlety in the notation of Eq.~\eqref{eq:projectedBCS_deltafunction}: the operators $\hat{U}_k$, $\hat{V}_k$, and $\hat{W}_k$ are meant to act on the entire product. That is, one should first take the product as written and then reorder the resulting expression, shifting all operators to the left so that each $\hat{\varphi}_j$ acts appropriately on the basis states.

\subsection{Classical BCS dynamics}
To compare local observables with the mean-field predictions, we require the mean-field wavefunction from Eq. \eqref{eq:mainBCS}. In the mean-field picture, each quantum spin evolves independently under an effective magnetic field. Each spin is in a coherent state and remains so at all times. The wavefunction is then written similarly to a BCS product state, in terms of coherent states at each site:
\begin{subequations}
\begin{align}
\label{eq:mfbcss1}\ket{\Psi_{\text{mf}}(t)} &= \bigotimes_{k=1}^{N}\left(u_k\ket{-1}+v_k\ket{0}+w_k\ket{1}\right), \\  
u_k &= e^{i\phi_k(t)}\sin\left(\frac{\theta_k(t)}{2}\right)^2,\\
v_k &= \frac{1}{\sqrt{2}}\sin(\theta_k(t))\,,\\
w_k &= e^{-i\phi_k(t)}\cos\left(\frac{\theta_k(t)}{2}\right)^2.
\end{align}
\end{subequations}
The angles $\phi_k(t), \theta_k(t)$ are time-dependent. For the normalization,
$|u_k|^2+|v_k|^2+|w_k|^2=1$. The remaining task is to determine the angles from the mean-field average of $\hat{\boldsymbol{s}}_k$ for $s=1$ as calculated in \cite{zabalo_nonlocality_2022}:
\begin{subequations}
\begin{align}
&\braket{\hat{o}_k}_{\text{mf}}\equiv\bra{\Psi_{\text{mf}}(t)}\hat{o}_k\ket{\Psi_{\text{mf}}(t)},\\ &\;\zeta_k\equiv \pi(k-\mu)/\nu,\\
&\label{eq:mfszexpr}\braket{\hat{s}_k^z}_{\text{mf}} = \cos(\theta_k(t)) = -\tanh\left(2\zeta_k\right),\\
&\braket{\hat{s}_k^-}_{\text{mf}} = e^{-i\phi_k(t)}\sin(\theta_k(t)) = \frac{e^{-2i\varepsilon_k t + i\varphi_k}}{\cosh(2\zeta_k)},\\&\braket{\hat{s}_k^+}_{\text{mf}} = \braket{\hat{s}_k^-}_{\text{mf}}^{*},\\&\;\varphi_k=\frac{2}{\nu}\sum_{j\neq k}\braket{\hat{s}_j^{z}}_{\text{mf}}\ln|\varepsilon_j-\varepsilon_k|.
\end{align}
\end{subequations}
The chemical potential $\mu$ is determined from the below constraint that 
\begin{equation}
-\sum_{k=1}^{N}\tanh\left(\frac{2\pi(k-\mu)}{\nu}\right) = J^{z}   
\end{equation}
for some fixed total spin $J^{z}$. Finally we provide the expressions for the angles $\theta_k(t)$ and $\phi_k(t)$ that complete the expression of Eq. \eqref{eq:mfbcss1}:
\begin{subequations}
\begin{align}
&\theta_k(t) = \pi-\cos^{-1}\left(\tanh(2\zeta_k\right)), \\
&\phi_k(t) = 2\varepsilon_k t - \varphi_k.
\end{align}    
\end{subequations}

\subsection{Local observables}

\subsubsection{Primary considerations}
In the thermodynamic limit, the Zeeman term in the Hamiltonian~\eqref{eq:mainBCS} scales linearly with the system size \( N \), while the interaction term \( \sum_{j,k} \hat{s}^{+}_j \hat{s}^{-}_k \) scales as \( N^2 \). To obtain a nontrivial and well-defined thermodynamic limit, the coupling constant \( g(t) = \frac{1}{\nu t} \) must be scaled to offset the \( N^2 \) scaling of the interaction. This motivates us to set
\begin{equation}
\label{eq:nuetaform}
\nu = \frac{N}{\eta},\;\;\eta>0,
\end{equation}
where $\eta$ is independent of $N$. 

In order to obtain properly normalized quantities, we evaluate $\braket{\Psi_{\scale[0.75]{\infty}}^{\scale[0.55]{(N_+)}}\scale[0.85]{(t)}|\Psi_{\scale[0.75]{\infty}}^{\scale[0.55]{(N_+)}}\scale[0.85]{(t)}}$. Since the operators $\hat{U}_{k},\hat{V}_{k}$ and $\hat{W}_{k}$ all commute with each other for arbitrary sites and
\begin{equation}
\begin{aligned}
\hat{U}_{k}^{\dagger}\hat{U}_{k}=&\,\mathbf{I},\\ \hat{V}_{k}^{\dagger}\hat{V}_{k}=&\,2\cosh\left(\frac{\pi}{\nu}\right)e^{-\frac{4\pi k}{\nu}}\mathbf{I},\\\hat{W}_{k}^{\dagger}\hat{W}_{k}=&\,e^{-\frac{8\pi k}{\nu}}\mathbf{I},   
\end{aligned}
\end{equation}
we find \footnote{note the change of variables $\xi =\phi -\phi'$ where $\phi'$ comes from $\bra{\Psi_{\scale[0.75]{\infty}}^{\scale[0.55]{(N_+)}}\scale[0.85]{(t)}}$}
\begin{equation}
\label{eq:overlapspint}
\braket{\Psi_{\scale[0.75]{\infty}}^{\scale[0.55]{(N_+)}}\scale[0.85]{(t)}|\Psi_{\scale[0.75]{\infty}}^{\scale[0.55]{(N_+)}}\scale[0.85]{(t)}} = \frac{1}{2\pi}\smashoperator{\int^{2\pi}_{-2\pi}}d\xi\exp\left[NG(\xi)\right],   
\end{equation}
where
\begin{equation}
\label{eq:overlapGfunc}
\begin{split}
&G(\xi) = i\xi \frac{N_{+}}{N}+\\&\frac{1}{N}\sum_{j=1}^{N}\ln\left[1+e^{-i\xi}\left(e^{\frac{\pi}{\nu}}+e^{-\frac{\pi}{\nu}}\right)e^{-\frac{4\pi j}{\nu}}+e^{-2i\xi}e^{-\frac{8\pi j}{\nu}}\right].
\end{split}
\end{equation}
This integral further simplifies into a summation:
\begin{equation}
\smashoperator{\sum_{\substack{N_1+2N_2\\=N_{+}}}} 2^{N_1} \cosh\left(\frac{\pi}{\nu}\right)^{N_1} 
\smashoperator{\sum_{\substack{\text{\tiny$\left|\{\alpha\}\right| = N_1$}\\\text{\tiny$\left|\{\beta\}\right| = N_2$}}}} 
e^{-\frac{4\pi}{\nu}\left( \sum_{j\in\{\alpha\}} j + 2 \sum_{k\in\{\beta\}} k \right)},
\end{equation}
which, however, cannot be evaluated exactly. We therefore apply the saddle point method to compute the original integral~\eqref{eq:overlapspint}. The corrections to this approximation are always of order \( 1/N \) or smaller~\cite{wong_asymptotic_2001}, and vanish as \( N \to \infty \). Hence, the saddle point evaluation becomes exact in the thermodynamic limit. The stationary point of the integrand in~\eqref{eq:overlapspint} is given by
\begin{equation}
\label{eq:statpoints1}
\xi_0^{(1)} = i \ln\left[ \frac{1 - \exp\left(\frac{2\pi N_{+}}{\nu}\right)}{\exp\left(-\frac{2\pi(2N - N_{+})}{\nu}\right) - 1} \right].
\end{equation}

\subsubsection{Multi-site operators}
Consider the following notation for an arbitrary (bi-linear) single-site operator:
\begin{equation}
\hat{s}^{p}_{i} \hat{s}^{q}_{i}, \quad p,q \in \{0,+,-,z\}, \quad \hat{s}^{0} = \mathbf{I}_{3\times 3}.
\end{equation}
Using this, we define an \( n \)-local operator \( \hat{\mathcal{B}} \) for some fixed positive integer \( n \) as
\begin{equation}
\label{eq:multisitelocalobservables}
\hat{\mathcal{B}} = \prod_{i=1}^{n} \left( \hat{s}^{p_i}_{i} \hat{s}^{q_i}_{i} \right), \quad p_i, q_i \in \{0,+,-,z\}.
\end{equation}
It turns out that the specific choice of sites \( i \) used to construct \( \hat{\mathcal{B}} \) is immaterial. Without loss of generality, we may choose the ordered set \( i \in \{1, \dots, n\} \) for convenience.

We are interested in evaluating the normalized non-zero matrix elements of \( \hat{\mathcal{B}} \) between the states given in Eq. \eqref{eq:ProjectedBCSs1asym} for different $N_+$ 
\begin{equation}
\label{eq:nonzeromatrixelm}
\mathcal{N}_{\Delta N_+}\bra{\Psi_{\scale[0.75]{\infty}}^{\scale[0.55]{(N_+\!+\!\Delta N_+)}}\scale[0.85]{(t)}} \hat{\mathcal{B}} \ket{\Psi_{\scale[0.75]{\infty}}^{\scale[0.55]{(N_+)}}\scale[0.85]{(t)}}  
\end{equation}
in the limit \( N \to \infty \). $\Delta N_+$ is defined as the difference between the number of raising operations and lowering operations involved within \( \hat{\mathcal{B}} \). The $\mathcal{N}_{\Delta N_+}$ term accounts for the normalization of the bra/ket states used in \eqref{eq:nonzeromatrixelm}. Since the noncommutativity of \( \hat{\varphi}_j \) with \( \hat{s}_{i}^{+,-} \) contributes only at order \( \mathcal{O}(1/N) \), we are justified in replacing the operators \( \hat{s}_{i}^{z} \) and \( (\hat{s}^{z}_{i})^2 \) in the exponents of~\eqref{eq:UVW} with their expectation values:
\begin{equation}
\hat{s}_{i}^{z} \rightarrow \braket{\hat{s}_{i}^{z}}, \quad (\hat{s}^{z}_{i})^2 \rightarrow \braket{(\hat{s}^{z}_{i})^2}.
\end{equation}
These substitutions effectively replace the \( \hat{s}^{z} \) operators and their powers in the wavefunction~\eqref{eq:ProjectedBCSs1asym} with scalars, thereby eliminating the operator ordering issues that arise due to the noncommutativity with \( \hat{\varphi} \).

To calculate the non-zero matrix element of any single-site operator $ \hat{s}_j^{p} \hat{s}_j^{q} $, we first evaluate the below (unnormalized) matrix element
\begin{equation}
\label{eq:gensinglesiteoverlap}
\begin{split}
&\braket{\Psi_{\scalebox{0.75}{$\infty$}}^{\scalebox{0.55}{$(N_+\!+\!\Delta{(p,q)})$}}(t)|\hat{s}_j^{p} \hat{s}_j^{q}|\Psi_{\scalebox{0.75}{$\infty$}}^{\scalebox{0.55}{$(N_+)$}}(t)} 
=\\ &\frac{1}{2\pi} \int_{-2\pi}^{2\pi} d\xi \, \exp\left[N G(\xi)\right] \, g^{(p,q)}_{j}(\xi).
\end{split}
\end{equation}
We refer to $ g^{(p,q)}_{j}(\xi) $ as a \textit{$g$-function} and denote by $\Delta{(p,q)}$ the difference between the number of raising ($+$) and lowering ($-$) symbols in $\{p,q\}$. These $g$-functions depend on the specific local operator under consideration. The unnormalized matrix element for the operator $ \hat{\mathcal{B}} $ defined in Eq.~\eqref{eq:nonzeromatrixelm}, without $\mathcal{N}$, is written analogously to~\eqref{eq:gensinglesiteoverlap}, except that the $g$-function in the integrand becomes a product:
\begin{equation}
g(\xi) = \prod_{i=1}^{n} g_{i}^{(p_i,q_i)}(\xi).
\end{equation}

Note that equations~\eqref{eq:overlapspint} and~\eqref{eq:gensinglesiteoverlap} share the same stationary point $ \xi_0^{(1)} $ given in~\eqref{eq:statpoints1}. This allows us to immediately evaluate Eq. \eqref{eq:nonzeromatrixelm} to order $ \mathcal{O}(1/N) $ as follows:
\begin{equation}
\label{eq:evmultisiteobs}
\begin{aligned}
&\;\;\mathcal{N}_{\Delta N_+}\bra{\Psi_{\scale[0.75]{\infty}}^{\scale[0.55]{(N_+\!+\!\Delta N_+)}}\scale[0.85]{(t)}} \hat{\mathcal{B}} \ket{\Psi_{\scale[0.75]{\infty}}^{\scale[0.55]{(N_+)}}\scale[0.85]{(t)}}
\\&= e^{-\frac{i}{2}\xi_0^{(1)}\Delta N_+}\prod_{i=1}^{n} g_{i}^{(p_i,q_i)}(\xi_0^{(1)}) \\  &=\prod_{i=1}^{n} 
e^{-\frac{i}{2}\xi_0^{(1)}\Delta(p_i,q_i)}g_{i}^{(p_i,q_i)}(\xi_0^{(1)})\\  &=\prod_{i=1}^{n} 
\mathcal{N}_{\Delta(p_i,q_i)}\bra{\Psi_{\scale[0.75]{\infty}}^{\scale[0.55]{(N_+\!+\!\Delta{(p_i,q_i)})}}\scale[0.85]{(t)}} \hat{s}_i^{p_i} \hat{s}_i^{q_i} \ket{\Psi_{\scale[0.75]{\infty}}^{\scale[0.55]{(N_+)}}\scale[0.85]{(t)}}.
\end{aligned}
\end{equation}

Thus, the non-zero matrix element of any $ n $-local operator from Eq. \eqref{eq:multisitelocalobservables} is given by the product of the same for individual single-site operators. Corrections to the result in~\eqref{eq:evmultisiteobs} are of order $ n/N $ or smaller, and therefore vanish in the thermodynamic limit $ N \to \infty $, since $ n $ is a fixed positive integer.

\subsubsection{Single-site observables}
\begin{figure}[ht]
\centering
\includegraphics[width=\linewidth]{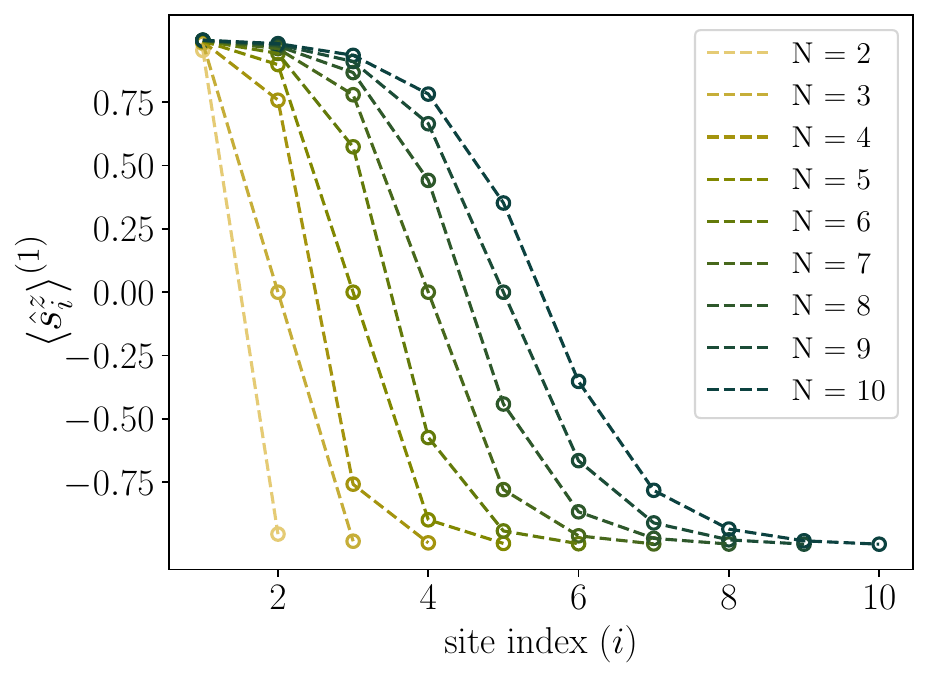}
\caption{Comparison of the expectation values $\left<\hat{s}_i^z\right>^{\!\scale[0.60]{(1)}}$ computed numerically at $t = 10^3$
    and  evaluated in the asymptotic state~\eqref{eq:s1asympsol}    as  functions of $N$ for $J^z = 0$. The dashed lines describe the asymptotic predictions, whereas the circles represent the outputs of numerical simulations. The system parameters  are the same as in Fig.~\ref{fig:overlap_incorrect}.}\label{fig:Numerical_Asymptotic_ExpectValue_Comparison}
\end{figure}

Consider a single-site spin observable $\hat{s}_{j}^{z}$. Its unnormalized expectation value in the state~\eqref{eq:ProjectedBCSs1asym} is 
\begin{equation}
\label{eq:szoverlapspint}
\begin{split}
\braket{\Psi_{\scale[0.75]{\infty}}^{\scale[0.55]{(N_+)}}\scale[0.85]{(t)}|\hat{s}_{j}^{z}|\Psi_{\scale[0.75]{\infty}}^{\scale[0.55]{(N_+)}}\scale[0.85]{(t)}} &= \frac{1}{2\pi}\int^{2\pi}_{-2\pi}d\xi\times\\&\exp\left[NG(\xi)\right] g^{(z,0)}_{j}(\xi),
\end{split}
\end{equation}
where
\begin{equation}
g^{(z,0)}_{j}(\xi) = -\frac{\sinh \left(\frac{4 \pi  j}{\nu }+i \xi \right)}{\cosh \left(\frac{\pi }{\nu }\right)+\cosh \left(\frac{4 \pi  j}{\nu }+i \xi \right)}.
\end{equation}    
Both integrals~\eqref{eq:overlapspint} and~\eqref{eq:szoverlapspint}, share the same stationary point $\xi_{0}^{(1)}$ given in~Eq.~(\eqref{eq:statpoints1}). Thus, in the thermodynamic limit $N \rightarrow \infty$, we evaluate the expectation value of $\hat{s}_{j}^{z}$ as
\begin{equation}
\label{eq:ThermLimit_Spin-1_sz_ExpectationValue}
\begin{aligned}
\braket{\hat{s}^{z}_{j}}^{\!\scalebox{0.60}{$(1)$}} 
&= \frac{\braket{\Psi_{\scalebox{0.75}{$\infty$}}^{\scalebox{0.55}{$(N_+)$}}(t)|\hat{s}_{j}^{z}|\Psi_{\scalebox{0.75}{$\infty$}}^{\scalebox{0.55}{$(N_+)$}}(t)}}{\braket{\Psi_{\scalebox{0.75}{$\infty$}}^{\scalebox{0.55}{$(N_+)$}}(t)|\Psi_{\scalebox{0.75}{$\infty$}}^{\scalebox{0.55}{$(N_+)$}}(t)}} \\
&= -\tanh\left( \frac{2\pi (j - \mu_{(1)})}{\nu} \right),
\end{aligned}
\end{equation}
where we identify the chemical potential $\mu_{(1)}$ as
\begin{equation}
\label{eq:s1chemicalpotential}
\mu_{(1)} = -\frac{i\nu \xi_0^{(1)}}{4\pi}.
\end{equation}
This result coincides with the mean-field expression in Eq.~\eqref{eq:mfszexpr} and becomes exact in the thermodynamic limit. The superscript $^{(1)}$ and subscript $_{(1)}$ indicate that these results pertain specifically to the spin-1 problem.

As a consistency check for the chemical potential, we verify that
\begin{equation}
\begin{aligned}
\sum_j \braket{\hat{s}_{j}^{z}}^{\!\scalebox{0.60}{$(1)$}} 
&\approx -N \int_{0}^{1} \tanh\left( 2\pi \eta y - \frac{2\pi \mu_{(1)}}{\nu} \right) dy \\
&= N_{+} - N,
\end{aligned}
\end{equation}
which correctly reproduces the $z$-component $J^z$ of the total spin.

These results are also confirmed numerically.
Fig.~\ref{fig:Numerical_Asymptotic_ExpectValue_Comparison} shows that the expectation values computed from the asymptotic wavefunction~\eqref{eq:s1asympsol} agree with those from numerical simulations. The same numerical simulation is then compared with the thermodynamic limit prediction in Fig.~\ref{fig:thermlimit_barfig_complete}. Finally, the scaling of corrections to the expression in~Eq.~\eqref{eq:ThermLimit_Spin-1_sz_ExpectationValue} with increasing system size $N$ is shown in Fig.~\ref{fig:Thermlimit_Proportionality}.

We now evaluate the expectation value $\braket{(\hat{s}_{j}^{z})^2}^{\!\scalebox{0.60}{$(1)$}}$. Following the same procedure as for $\braket{\hat{s}_{j}^{z}}^{\!\scalebox{0.60}{$(1)$}}$, we obtain the $g$-function in the integrand of~\eqref{eq:szoverlapspint} as
\begin{equation}
\label{eq:gzz}
g^{(z,z)}_{j}(\xi) = \frac{1}{1 + \cosh\left( \frac{\pi}{\nu} \right) \, \text{sech}\left( \frac{4\pi j}{\nu} + i\xi \right)}.
\end{equation}
Evaluating this at the saddle point $\xi = \xi_0^{(1)}$ then gives
\begin{equation}
\braket{(\hat{s}^{z}_{j})^2}^{\!\scalebox{0.60}{$(1)$}} = \frac{1}{1 + \text{sech}\left( \frac{4\pi (j - \mu_{(1)})}{\nu} \right)}.
\end{equation}

This result also agrees with the mean-field prediction, given by $\braket{(\hat{s}^{z}_j)^2}_{\text{mf}}$ computed from Eq.~\eqref{eq:mfbcss1}. Similarly, all appropriately normalized nonzero matrix elements of single-site operators match the corresponding averages calculated from the mean-field wavefunction in the thermodynamic limit. Additional examples are provided in Appendix \ref{sec_App:matrix_elements}.

\subsubsection{Agreement with mean-field predictions}
Returning to the general $n$-local operator $\hat{\mathcal{B}}$ in \eqref{eq:multisitelocalobservables} and applying the result of Eq.~\eqref{eq:evmultisiteobs}, we obtain
\begin{equation}
\mathcal{N}_{\Delta N_+}\bra{\Psi_{\scale[0.75]{\infty}}^{\scale[0.55]{(N_+\!+\!\Delta N_+)}}\scale[0.85]{(t)}} \hat{\mathcal{B}} \ket{\Psi_{\scale[0.75]{\infty}}^{\scale[0.55]{(N_+)}}\scale[0.85]{(t)}} = \prod_{i=1}^{n}\braket{\hat{s}^{p}_{i} \hat{s}^{q}_{i}}_{\text{mf}}.
\end{equation}
This result is valid up to order $\mathcal{O}(n/N)$ and becomes exact in the thermodynamic limit. 

Similar procedures for calculating expectation values fail for any \textit{non-local} operators because the corrections neglected in the saddle-point analysis then contribute at order $\mathcal{O}(1)$ and therefore become relevant. 
Mean-field predictions for non-local observables often differ from quantum-mechanical results. For example, the Loschmidt echo, as calculated from the classical picture, differs from its quantum counterpart \cite{gaur_singularities_2022}. Further investigations of non-local observables (such as the Loschmidt echo) are beyond the scope of this paper.
\subsubsection{Variance and higher moments of \texorpdfstring{$\hat{s}^{z}_j$}{szj} for large spins}\label{sec:highmomentssz}
In investigating spin-1 operators, we also address general spin-$s$ single-site observables to understand the $s \gg 1/2$ regime in the thermodynamic limit. We first consider the form of $\omega_j$ in \eqref{eq:genasympsolgamma} in this limit. Recalling that $\nu = N / \eta$ [Eq.~\eqref{eq:nuetaform}], we  readily determine $\omega_j$ from \eqref{eq:BCSgsspins} in the $\eta \to 0^+$ (stiff-ramp/diabatic) limit for any spin-$s$ case.
\begin{equation}
\label{eq:diabaticlimit}
\lim_{\eta\rightarrow 0^+}\omega_j(\nu) = -N_j\ln{\left(\sqrt{\begin{pmatrix}
2s \\ j    
\end{pmatrix}}\,\right)}
\end{equation}
We reiterate that exact expressions for $\omega_j(\nu)$ are available only for the spin-1 \eqref{eq:correctedgammafunc} and spin-3/2 \eqref{eq:correctedgammafunc32} cases. However, in general, the exact expression is not required to compute expectation values in the thermodynamic limit. As we consider the $\eta \to 0^+$ limit, it follows from \eqref{eq:nuetaform} that $\nu \to \infty$, which is equivalent to taking the $N \to \infty$ limit with fixed $\eta$. Since all $\omega_j$ are independent of the system size $N$, Eq.~\eqref{eq:diabaticlimit} also holds in the thermodynamic limit. We can thus compute the expectation values of $\hat{s}_j^{z}$ and $(\hat{s}_j^{z})^2$ for general spin-$s$.\begin{subequations}
\label{eq:s_moments}
\begin{align}
\braket{\hat{s}^{z}_{j}}^{\!(s)} &= -s\tanh\left(\frac{2s\pi (j-\mu_{(s)})}{\nu}\right), \label{eq:firstcumulant}\\
\braket{(\hat{s}^{z}_{j})^2}^{\!(s)}&=s^2+\frac{s-2 s^2}{1+\cosh \left(\frac{4s \pi (j-\mu_{(s)})}{\nu} \right)},\\
    \mu_{(s)} &= -\frac{i\nu\xi_0^{(s)}}{4\pi s}, \label{eq:mu_s}\\
\xi_0^{(s)} &=  i\ln\left[\frac{1-\exp\left(\frac{2 \pi N_{+}}{\nu}\right)}{\exp\left(-\frac{2\pi\left(2Ns-N_{+}\right)}{\nu}\right)-1}\right]. \label{eq:xi_s}
\end{align}
\end{subequations}
The quantum corrections to the above results are suppressed at order $1/N$ or smaller. Next, we define cumulants for a quantum observable $\hat{X}$ on the state $\ket{\phi}$ \cite{li_generalized_2021}. Consider the following expectation value of $\exp(s \hat{X})$,
\begin{equation}
\label{eq:cumulantoverlap}
\braket{\phi|\exp(s\hat{X})|\phi} = \sum_{n=0}^{\infty}\braket{\hat{X}^{n}}\frac{s^n}{n!},\;\;s\in\mathrm{C}.
\end{equation}
The level-$m$ cumulant $\kappa_{m}(\hat{X})$ of $\hat{X}$ in the state $\ket{\phi}$ is generated by the (natural) logarithm of  \eqref{eq:cumulantoverlap}
\begin{equation}
\log\left(\braket{\phi|\exp(s\hat{X})|\phi}\right) = \sum_{m=1}^{\infty}\kappa_{m}(\hat{X})\frac{s^m}{m!}.
\end{equation}
The first four cumulants are given as follows
\begin{subequations}
\begin{align}
\kappa_1(\hat{X}) &= \braket{\hat{X}}, \\
\kappa_2(\hat{X}) &= \braket{\hat{X}^2} - \braket{\hat{X}}^{\!2}, \\
\kappa_3(\hat{X}) &= \braket{\hat{X}^3} - 3\braket{\hat{X}^2}\braket{\hat{X}}+2\braket{\hat{X}}^{\!3}, \\
\kappa_4(\hat{X}) &= \braket{\hat{X}^4} - 4\braket{\hat{X}^3}\braket{\hat{X}} -3\braket{\hat{X}^2}^{\!2} \\&+ 12\braket{\hat{X}^2}\braket{\hat{X}}^{\!2} - 6\braket{\hat{X}}^{\!4}.\nonumber  
\end{align}   
\end{subequations}
We are interested in computing the cumulants of the observable $\hat{s}^{z}_j$ in the state~\eqref{eq:ProjectedBCSs1asym} in the thermodynamic limit. The first cumulant, i.e., the expectation value (mean), has already been evaluated in~\eqref{eq:firstcumulant}. Using~\eqref{eq:s_moments}, the second cumulant, corresponding to the variance $\text{Var}(\hat{s}_j^{z})$, is given by:
\begin{equation}
\label{eq:genspinvariance}
\text{Var}(\hat{s}_j^{z}) 
= \frac{s}{1+\cosh\left(\frac{4s\pi (j-\mu_{(s)})}{\nu}\right)}.
\end{equation}
Similarly, the third and fourth cumulants, known as skewness $\text{Skew}(\hat{s}_j^{z})$ and kurtosis $\text{Kurt}(\hat{s}_j^{z})$ respectively are computed below:
\begin{subequations}
\begin{equation}
\text{Skew}(\hat{s}_j^{z}) = \frac{s\tanh\left(\frac{2s\pi (j-\mu_{(s)})}{\nu}\right)}{2\cosh\left(\frac{2s\pi (j-\mu_{(s)})}{\nu}\right)^2}, 
\end{equation}
\begin{equation}
\text{Kurt}(\hat{s}_j^{z}) =\frac{s\left(\cosh\left(\frac{4s\pi (j-\mu_{(s)})}{\nu}\right)-2\right)}{4\cosh\left(\frac{2s\pi (j-\mu_{(s)})}{\nu}\right)^4},       
\end{equation}
\end{subequations}
Since the skewness is nonzero, the probability distribution for measuring $\hat{s}_j^{z}$ in the state \eqref{eq:genasympsol} for general spin-$s$ is \textit{non-Gaussian}. 

Before considering the $s\rightarrow\infty$ limit, it is essential to recast $\nu$ as
\begin{equation}
\nu = \frac{sN}{\eta},\;\;\eta>0,
\end{equation}
to ensure proper scaling in the thermodynamic and $s\to\infty$ limits. For $s \gg 1/2$, the variance \eqref{eq:genspinvariance} grows proportionally to $s$. In the context of spin-coherent states, the relative uncertainty of their spin components is given by $s^{-1} \sqrt{\text{Var}(\hat{s}_j^z)} \propto s^{-1/2}$, consistent with the classical picture. A general expression for higher moments of $\hat{s}_j^z$ is given in \eqref{eq:generalszavg}. For $s \gg 1/2$ we have
\begin{equation}
\braket{(\hat{s}^{z}_{j})^n}^{\!(s)} \approx  \left(\braket{\hat{s}^{z}_{j}}^{\!(s)}\right)^n   
\end{equation}
for any non-negative integer $n$, with corrections vanishing in the $s \to \infty$ limit. Details of the evaluation are provided in Appendix \ref{sec_App:higher_moments}. Based on these moments, the probability distribution of $\hat{s}_j^z$ for $s \gg 1/2$ is represented by a Dirac delta function. In the classical spin model, the spin vector at each site $j$ has a definite $z$-component, specified by the delta function.

\begin{figure}[ht]
\centering
\includegraphics[width=\linewidth]{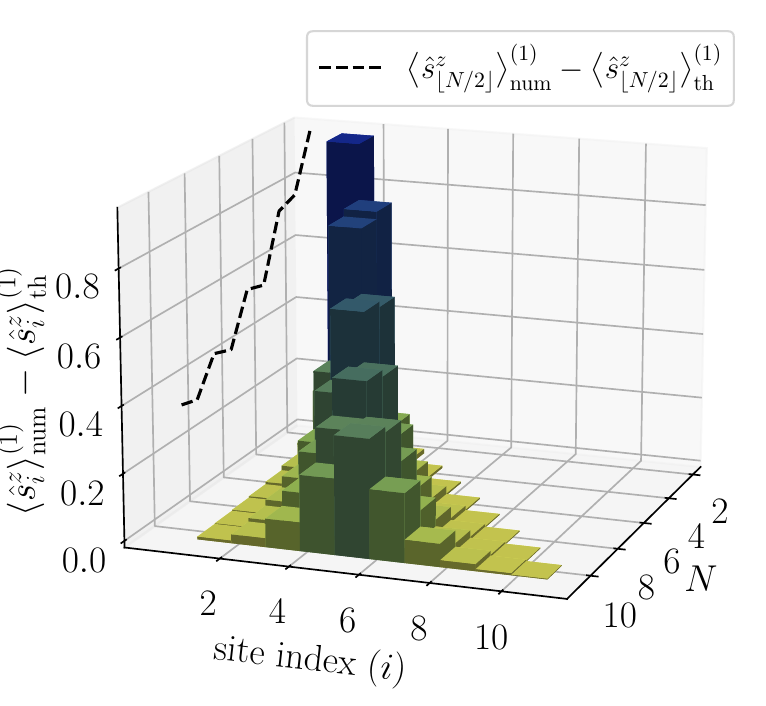}
\caption{Difference between numerically computed expectation values $\left<\hat{s}_i^z\right>_{\text{num}}^{\!\scale[0.60]{(1)}}$ (evaluated at $t = 10^3$) and the asymptotic prediction in the thermodynamic limit $\left<\hat{s}_i^z\right>_{\text{th}}^{\!\scale[0.60]{(1)}}$ from Eq.~\eqref{eq:ThermLimit_Spin-1_sz_ExpectationValue}. The color indicates the height of each column. The dashed line shows the result for site indices $i = \left\lfloor N/2 \right\rfloor$ across different values of $N$, extracted from the bar chart to highlight the decreasing trend with increasing $N$. Other parameters are identical to those in Fig. \ref{fig:overlap_incorrect}.}
\label{fig:thermlimit_barfig_complete}
\end{figure}

\begin{figure}[ht]
\centering
\includegraphics[width=\linewidth]{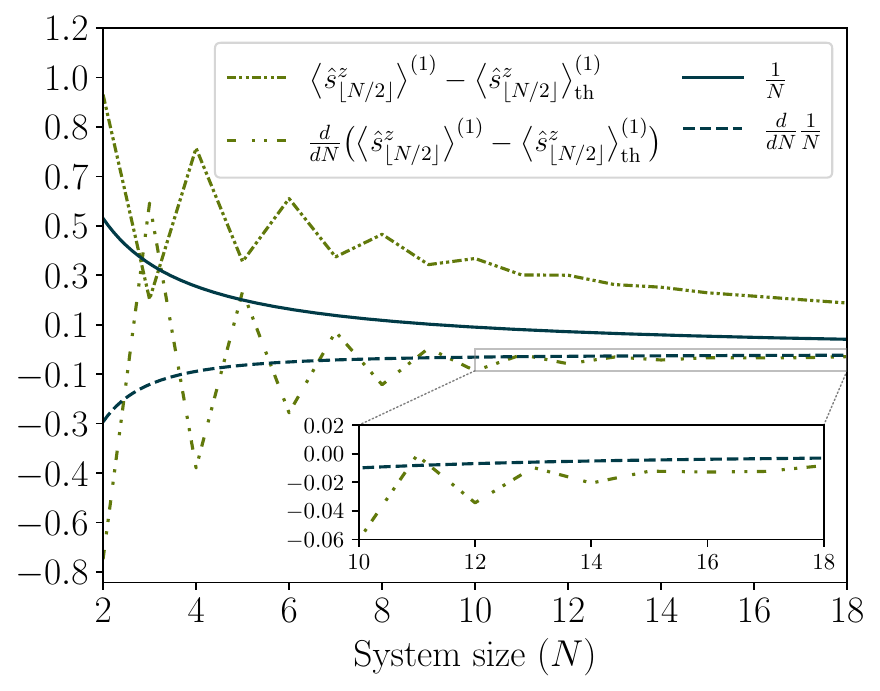}
\caption{Difference between $\left<\hat{s}_i^z\right>^{\!\scale[0.60]{(1)}}$, calculated from the asymptotic wavefunction \eqref{eq:s1asympsol} at $t = 10^3$, and the thermodynamic average $\left<\hat{s}_i^z\right>_{\text{th}}^{\!\scale[0.60]{(1)}}$ from Eq.~\eqref{eq:ThermLimit_Spin-1_sz_ExpectationValue}.  The (numerical) derivative of the differences are shown in dot-dashed lines. The correction to the thermodynamic limit scales as $\frac{1}{N}$ (solid black line). The system parameters are identical to those in Fig. \ref{fig:overlap_incorrect}.}
\label{fig:Thermlimit_Proportionality}
\end{figure}

\section{Steady State}\label{sec:steadystate}
Having addressed the thermodynamic limit of $n$-local observables, we now turn to the long-time behavior of Eq.~\eqref{eq:mainBCS} for spin-1, spin-3/2, and spin-2 systems. We investigate the steady-state properties of the RG  model in the $N \to \infty$ limit, highlight the discrepancies that arise with varying spin magnitudes, and relate these differences to the earlier claim of non-analyticity in the model.

\subsection{Probability distribution}
At large times ($t\gg 1$) the interaction term in \eqref{eq:mainBCS} becomes suppressed, which only leaves the Zeeman term, $2\sum_i\varepsilon_i\hat{s}_{i}^{z}$. The spin-$s$ steady state of interest is the asymptotic state given in Eq.~\eqref{eq:genasympsol}, in which each basis state can be rewritten as follows:
\begin{equation}
\smashoperator{\bigotimes_{\{\alpha^{(j)}\}\in\boldsymbol{\alpha}}}\ket{\{\alpha^{(j)}\}}\otimes\ket{\oslash} =  \ket{\{s^z\}},\;\;\ket{\{s^z\}} \equiv \bigotimes_{i = 1}^N\ket{s^{z}_i}.
\end{equation}
The \textit{configuration} of on-site magnetization $\{s^{z}\} \equiv \{s^{z}_1,\dots,s^{z}_N\}$, for each $\ket{\{s^z\}}$  is determined as follows: for all $m=1,\dots,2s$  and for every site $i\in\{\alpha^{(m)}\}$, $s^z_i = -s+m$; for the remaining sites $k$ not included in any $\{\alpha^{(m)}\}$, $s^z_k = -s$. With this, the probability distribution $P_{s}(\{s^{z}\})$ for the system having configuration $\{s^{z}\}$ is easily determined from the asymptotic wavefunction in Eq. \eqref{eq:genasympsol}. The general form of $P_s(\{s^z\})$ is 
\begin{equation}
\label{eq:Pszasymp}
P_s(\{s^{z}\})=\frac{1}{Z}e^{-\sum_j\left(\frac{4s\pi j s_j^z}{\nu}+\varrho_j^{(s)}(\nu)\right)}\!\delta\!\left(J^{z} -\sum_k s_k^{z}\right).
\end{equation}
Here, $Z$ is the partition function and $\varrho_j^{(s)}$ are the additional terms that are determined from the expression of $\gamma$ in Eq. \eqref{eq:genasympsolgamma}. We provide $\varrho_j^{(s)}(\nu)$ for different spin magnitudes $s$ from $1/2$ to $2$ as follows:
\begin{subequations}
\begin{equation}
\label{eq:varrhos0-5}
\varrho_j^{(1/2)}(\nu) = 0,    
\end{equation}
\begin{equation}
\label{eq:varrhos1}
\varrho_j^{(1)}(\nu) = (s_j^z)^2\ln\left(2\cosh\frac{\pi}{\nu}\right), \end{equation}
\begin{equation}
\varrho_j^{(3/2)}(\nu) = \frac{1}{2}(s_j^z)^2\ln \left[2 \cosh \left(\frac{2 \pi }{\nu }\right)+1\right],  
\end{equation}   
\begin{equation}
\begin{split}
\varrho_j^{(2)}(\nu) = &\left[\frac{7}{12}(s_j^z)^2-\frac{1}{12}(s_j^z)^4\right]\log \left( \frac{3^\frac{5}{2}}{4^\frac{5}{2}}\frac{\sinh\left(\frac{\sqrt{3} \pi }{\sqrt{8} \nu }\right)}{\sqrt{2}\sinh\left(\frac{\pi }{4 \nu }\right)}\right)+ \\&\frac{1}{2}(s_j^z)^2\ln\!\left[1+\frac{\sinh\left(\frac{5 \pi }{\nu }\right)}{\sinh\left(\frac{3 \pi }{\nu }\right)}\right]. 
\end{split}
\end{equation}
\end{subequations}
For $s=1/2$, we have used the asymptotic wavefunction from \cite{zabalo_nonlocality_2022}. For spin-$1$ and spin-$3/2$ we have utilized \eqref{eq:correctedgammafunc} and \eqref{eq:correctedgammafunc32}, respectively. Finally, for $s=2$, we used our conjectured form of the unknown function $h(\nu)$ in Eq.~\eqref{eq:s2hnuguess}, together with Eq.~\eqref{App_eq:Gamma_Spin2_Guess}. The partition function $Z$ is determined from the constraint that the sum of the probabilities over all possible spin configurations must equal unity. The delta function appears due to conservation of the total spin of the system. In the mean-field picture, where one directly deals with a product state, the probability is given by the same function, but without the delta function. 

In the thermodynamic limit, however, the probability distribution corresponding to both the asymptotic solution and the mean-field solution become identical up to $\mathcal{O}(1/N)$ corrections, as demonstrated in Figs. \ref{fig:thermlimit_barfig_complete} and \ref{fig:Thermlimit_Proportionality}. Focusing specifically on the case $s = 1$, the distribution becomes
\begin{equation}
\label{eq:Pszthermlimit}
P_1(\{s^z\}) \approx C_p\smashoperator{\exp}\left[-\smashoperator{\sum}_{j}\left(\frac{4\pi (j-\mu_{(1)}) s_j^{z}}{\nu}+\varrho_{j}^{(1)}(\nu) \right)\right],
\end{equation}
where $C_p$ is a normalization constant.

\subsection{Non-GGE predictions}
At $t\rightarrow\infty$, the eigenstate of the Hamiltonian \eqref{eq:mainBCS} is given by
\begin{equation}
\ket{\psi(t)} = \sum_{\{s^{z}\}}C_{\{s^{z}\}}e^{-iE_{\{s^{z}\}}t}\ket{\{s^z\}},  
\end{equation}
where $E_{\{s^z\}}$ is the eigenvalue of the  Zeeman term corresponding to the eigenvector $\ket{\{s^{z}\}}$.
The expectation value of any observable $\hat{\mathcal{B}}$ is written as
\begin{equation}
\label{eq:diagdistr}
\braket{\hat{\mathcal{B}}} = \sum_{\{s^{z}\}}\left|C_{\{s^{z}\}}\right|^2\braket{\{s^{z}\}|\hat{\mathcal{B}}|\{s^z\}}, 
\end{equation}
where the diagonal ensemble $\left|C_{\{s^{z}\}}\right|^2$ is given by $P_s({\{s^z\}})$ from Eq. ~\eqref{eq:Pszasymp}. Hence, $(2s+1)^N$ coefficients are required to fully determine the observable. At this point, we construct a statistical ensemble $\hat{\rho}_{\text{s.t.}}^{(s)}$ as a density matrix, that the system conforms to as $t\rightarrow\infty$. This ensemble incorporates a finite set of integrals of motion of the system, together with \textit{their integer powers}. Focusing on the $s=1$ case, these are $s_k^{z}$ and $(s_k^{z})^2$ for every site index $k$. The construction of the ensemble is given below:
\begin{equation}
\label{eq:distr}
\hat{\rho}_{\text{s.t.}}^{(1)} \propto \exp\left\{-\sum_k[\vartheta_{k,1}\hat{s}^{z}_k+\vartheta_{k,2}(\hat{s}^{z}_k)^2]\right\},   
\end{equation}
where $\vartheta_{k,1}, \vartheta_{k,2}\in\mathrm{C}$. In its diagonal basis, which are the eigenstates $\ket{\{s^z\}}$, the diagonal elements of \eqref{eq:distr} are
\begin{equation}
\label{eq:diagelm}
\rho_{\text{s.t.}}^{(1)}(\{s^z\}) = C_{\rho}e^{-\sum_k(\vartheta_{k,1}s^{z}_k+\vartheta_{k,2}(s^{z}_k)^2)}    
\end{equation}
where $C_{\rho}$ is a normalization constant. Here we require at most $2N$ coefficients $\vartheta$ to determine the distribution, against the $(2\times 1+1)^N$ coefficients $\left|C_{\{s^{z}\}}\right|^2$ in Eq.~\eqref{eq:diagdistr}. Comparing \eqref{eq:diagelm} with \eqref{eq:Pszthermlimit}, we find 
\begin{equation}
\begin{aligned}
\vartheta_{k,1} =&\frac{4\pi(k-\mu_{(1)})}{\nu},\\
\vartheta_{k,2}=&\ln\left(2\cosh\frac{\pi}{\nu}\right).
\end{aligned}
\end{equation}
This result is different from the case of $s=1/2$ described in \cite{zabalo_nonlocality_2022} where it was found that
\begin{equation}
\label{eq:s12distr}
\hat{\rho}_{\text{s.t.}}^{(1/2)} \propto \exp\left(-\sum_{k}\frac{2\pi(k-\mu_{(1/2)})}{\nu}\hat{s}^{z}_{k}\right).
\end{equation}
which is equivalent to the Generalized Gibbs Ensemble $\hat{\rho}_{\text{GGE}}$ for a integrable Hamiltonian $\hat{H}$, which is
\begin{equation}
\label{eq:GGEdist}
\begin{split}
&\hat{\rho}_{\text{GGE}} \propto \exp\left(-\sum_{k=1}^{n}\lambda_k \hat{I}_k\right)\\&[\hat{H},\hat{I}_k]=0,\;\; [\hat{I}_j,\hat{I}_k]=0,\;\;\lambda_k\in\mathrm{C}.
\end{split}
\end{equation}
In the spin-$1/2$ problem, the conserved quantities are $\hat{I}_k=\hat{s}^{z}_k$ for all $k$ and $\hat{H}$ is given by Eq. ~\eqref{eq:mainBCS} with the spin-$1/2$ operators. The authors in \cite{zabalo_nonlocality_2022} identify the distribution \eqref{eq:s12distr} for a system that starts at $t=0^+$ in its lowest energy state, which is Eq. ~\eqref{eq:BCSgsspins} with $s=1/2$, and evolves for a long time thereafter. Accordingly, they refer to this distribution as an ``emergent GGE'' to emphasize the dynamical context in which it arises. A key distinction in the spin-$1$ case is the presence the quadratic terms in Eq.~\eqref{eq:diagelm}, which means the system \textit{no longer conforms} to the form of the distribution given in Eq.~\eqref{eq:GGEdist}. The same deviation holds for the $s = 3/2$ and $s = 2$ cases. 

\subsection{Discrepancy from the spin 1/2 problem}\label{sec:Discrepancy}
\begin{figure}
\centering
\includegraphics[width = \linewidth]{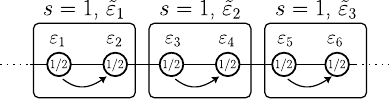}
\caption{Pictorial representation of combining spin-$1/2$ operators along a chain by deforming their local Zeeman fields to be equal. By introducing such degeneracies, a spin-$1$ RG Hamiltonian is constructed out of a spin-$1/2$ Hamiltonian.}
\label{fig:combining_spins}
\end{figure}

With the available expressions for the probability distributions $P_s({s^z})$ at different spin magnitudes $s$, we now comment on the deviations observed between the solutions of the spin-$1/2$ RG Hamiltonian and those for Hamiltonians with $s > 1/2$. To illustrate this, we revisit the example mentioned in the Introduction. Consider \eqref{eq:mainBCS} for an arbitrary spin magnitude $s$, but now with a \textit{time-independent} coupling $g$. Starting with $s=1/2$, after we introduce the same Zeeman magnetic field $\varepsilon_i$ for the site $i$ and its neighboring site $i+1$ in a pairwise manner, they can be combined into a single larger-spin using the standard Clebsch-Gordon decomposition. An example of this procedure is given below, where a pair of spin-$1/2$ $\hat{s}^{z}$ operators is transformed into a spin-$1$ $\hat{s}^{z}$ via a local unitary transformation $U$:
\begin{equation}
\begin{split}
&U\left(\frac{1}{2}\begin{bmatrix}1&\cdot\\\cdot&-1\end{bmatrix}\otimes\mathrm{I}_{2\times 2} +\mathrm{I}_{2\times 2}\otimes\frac{1}{2}\begin{bmatrix}1&\cdot\\\cdot&-1\end{bmatrix}\right)U^{-1} = \\&\hphantom{=}\begin{bmatrix}
    1&\cdot&\cdot\\\cdot&0&\cdot\\\cdot&\cdot&-1
\end{bmatrix} \oplus \mathbf{0}_{1\times 1};\;\;U= \begin{bmatrix}
\,1&\cdot&\cdot&\cdot\,\\\cdot&\frac{1}{\sqrt{2}}&\frac{1}{\sqrt{2}}&\cdot\\\cdot&\cdot&\cdot&1\\\cdot&\frac{1}{\sqrt{2}}&\frac{-1}{\sqrt{2}}&\cdot   
\end{bmatrix}.
\end{split}    
\end{equation}
Here $\mathbf{0}_{1\times 1}$ denotes the spin-$0$ (singlet) sector. The same procedure applies to $\hat{s}^\pm$ operators. By making disjoint pairings of all the sites (provided they are also even in number to begin with), we obtain the spin-$1$ Hamiltonian after applying local unitary transformations to every pair. We neglect the spin-$0$ components of the model, as the initial spin-$1/2$ state comprises only of triplet states which do not couple with singlet sectors. We explicitly consider an initial condition where all $\varepsilon_j$ are distinct, so that no degeneracies are present. However, one can imagine slowly tuning two adjacent Zeeman fields $\varepsilon_j$ and $\varepsilon_{j+1}$ to become equal, as illustrated schematically in Fig.~\ref{fig:combining_spins}. In this way, we effectively recover the full spin-$1$ problem. 

However, in the setting of explicit time dependence of the Hamiltonian in Eq.~\eqref{eq:mainBCS}, the same cannot be said for their dynamics, which is evident from the steady-state behavior of the spin-$1/2$ and the spin-$1$ systems. Starting with $P_{1/2}(\{s^z\})$, from Eq.~\eqref{eq:Pszasymp} with $s=1/2$ we find that
\begin{equation}
\begin{split}
\sum_{j}\frac{2\pi js^{z}_j}{\nu} &= \sum_{k}\left(\frac{2\pi (2k-1)s^{z}_{2k-1}}{\nu}+\frac{2\pi (2k)s^{z}_{2k}}{\nu}\right) \\
&= \sum_k\frac{4\pi k (s^{z}_{2k-1}+s^{z}_{2k})}{\nu}-\underbrace{\sum_k\frac{2\pi s_{2k-1}^{z}}{\nu}}_{\propto J^z}.
\end{split}
\end{equation}
Apart from a common prefactor---proportional to the total magnetization---we only recover the linear component of $P_{1}(\{s^z\})$ as given in Eq.~\eqref{eq:Pszasymp} for $s=1$. Crucially, there is no way to recover the quadratic components in $\varrho_j^{(1)}(\nu)$ \eqref{eq:varrhos1} from $P_{1/2}(\{s^z\})$. This implies that higher-spin dynamics cannot be inferred solely from lower-spin systems. Each spin-$s$ problem must therefore be treated independently.

\begin{figure*}
\centering
\includegraphics[width=\linewidth]{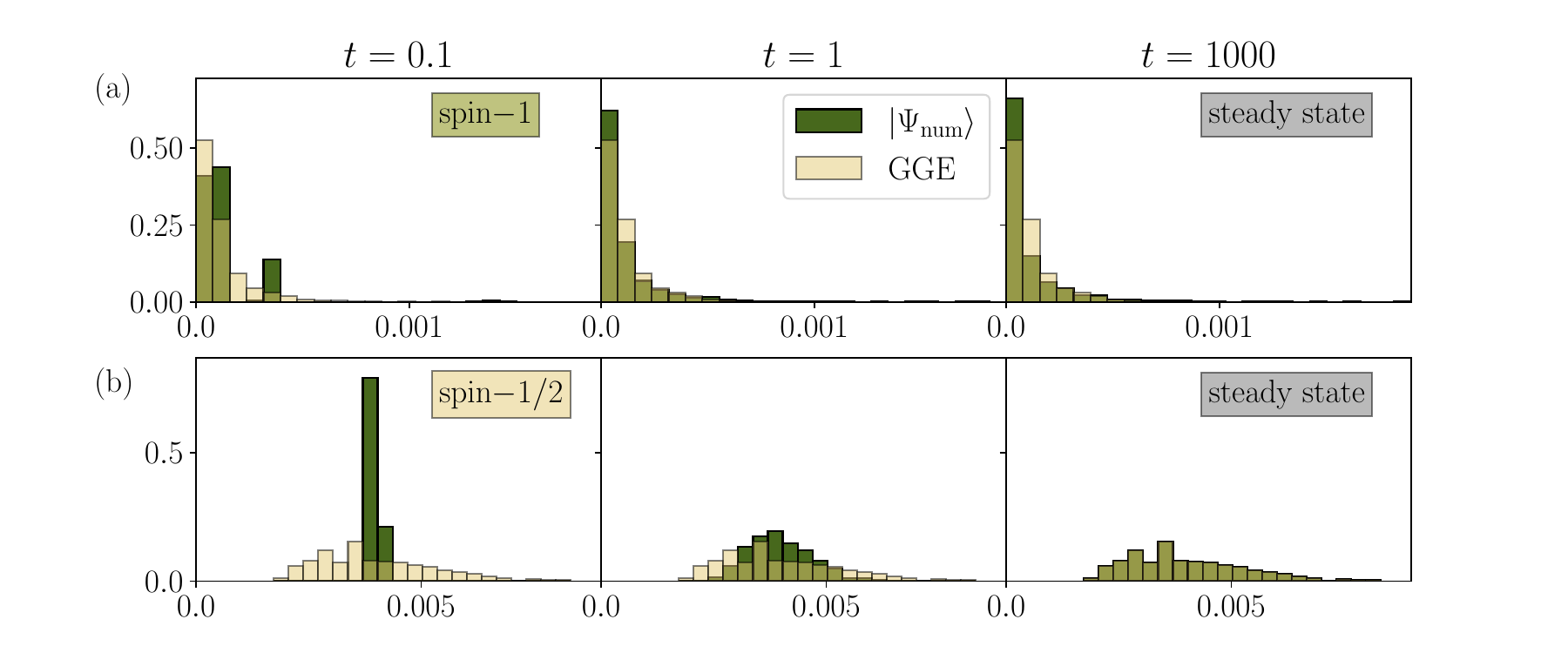}
\caption{Probability distribution of a numerically simulated wavefunction (green bars), evaluated at times $t = \{0.1,1,1000\}$ overlaid with the natural  asymptotic  GGE distribution given by 
Eq.~\eqref{eq:Pszasymp} with $\varrho_j^{(s)} = 0$ (semitransparent yellow). The top row (a) shows distributions for a spin-$1$ RG Hamiltonian with $N = 10$, $J^z = 0$, and $\nu = 70$, i.e., in the ‘intermediate-ramp’ regime. Bin width = $7.9\times 10^{-5}$. The bottom row (b) shows the same as the top row but for spin-$1/2$. Note how the spin-$1/2$ system conforms to the GGE distribution in the steady state, unlike the spin-$1$ system. Bin width = $3.3\times 10^{-4}$.}
\label{fig:numvsGGE}
\end{figure*}
This discrepancy remains for any given finite time. Ultimately, the exact wavefunctions of the spin-$1/2$ system do not account for the degenerate cases where $\varepsilon_j=\varepsilon_k$ for some site $j$ and $k$, due to the implicit assumption that all $\varepsilon_i$ are distinct. A direct consequence of this discrepancy is the non-analyticity of the probability transitions, as described in the ``Non-analytic behavior'' subsection of Sec. \ref{sec:analyticitybreakdown}. We illustrate this in the two site spin-$1/2$ problem in Fig. \ref{fig:Transition_Probabilities}. Recalling $\Delta=\varepsilon_2-\varepsilon_1$ and the generalized coupling $g(t)=1/(\nu t^{\alpha})$, we observe that as $\Delta\rightarrow0$, that is, $\varepsilon_2\to\varepsilon_1$, the transition probability for the $\alpha=1$ case becomes discontinuous. At $\Delta=0$, the system effectively reduces to a single-site spin-$1$ problem.

We also employ numerical simulations to determine the expected values of local magnetizations for the case of $\alpha \neq 1$, as shown in Fig. \ref{fig:Integrability_Check}, using various sets of $\varepsilon$ and spin magnitudes. For $\alpha \neq1$, these averages depend on the choice of the Zeeman term, whereas they remain unchanged for $\alpha=1$. This also demonstrates that the probability distributions, and thus the transition probabilities, are sensitive to $\varepsilon$ when the system deviates from the integrable point. Consequently, the transition probability curves for $\alpha\neq1$ in Fig. \ref{fig:Transition_Probabilities} vary continuously with $\Delta$.

\subsection{Experimental predictions}\label{sec:predictions} 
We finally summarize some of the qualitative predictions   that we expect to observe in appropriate experimental setups. 

\begin{figure}[ht]
\centering
\includegraphics[width=\linewidth]{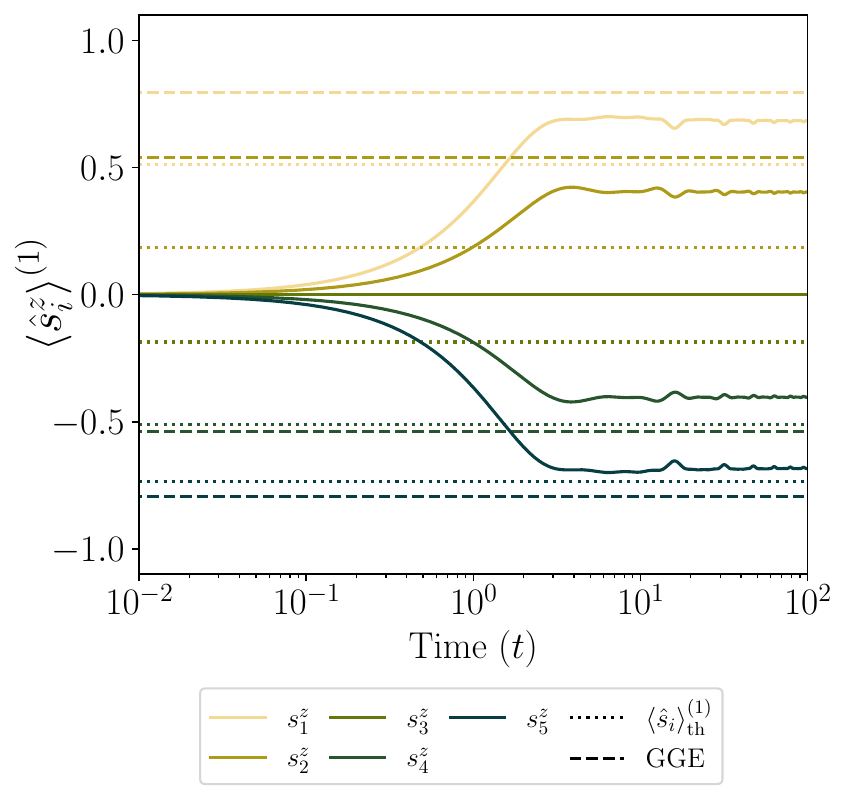}
\caption{Expectation values $\left<\hat{s}_i^z\right>^{\!\scale[0.60]{(1)}}$ for the $N=5$ spin$-1$ RG Hamiltonian as a function of time. Solid lines are numerical simulations. Dotted lines are thermodynamic limit predictions \eqref{eq:ThermLimit_Spin-1_sz_ExpectationValue} and dashed lines are the GGE  [Eq.~\eqref{eq:Pszasymp} with $\varrho^{(s)} = 0$] predictions. For this figure, $\nu = 5/0.3$, with all other parameters identical to those in Fig.~\ref{fig:overlap_incorrect}.
}
\label{fig:sz_expectation_values}
\end{figure}

\begin{figure}
\centering
\includegraphics[width=\linewidth]{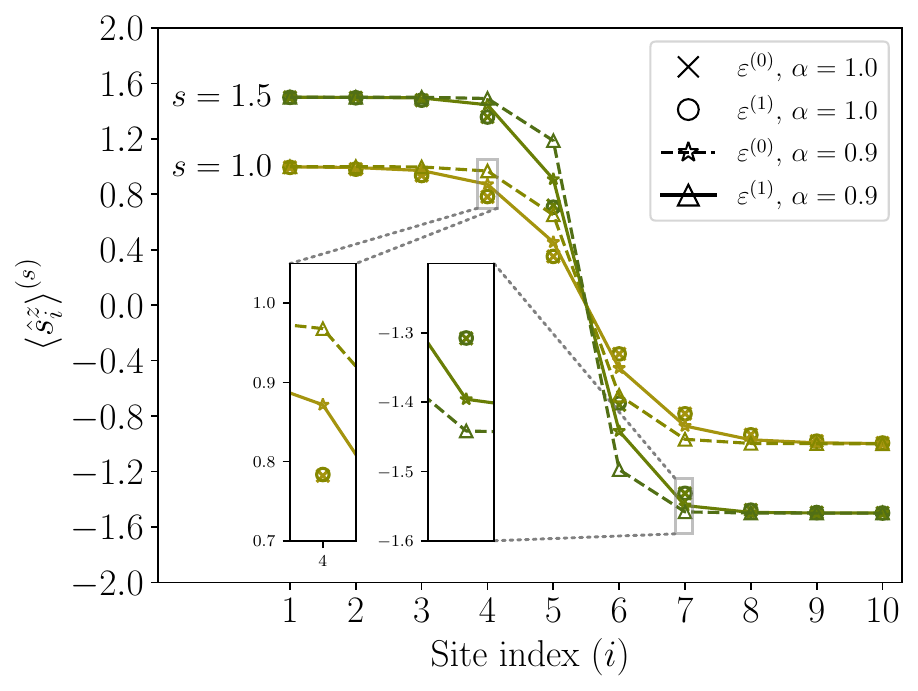}
\caption{
Numerical simulation of local magnetization per site for different spin magnitudes and integrability-breaking parameter $\alpha$, with the coupling $g(t)=1/(\nu t^{\alpha})$. For $\alpha = 1$, our system is integrable, and observables are independent of the specific choice of $\varepsilon$ (with the exception of degeneracies), as seen from \eqref{eq:ThermLimit_Spin-1_sz_ExpectationValue}. In this figure, $N=10$, $\nu=N/\eta$ with $\eta=1$, $\varepsilon_i^{(0)} = \frac{i}{N}$ and $\varepsilon_i^{(1)} = 0.49 + 0.002 \frac{i}{N}$ and both $\varepsilon^{(0)}_1 = \varepsilon^{(1)}_1 = 0.1$. For $\alpha \neq 1$, the expectation values vary depending on the choice of energies. All data points are evaluated at $t = 10^{4}$. For $s=1/2$, see 
Fig.~2 in \cite{zabalo_nonlocality_2022}.}
\label{fig:Integrability_Check}
\end{figure}

\begin{enumerate}
\item The  single-site magnetization   in the asymptotic state has a non-Gaussian distribution, as previously discussed in Sec. \ref{sec:highmomentssz}.  For higher spin values, the distribution becomes increasingly peaked, approaching a  delta function. Numerical results for a 5-site spin$-1$ RG Hamiltonian are compared against the thermodynamic and GGE predictions in Fig. \ref{fig:sz_expectation_values}.
\item For $s>1/2$, the probability distribution of a given configuration of local magnetizations \eqref{eq:Pszasymp} no longer conforms to the natural generalized Gibbs ensemble at late times, unlike for the $s=1/2$ case. We illustrate the expected results in Fig. \ref{fig:numvsGGE}.
\item These local magnetizations for the given $1/t$ coupling used in \eqref{eq:mainBCS} are special in that they remain invariant under variations of $\varepsilon$ in the Zeeman term. For non-integrable couplings of the form $g(t)\propto1/t^{\alpha},\;\alpha\neq1$, the local magnetization exhibit a dependence on $\varepsilon$, as we demonstrate in Fig. \ref{fig:Integrability_Check}.
\item The transition probabilities of the model exhibit discontinuities when $\varepsilon_j=\varepsilon_k,\;j\neq k$. In contrast, we suggest  that such discontinuities do not arise for non-integrable couplings. The integrable point at $\alpha = 1$ distinctly exhibits this discontinuous behavior.
\end{enumerate}

\section{Towards experimental realizations}\label{sec:Experimental_Realizations}

In this section we sketch possible pathways towards an experimental realization of the Hamiltonian central to this work. For convenience we repeat the definition of the Hamiltonian below:
\begin{subequations}
\begin{align}
\hat{H}(t) &= \hat{H}_z + \hat{H}_{\rm int} \\
\hat{H}_z & = 2\sum_{j=1}^{N}\varepsilon_{j}\hat{s}_{j}^{z} \\
\hat{H}_{\rm int} &= -g(t)\sum_{j,k=1}^{N}\hat{s}_{j}^{+}\hat{s}_{k}^{-},\;\;g(t)=\frac{1}{\nu t}.
\end{align}
\label{App_eq:Hamiltonian}
\end{subequations}
One possibility for simulating the time-dependent Hamiltonian (\ref{App_eq:Hamiltonian}) is on a quantum computer. To this end, we note a recent experiment with superconducting circuits, in which ground states of the RG model were prepared \cite{OBrien_2023_NatPhys}. Given the all-to-all nature of the spin-flip interaction, other candidate platforms include cold atoms coupled to a cavity or spin Hamiltonians engineered in trapped-ion systems. 

\subsection{General considerations}
First, we address the question of the relevant energy and time scales. The energy scale of the spin-flip interaction term in the Hamiltonian is given by $g(t)$. To characterize the energy scale of the Zeeman term, we consider the standard deviation of the set of Zeeman fields $\{\varepsilon_1,\dots,\varepsilon_N\}$ 
\begin{equation}
\Delta \varepsilon = \left[ \frac{1}{N} \sum_{i=1}^N \left( \varepsilon_i - \bar{\varepsilon} \right)^2 \right]^\frac{1}{2},
\end{equation}
where $\bar{\varepsilon}$ is the mean energy. For the linear ramping of the Zeeman fields, $\varepsilon_i = i/N$, as considered in this article, we obtain
\begin{equation}
\Delta \varepsilon^2 = \frac{1}{12} \frac{(N+1) N (N-1)}{N^3},
\end{equation}
which in the large $N$ limit sets $\Delta \varepsilon^2 \approx 1/12$. With these definitions at hand we define a scaled (dimensionless) time as
\begin{equation}
\tilde{t} = \frac{\Delta \varepsilon}{g(t)} = \Delta \varepsilon \nu t.
\label{eq:scaled_time}
\end{equation}
In our numerical simulation (not shown), we observe that the scaled time $\tilde{t}_{\rm ss}$, after which the steady state is reached to a reasonable accuracy, only depends weakly on the values of $\Delta \varepsilon$ and $\nu$ such that $\tilde{t}_{\rm ss} \approx {\rm const.}$ for a given $N$. Specifically, for $N \lesssim 10$ we find that $\tilde{t}_{\rm ss} \sim O(100)$ and that it increases with $N$. This is consistent with the results of Ref. \cite{zabalo_nonlocality_2022}, which analyzed the spin-$1/2$ model in large-$N$ and mean-field limits, finding that for $\tilde{t} \sim N^2$ one achieves steady state predictions for $\braket{\hat{s}^z_j}$ with considerable accuracy.

Based on these estimates, we find that the times required to reach the steady state (or even to observe non-trivial transient dynamics) are prohibitive in the cavity setting, which typically involves a large number of atoms ($10^5$). We refer the reader to Appendix~\ref{app:Cavity} for further details. There, we also briefly discuss the possibility of using the cavity setup to probe a macroscopic spin limit of the model~(\ref{App_eq:Hamiltonian}). For the remainder of this section, we therefore focus on trapped-ion setups, which permit controlled manipulation of between 1 and $\sim 100$ ions.

\subsection{Trapped ions}\label{sec:Ions}
One of the approaches to implement \eqref{App_eq:Hamiltonian} is using trapped ions. To do this, three key ingredients are needed, all of which have been demonstrated experimentally. These are
\begin{enumerate}[(i)]
\item all-to-all \emph{equal} interactions,
\item operate spin-1 qudits instead of qubits,
\item being able to simulate the evolution under the full Hamiltonian \eqref{App_eq:Hamiltonian} which contains both the spin-exchange interaction ($\propto \hat{s}^+ \hat{s}^-$) and the transverse field ($\propto \hat{s}^z$).
\end{enumerate}

The common starting point for deriving effective spin Hamiltonians in 
trapped ion experiments is the native Hamiltonian that describes the interaction of a system of $N$ trapped ion qubits with a laser light of frequency $\omega_L$ and $k$-vector $k_L$ \cite{monroe_programmable_2021, schneider_experimental_2012,porras_effective_2004} (In this Section we use ``qubits" for the spin-1/2 systems and ``qudits" for higher spin systems. The usual Pauli matrices $\sigma_{x,y,z}$ used below pertain to the qubit basis.). Setting $\hbar=1$, we have
\begin{equation}
\hat{H} = \sum_{i=1}^N \Omega_i \cos(\mu t + \varphi_i) (\vec{k} \cdot \vec{x}_i) \, \sigma^i_{\theta_i},
\label{App_eq:H_ion}
\end{equation}
where $\Omega_i$ is the Rabi frequency coupling the two qubit levels of the $i$-th ion, $\mu = \omega_L - \omega_0$ with $\omega_0$ the qubit transition frequency, $\varphi_i$ is an arbitrary offset phase,
\begin{equation}
\sigma^i_{\theta_i} = \sigma^x_i \cos(\theta_i) + \sigma^y_i \sin (\theta_i)
\end{equation}
and
\begin{equation}
\vec{k} \cdot \vec{x}_j = \sum_{m} \eta_{jm} \left( a^\dag_m {\rm e}^{i \omega_m t} + a_m {\rm e}^{-i \omega_m t} \right).
\end{equation}
Here
\begin{equation}
\eta_{im} = |\vec{k}_L| b_{im} \sqrt{\frac{1}{2M\omega_m}}
\end{equation}
is the so-called Lamb-Dicke parameter corresponding to the coupling of the $m$-th motional eigenmode of frequency $\omega_m$ of the ion crystal to the motion of the $i$-th ion $\vec{x}_i$, and $M$ is the ion mass. A Magnus expansion is used to derive the unitary evolution operator $U(t) = {\cal T} {\rm exp}(-\int^t {\rm d}t' \, H(t'))$ corresponding to the (explicitly time-dependent) Hamiltonian \eqref{App_eq:H_ion}. After what are by now standard derivations and approximations, in particular keeping only the secular terms in $U(t)$, in case of \eqref{App_eq:H_ion} this leads to \cite{kotibhaskar_programmable_2024}
\begin{equation}
\hat{H}_{\rm eff} = \sum_{i<j} J_{ij} \sigma^i_{\theta_i} \sigma^j_{\theta_j},
\label{App_eq:Heff}
\end{equation}
with 
\begin{equation}
J_{ij} = k^2_L \frac{\Omega_i \Omega_j}{2M} \sum_m \frac{b_{im} b_{jm}}{\mu^2 - \omega_m^2}.
\label{App_eq:Jij}
\end{equation}
We are in the position to discuss the ingredients (i)--(iii) of the implementation. 

\begin{enumerate}[(i)]
\item It is well known that the range of effective interactions $J_{ij}$ follows a power-law decay, i.e., $J_{ij} \propto 1/|i-j|^\alpha$, where the exponent $\alpha$ lies within the range $0 \leq \alpha \leq 3$ and can be tuned via the laser de-tuning $\mu$. In particular, all-to-all equal couplings, corresponding to $\alpha = 0$, can be realized by tuning $\mu \approx \omega_{\text{COM}}$, close to the so-called center-of-mass (COM) mode, as demonstrated experimentally, e.g., in \cite{britton_engineered_2012}.
\item In principle, the above equations \eqref{App_eq:H_ion}–\eqref{App_eq:Jij} can be readily generalized to qudits, and specifically to spin-1. To this end, there is ongoing work to extend the ion-based quantum simulation toolbox to qudits, including the experimental demonstration of a universal gate set for qudits \cite{ringbauer_universal_2022}, as well as further theoretical developments of qudit gates for ions \cite{low_practical_2020}.
\item A more subtle point concerns the engineering of an analogue version of the Hamiltonian \eqref{App_eq:Hamiltonian}. While the Magnus expansion-based derivation of $H_{\rm eff}$ closes for spin-spin interaction Hamiltonians of the form \eqref{App_eq:Heff}, it includes higher-order contributions if an effective magnetic field \begin{equation}
    \hat{H}_B = \sum_{i=1}^N B_i \sigma^z_i
\end{equation}
is also included in \eqref{App_eq:H_ion}. Here, $B_i$ can be achieved by additional lasers resonant with the qubit transition \cite{porras_effective_2004, schneider_experimental_2012}. This would then lead to a correction term \cite{wang_intrinsic_2012}, $\hat{H}_{\text{Error}}$ to Eq.~\eqref{App_eq:Hamiltonian}, the effect of which on the time dynamics should be carefully examined and evaluated.

For this reason, a popular scheme is to implement time evolution using the Trotter-Suzuki expansion of the time-evolution operator. This approach has the advantage that one applies a sequence of spin-exchange Hamiltonians $\sum_{ij} \hat{s}^+_j \hat{s}^-_i$  interleaved with the transverse field ones $\sum_i \varepsilon_i \hat{s}^z_i$, which individually do not suffer from the additional errors $H_{\text{Error}}$. Experimental demonstrations of such a discretised time evolution include, for instance, the implementation of the transverse-field Ising model \cite{lanyon_universal_2011} and variational quantum simulations in quantum chemistry \cite{nam_ground-state_2020}. 
\end{enumerate}

\subsubsection*{Experimental estimates} 
\begin{figure*}[ht]
\centering
\includegraphics[width=\linewidth]{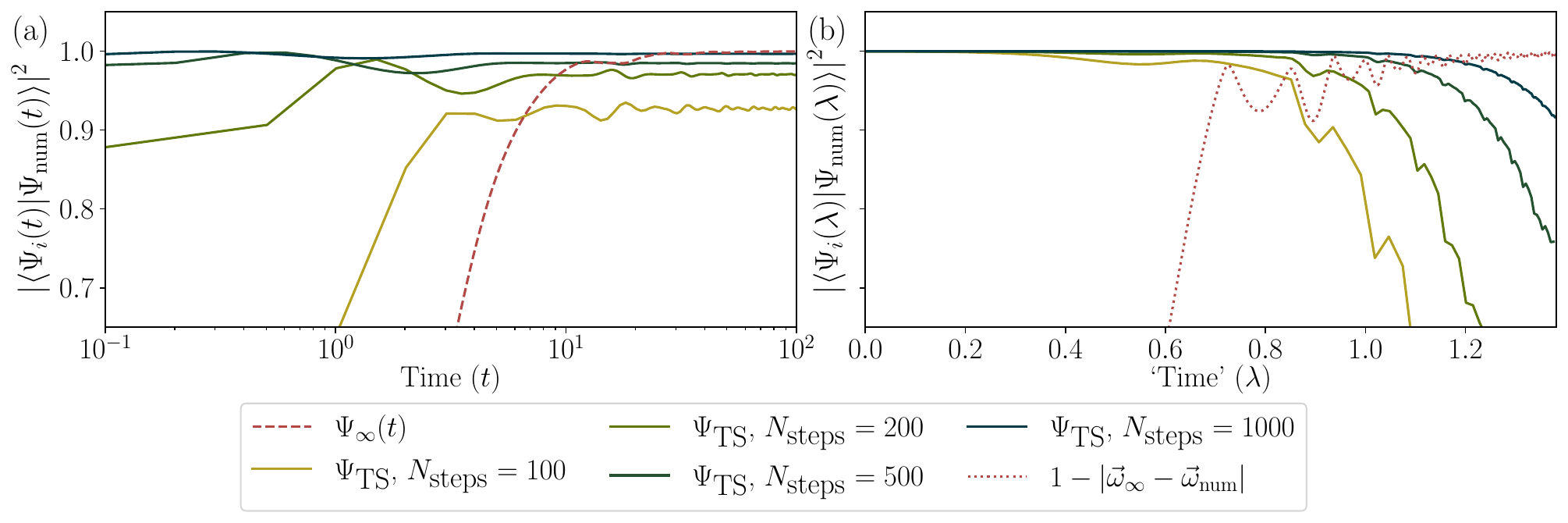}
\caption{Examples of Trotter-Suzuki (TS) implementation. (a) TS approximation of Eq.~\eqref{eq:Trotter} for varying amounts of $N_{\text{steps}}$. For comparison, the overlap with the asymptotic wavefunction is shown (dashed, red), indicating how long the TS protocol should be run to reach a steady state. (b) TS implementation of Eq.~\eqref{eq:tdbs_exponential}. Since no asymptotic wavefunction has been calculated for Eq.~\eqref{eq:tdbs_exponential}, we instead plot (in dotted, red) the quantity $1-|\vec{w}_\infty - \vec{w}_{\rm num}|$, which quantifies the similarity between the weights of $\ket{\Psi_\infty(t=10^3)}$ and $\ket{\Psi_{\rm num}(\lambda)}$. This measure reaches 1 when $\vec{w}_\infty = \vec{w}_{\rm num}$, indicating the onset of the steady state.
Here $|\cdot|$ denotes the vector norm and $w_j = |c_j|^2$ are the weights of wavefunction coefficients $\ket{\Psi} = \sum_j c_j \ket{b_j}$ with $\{ \ket{b_j} \}$ the set of basis states. $\ket{\Psi_{\rm num}(\lambda)}$ is the numerically computed wavefunction of \eqref{eq:tdbs_exponential}.
For these panels, $N = 5$ and other parameters are the same as in Fig \ref{fig:overlap_incorrect}. 
}
\label{fig:Trotter_combined}
\end{figure*}

For the experimental estimate, we use values from a recent experiment \cite{kotibhaskar_programmable_2024} that realized an instance of an XY Heisenberg model. For isotropic couplings $J_x = J_y = J \approx 2\pi \times 100 , {\rm Hz}$, the model reduces to the spin-exchange Hamiltonian $\hat{H} = J \sum_{i \neq j} \sigma^+_i \sigma^-_j$. In what follows, we consider the spin-1 case. In Fig.~\ref{fig:Trotter_combined}(a), we show the Trotterized time evolution according to
\begin{equation}
\ket{\Psi(t)} = \prod_{j=1}^{N_{\rm steps}} {\rm e}^{-i \hat{H}_z(t_j) \Delta t} {\rm e}^{-i \hat{H}_{\rm int}(t_j) \Delta t} \ket{\Psi(0)},
\label{eq:Trotter}
\end{equation}
where $\ket{\Psi(0)}$ is the initial BCS state given in Eq.~\eqref{eq:BCSgsspin1}. For simplicity, we consider $N_{\rm steps}$ Trotter steps of constant duration $\Delta t = t/N_{\rm steps}$, such that $t_j = j \Delta t$. In practice, we consider an initial time $t_i > 0$, so that $t_j = j \Delta t + t_i$, and the step duration becomes $\Delta t = (t - t_i)/N_{\rm steps}$. We observe that, for the studied example of $N=5$ spins, a steady state is reached with good accuracy using $N_{\rm steps} \approx 1000$. In numerical simulations throughout this paper, the time $t$  was implicitly expressed in units of 
$1/\varepsilon_N$ (with $\varepsilon_N = N/N = 1$), which corresponds (for large $N$) to the bandwidth $W_\varepsilon = \varepsilon_N - \varepsilon_1 \approx \varepsilon_N$.

To make a connection with the experiment, we consider a maximum duration of $T = 50\; {\rm ms}$ \cite{Neyenhuis_2017_SciAdv}, which, for $N_{\rm steps} = 1000$, yields a time step of $\Delta t = 50\; \mu{\rm s}$ and sets the corresponding energy scale $\varepsilon_N \approx 2\pi \times 30\; {\rm Hz}$---on the same order as $J$. For the evolution  in the time interval $t \in [10^{-3}, 10]$, which is sufficient to  reach the steady state, and for $\nu = N$, the corresponding maximum required value of $g(t)$ is $g(t_i) \approx 2\pi \times 6.4\; {\rm kHz}$. This value can be further reduced by taking larger initial times $t_i$.

Before commenting on these results, we analyze one more scenario. Instead of varying $g$, it may be more practical in experiments to tune the energies $\varepsilon_i$ while keeping the interaction strength $g$ constant. To this end, consider the variable transformation $t = e^{\nu \lambda}$. In this case, the Schrödinger equation for the Hamiltonian in Eq.~\eqref{eq:mainBCS} becomes
\begin{equation}
\label{eq:tdbs_exponential}
i\frac{\partial\ket{\psi(\lambda)}}{\partial \lambda} = \left[2\nu e^{\nu \lambda}\sum_{i=1}^{N}\varepsilon_i\hat{s}^{z}_i-\sum_{i,j=1}^{N}\hat{s}_j^+\hat{s}_k^{-}\right]\ket{\psi(\lambda)}.
\end{equation}
With this change, the time dependence effectively shifts to the Zeeman term, such that the interval $t\in(0^+,\infty)$ becomes $\lambda\in(-\infty,\infty)$. Similar to the simulation of Eq.~\eqref{eq:Trotter}, in Fig.~\ref{fig:Trotter_combined}(b) we consider the Trotterized time evolution, but now according to the Hamiltonian given in Eq.~\eqref{eq:tdbs_exponential}. Here, we restore the physical dimensions by rescaling $\lambda \rightarrow \lambda/J$, where $J$ is the (constant) interaction energy scale.  Considering the time evolution over the interval $t \in [10^{-3}, 10]$ and using the substitution $t = {\rm e}^{\nu \lambda}$, we obtain the corresponding dimensional Trotter time step as $\Delta \lambda = {\rm span}(\lambda)/N_{\rm steps}/J \approx 3 \; \mu{\rm s}$, with the total duration $3 \; {\rm ms}$. The respective minimum and maximum values of $\varepsilon_N = 2 \nu J {\rm e}^{\nu \lambda} $ are $2 \pi \times 1 \; {\rm Hz}$ and $2\pi \times 10 \; {\rm kHz}$. 

Based on the above estimates, we find that simulating the time evolution of the Hamiltonian \eqref{eq:mainBCS} [or its variant \eqref{eq:tdbs_exponential}] is within reach of current trapped ion experiments. One important remark is in order. In the above analysis, we have assumed a constant time step while dealing with a time-dependent Hamiltonian. A direct consequence of this choice is a deviation from the exact numerical solution of the Schrödinger equation for the Hamiltonians \eqref{eq:mainBCS} and \eqref{eq:tdbs_exponential}---in the former case, rapid changes in the Hamiltonian occur early in the dynamics, whereas in the latter, they occur at late times. This is reflected in the discrepancies between the Trotterized and exact evolutions shown in Figs.~\ref{fig:Trotter_combined}a and \ref{fig:Trotter_combined}b. A significant improvement in accuracy can be achieved by using an adaptive Trotter scheme, in which the time sampling and step durations are optimized based on the structure of the Hamiltonian \cite{Zhao_2024_PRL, Zhao_2023_PRXQuantum}. We leave a detailed investigation of this promising approach for future work.

\section{Conclusion and Discussion}
In this work, we have investigated the late-time dynamics of a time-dependent Richardson-Gaudin model with interaction strength inversely proportional to time, for arbitrary spin $s > 1$. Starting from the off-shell Bethe ansatz solution of the model in terms of a multiple contour integral, we applied the saddle point method to obtain the asymptotic wavefunction and subsequently identified a correction term, which we verified numerically.

We explicitly derived the full many-body asymptotic wavefunctions for spin-$1$ and spin-$3/2$ in terms of elementary functions and the gamma function $\Gamma(x)$. For the general spin-$s$ case, we obtained the solution up to a time-independent correction term. Determining this term requires solving the two-site Richardson-Gaudin model for spin-$s$. For spin-$2$, we proposed a candidate correction term in Appendix~\ref{sec_App:highspinrep} that appears to be numerically exact, while obtaining an explicit expression for the correction at $s > 2$ remains an open problem.

We used the asymptotic wavefunction to derive several results and make a range of observations. One of the central findings is that, in contrast to the time-independent case, higher-spin solutions of time-dependent Richardson-Gaudin Hamiltonians cannot be generated by fusing lower-spin degrees of freedom. For instance, the asymptotic probability distribution of the local magnetization, \( P_s\left(\{s_j^z\}\right) \), includes quadratic terms \( (s_j^z)^2 \) for \( s = 1 \) and \( s = 3/2 \), and a quartic term \( (s_j^z)^4 \) for \( s = 2 \), whereas in the spin-\(1/2\) case the distribution is linear in \( s_j^z \). These nonlinear contributions cannot be reproduced by combining neighboring spin-\(1/2\) sites. This nonanalytic behavior reflects the intrinsic complexity of the time-dependent problem and emphasizes the necessity of analyzing each spin-\(s\) model independently, rather than expecting higher-spin dynamics to emerge from lower-spin cases.

The validity of the mean-field approximation depends on the type of observable under consideration. A distinction must be made between local and nonlocal observables. Local observables are defined as sums of products of spin operators in which the number of operators per term remains finite in the thermodynamic limit \( N \to \infty \), where \( N \) is the total number of spins. We have shown that all such local observables agree exactly with the corresponding mean-field predictions to leading order in \( 1/N \). In contrast, the mean-field approximation fails for nonlocal observables, such as the Loschmidt echo and bipartite entanglement entropy.

At large times, the interaction term vanishes and the \( z \)-components of the spins, \( \hat{s}_j^z \), emerge as natural integrals of motion. In the spin-\(1/2\) case, it has been previously shown that the steady state is described by a generalized Gibbs ensemble (GGE) constructed from these emergent conserved quantities~\cite{zabalo_nonlocality_2022}. However, we explicitly demonstrated that this description breaks down for \( s > 1/2 \), where the steady-state density matrix contains higher powers of \( \hat{s}_j^z \), indicating a more intricate structure that lies beyond the standard GGE framework.

Finally, we discussed possible avenues for experimentally realizing the spin-$s$ Richardson-Gaudin model. While our analysis of these implementations is rudimentary at best, we are confident that the predictions made in this work can be tested using one or more of the proposed experimental platforms. Given the scarcity of exact solutions for interacting, time-dependent quantum models, the model we present offers an interesting testing ground for exploring time-dependent many-body quantum mechanics.

\begin{acknowledgments}
The work of S.B. and V.G. is partially supported by the Delta Institute for Theoretical Physics (DITP). The DITP consortium, a program of the Netherlands Organization for Scientific Research (NWO), is funded by the Dutch Ministry of Education, Culture and Science (OCW). The work of L.B. is partially supported by the Institute of Theoretical Physics Amsterdam (ITFA) at the University of Amsterdam (UvA). J.M. is supported by the Dutch Research Council (NWO/OCW) as part of the Quantum Software Consortium program (Project No. 024.003.037), Quantum Delta NL (Project No. NGF.1582.22.030), and ENW-XL grant (Project No. OCENW.XL21.XL21.122). The authors also thank the Delta Institute for Theoretical Physics and the Institute of Physics at the University of Amsterdam for graciously hosting E.Y., which ultimately resulted in this work. Finally, the authors are grateful to E. Demler for his insights on experimental realizations of the present work and P. Corboz for providing access to the high-performance workstation `Galatea'. 
\end{acknowledgments}

\section*{Data availability}
All code used to generate the figures in this work is publicly available at~\cite{bakker_exact_2024}.

\appendix
\section{Exact solution of the two-site RG model}
In this appendix, we present exact solutions to the Schr\"odinger equation for Eq.~\eqref{eq:mainBCS} with two sites, $N = 2$. For spin-1, the full set of solutions is provided, organized according to the sectors in the block-diagonalized Hamiltonian. For spin-3/2, we give only the asymptotic solutions in the two relevant magnetization sectors. We will use the notation
\begin{equation}
\begin{split}
\tau = (\varepsilon_2 - \varepsilon_1)t,\\
r = \frac{\varepsilon_1 + \varepsilon_2}{\varepsilon_2 - \varepsilon_1}.
\end{split}
\end{equation}

\subsection{Spin 1}\label{sec:Appendix_N=2DiffEqsSol}
For $J^z = -2$ magnetization sector, the solution $\ket{\psi_{J^z}^{(N,s)}}$ is trivial:
\begin{subequations}
\begin{equation}
\ket{\psi_{-2}^{\scale[0.55]{(2,1)}}(t)} = \exp{\left[2 i (\varepsilon _1+ \varepsilon _2)t\right]}\ket{-1,-1}.    
\end{equation}   

For $J^z = \pm 1$, the solutions are:
\begin{equation}\label{App_eq:2-site-jz=-1}
\begin{split}
\ket{\psi_{-1}^{\scale[0.55]{(2,1)}}(t)} &= \left[\frac{\pi}{2\cosh\left(\frac{2\pi}{\nu}\right)}\right]^{\frac{1}{2}}  e^{i r \tau} \tau^{\frac{1}{2}+\frac{2 i}{\nu }}\\& \left[J_{\frac{1}{2}+\frac{2 i}{\nu }}(\tau)\ket{2} - i J_{-\frac{1}{2}+\frac{2 i}{\nu }}(\tau) \ket{1} \right],
\end{split}    
\end{equation}
\begin{equation}\label{App_eq:2-site-jz=+1}
\begin{split}
\ket{\psi_{1}^{\scale[0.55]{(2,1)}}(t)} &= \left[\frac{\pi}{2\cosh\left(\frac{2\pi}{\nu}\right)}\right]^{\frac{1}{2}}  e^{-i r \tau} \tau^{\frac{1}{2}+\frac{4 i}{\nu }} \\&\left[J_{\frac{1}{2}+\frac{2 i}{\nu }}(\tau)\ket{2} + i J_{-\frac{1}{2}+\frac{2 i}{\nu }}(\tau) \ket{1} \right].
\end{split}
\end{equation}

In Eqs. \eqref{App_eq:2-site-jz=-1} and \eqref{App_eq:2-site-jz=+1}, we introduce the following basis states:
\begin{equation}
\begin{aligned}
    \ket{1} = & \frac{1}{\sqrt{2}}\left(\ket{\pm 1,0} + \ket{0,\pm1} \right), \\
    \ket{2} = &\frac{1}{\sqrt{2}}\left(\ket{\pm 1,0} - \ket{0,\pm1} \right),
\end{aligned}
\end{equation}
where the $\pm$ refers to the basis states in \eqref{App_eq:2-site-jz=-1} and \eqref{App_eq:2-site-jz=+1} respectively.
Then we have for $J^z = 2$:
\begin{equation}
\ket{\psi_{2}^{\scale[0.55]{(2,1)}}(t)} = \exp{\left[-2it \left(\varepsilon _1+\varepsilon _2\right)+\frac{4 i\log (t)}{\nu}\right]}\ket{1,1}. 
\end{equation}   
\end{subequations}

The solution for $J^z=0$ wave function is given using the following basis
\begin{equation}
\begin{split}
\ket{1} &= \frac{1}{\sqrt{6}}\left(\ket{-1,1}+2\ket{0,0}+\ket{1,-1}\right), \\
\ket{2} &= \frac{1}{\sqrt{2}}\left(\ket{1,-1}-\ket{-1,1}\right), \\
\ket{3} &= -\frac{1}{\sqrt{3}}\left(\ket{-1,1}-\ket{0,0}+\ket{1,-1}\right). \\
\end{split}
\end{equation}
This allows us to write
\begin{equation}
\ket{\psi_{0}^{\scale[0.55]{(2,1)}}(t)} = \phi_{1}(t)\ket{1}+\phi_{2}(t)\ket{2}+\phi_{3}(t)\ket{3},
\end{equation}
with
\begin{equation}
\begin{split}
\phi_1(\tau) =& \tau^{\frac{6i}{\nu}} \;_1F_2\left(\text{\footnotesize$\frac{i}{\nu};\frac{1}{2}+\frac{2i}{\nu},\frac{3i}{\nu};-\tau^2$}\right), \\  
\phi_2(\tau) =&  c_1 \tau^{\frac{6i}{\nu}}\;_1F_2\left(\text{\footnotesize $\frac{\nu+i}{\nu};\frac{3}{2}+\frac{2i}{\nu},1+\frac{3i}{\nu};-\tau^2$}\right),  \\ 
\phi_3(\tau) =& \frac{\tau^{\frac{6i}{\nu}}}{\sqrt{2}}\;_1F_2\left(\text{\footnotesize$\frac{i}{\nu};\frac{1}{2}+\frac{2i}{\nu},\frac{3i}{\nu};-\tau^2$}\right)\\&-\frac{\tau^{\frac{6i}{\nu}}}{\sqrt{2}}\,_1F_2\left(\text{\footnotesize$\frac{\nu + i}{\nu};\frac{3}{2}+\frac{2i}{\nu},1+\frac{3i}{\nu};-\tau^2$}\right)\\ &+c_2 \tau^{\frac{6i}{\nu}}\;_1F_2\left(\text{\footnotesize$\frac{2\nu + i}{\nu};\frac{5}{2}+\frac{2i}{\nu},2+\frac{3i}{\nu};-\tau^2$}\right), \\
c_1 =& \frac{2\nu i \tau}{(4i+\nu)\sqrt{3}},\\c_2=&\frac{2\sqrt{2}\nu^2(\nu+i)\tau^2}{(\nu+3i)(\nu+4i)(3\nu+4i)}.
\end{split}    
\end{equation}
\subsection{Spin 3/2}\label{sec_App:s32N2asym}
Given the already lengthy presentation of the $J^z=0$ solution to two-site $s=1$ problem, we only present the asymptotic solutions for $J^z=\pm 1$:
\begin{subequations}
\begin{equation}
\begin{split}
&\lim_{t\rightarrow\infty}\ket{\psi_{-1}^{(2,\sfrac{3}{2})}(t)} = \mathcal{N}(\nu)\mathcal{C}_1(t,\nu,\boldsymbol{\varepsilon})\times\\
&\qquad e^{-\gamma_{0,1}}\bigg[e^{-\frac{6\pi}{\nu}}e^{-4i\varepsilon_{1}t}\ket{\sfrac{1}{2},-\sfrac{3}{2}}\\
&\qquad +e^{-(\gamma_{2,0}-\gamma_{0,1})}e^{-\frac{9\pi}{\nu}}e^{-2i(\varepsilon_1+\varepsilon_{2})t}\ket{-\sfrac{1}{2},-\sfrac{1}{2}}\\
&\qquad +e^{-\frac{12\pi}{\nu}}e^{-4i\varepsilon_{2}t}\ket{-\sfrac{3}{2},\sfrac{1}{2}}\bigg],\\
&\mathcal{C}_1(t,\nu,\boldsymbol{\varepsilon}) = t^{\frac{2i}{\nu}}e^{2i(\varepsilon_1+\varepsilon_2)t}e^{\frac{9\pi }{\nu }},
\end{split}
\end{equation}  
\begin{equation}
\begin{split}
&\lim_{t\rightarrow\infty}\ket{\psi_{1}^{(2,\sfrac{3}{2})}(t)} = \mathcal{N}(\nu)\mathcal{C}_2(t,\nu,\boldsymbol{\varepsilon})\times\\
&\qquad e^{-\gamma_{1,0}}\bigg[e^{-\frac{15\pi}{\nu}}e^{-6i\varepsilon_{1}t-2i\varepsilon_{2}t}\ket{\sfrac{3}{2},-\sfrac{1}{2}}\\
&\qquad +e^{-(\gamma_{0,2}-\gamma_{1,0})}e^{-\frac{18\pi}{\nu}}e^{-4i(\varepsilon_1+\varepsilon_{2})t}\ket{\sfrac{1}{2},\sfrac{1}{2}}\\
&\qquad +e^{-\frac{21\pi}{\nu}}e^{-2i\varepsilon_{1}t-6i\varepsilon_{2}t}\ket{-\sfrac{1}{2},\sfrac{3}{2}}\bigg],\\
&\mathcal{C}_2(t,\nu,\boldsymbol{\varepsilon}) = t^{\frac{4i}{\nu}}e^{4i(\varepsilon_1+\varepsilon_2)t}e^{\frac{18\pi }{\nu }},\\
\end{split}
\end{equation}  
\end{subequations}
with
\begin{equation*}
\mathcal{N}(\nu) = \frac{\Gamma \left(\frac{1}{2}+\frac{3 i}{\nu }\right) \Gamma \left(\frac{5 i+\nu }{\nu }\right)}{\sqrt{5 \pi } \Gamma \left(\frac{2 i+\nu }{\nu }\right)},
\end{equation*}
and the $\gamma$ functions in the asymptotics are defined in Eq.~\eqref{eq:correctedgammafunc32}.

\section{Spin 2 asymptotics}\label{sec_App:highspinrep}
We present here the expression for $\gamma_{N_1,N_2,N_3}^{\times}$  obtained directly from a saddle point calculation. This involves a numerical reduction of the roots of the Richardson equation and ultimately yields the following expression:
\begin{equation}\label{App_eq:Gamma_Spin2_Incorrect}
\begin{aligned} \gamma^{\times}_{N_1,N_2,N_3} = &-\frac{1}{\nu}(N_1 + \frac{4}{3}N_2 + N_3) \times\\
&\Bigg\{3i\ln(\nu t) + \frac{3\pi}{2} -i\left[\ln(32) -3\right]\Bigg\}\\
&+\frac{i}{\nu}N_2\ln\left(\frac{81 \sqrt[3]{2}}{64} \right).
\end{aligned}
\end{equation}
We also present a partial solution to $\gamma_{N_1,N_2,N_3}$ which we obtain from available exact results
\begin{equation}\label{App_eq:Gamma_Spin2_Guess}
\begin{split}
&\gamma^{\text{(partial)}}_{N_1,N_2,N_3} =
-\left(\frac{N_1}{2}+N_2+\frac{N_3}{2}\right) h(\nu) \\  
&-\frac{(N_1+N_3)}{2}\ln \left(\frac{2 \sqrt{6 \pi } \Gamma \left(\frac{3 i}{\nu }\right) \Gamma \left(\frac{7 i+\nu }{\nu }\right)t^{\frac{2i}{\nu}}}{7 \Gamma \left(\frac{1}{2}+\frac{i}{\nu }\right) \Gamma \left(\frac{7 i}{\nu }\right) \Gamma \left(\frac{4 i+\nu }{\nu }\right)}\right),
\end{split}
\end{equation}
wherein a combination of educated guesses with a numerical fit lead to the following expression of $h(\nu)$
\begin{equation}
\label{eq:s2hnuguess}
\begin{split}
h(\nu) &= \ln \left(\frac{\Gamma \left(\frac{1}{2}\right)^2\Gamma \left(1+\frac{3 i}{\nu }\right)^2 \Gamma \left(\frac{i}{4 \nu }\right)t^{\frac{4i}{\nu}}}{\Gamma \left(\frac{1}{2}+\frac{i}{\nu }\right)^2 \Gamma \left(1+\frac{4 i}{\nu }\right)^2 \Gamma \left(\frac{\sqrt{6} i}{4 \nu }\right)}\right)\\ &+\frac{i}{2\nu}\ln \left(\frac{2}{3}\right).
\end{split}
\end{equation}

The choices of the gamma functions in $h(\nu)$ are made so as to fit the pattern identified in the spin-$3/2$ and spin-$1$ cases. The function $\Gamma(1/2+i/\nu)$ in the denominator appears in all the given expressions. The form of $\gamma_{N_1,N_2}$ in the spin-$3/2$ asymptotics motivates the inclusion of a ratio of $\Gamma(1+ia/\nu)$ functions. The partial solution similarly suggests the ratio of $\Gamma(ia/\nu)$ functions. All integer parameters $a$ within the gamma functions are determined by brute force. The appearance of $\sqrt{6}$ is essential, as it corresponds to the weightings in the ground state in the stiff-ramp (diabatic) limit ($\eta \rightarrow 0$), recalling that $\nu = N/\eta$. The second term in the expression for $h(\nu)$ is taken from \eqref{eq:correctedgammafunc}, and its prefactor was obtained via numerical fitting. All exponents of the gamma functions are chosen such that the proposed ansatz performs well for higher site numbers. This guess has been numerically tested and appears  exact (see Fig. \ref{fig_App:N8Spin2}).

\section{Matrix elements of \texorpdfstring{$\hat{s}_j^{-}$}{s-j} and \texorpdfstring{$(\hat{s}_j^{-})^2$}{(s-j)**2}}\label{sec_App:matrix_elements}
We provide here the nonzero matrix elements of the spin-1 single-site lowering operators in the $N \rightarrow \infty$ limit, computed using Eq.~\eqref{eq:gensinglesiteoverlap}. The unnormalized matrix elements are listed below.
\begin{equation}
\label{eq:matrelemsminus}
\braket{\Psi_{\scalebox{0.75}{$\infty$}}^{\scalebox{0.55}{$(N_+\!\!-\!1)$}}(t)|\hat{s}_j^{-}|\Psi_{\scalebox{0.75}{$\infty$}}^{\scalebox{0.55}{$(N_+)$}}(t)} 
=\frac{1}{2\pi} \smashoperator{\int_{-2\pi}^{2\pi}}\!d\xi e^{N G(\xi)} \, g^{(-,0)}_{j}(\xi),
\end{equation}
\begin{equation}
\label{eq:matrelemsminussqr}
\braket{\Psi_{\scalebox{0.75}{$\infty$}}^{\scalebox{0.55}{$(N_+\!\!-\!2)$}}(t)|\scalebox{0.85}{$(\hat{s}_j^{-})^2$}|\Psi_{\scalebox{0.75}{$\infty$}}^{\scalebox{0.55}{$(N_+)$}}(t)} 
\!=\!\frac{1}{2\pi} \smashoperator{\int_{-2\pi}^{2\pi}}\!d\xi e^{N G(\xi)} \, g^{\scalebox{0.55}{$(-,-)$}}_{j}(\xi).    
\end{equation}
The \textit{g-functions} of the above integrals are 
\begin{equation}
g^{(-,0)}(\xi) = \frac{2 \left(\Gamma \left(\frac{\nu-2i}{2\nu}\right)\left(\frac{t}{2}\right)^{\frac{i}{\nu }} e^{\frac{4 \pi  j}{\nu }+i \xi }+\Gamma \left(\frac{\nu+2i}{2\nu}\right)\left(\frac{t}{2}\right)^{\frac{-i}{\nu }}\right)}{\sqrt{\pi }\,\text{sech} \left(\frac{\pi }{\nu }\right) \left(e^{\frac{8 \pi  j}{\nu }+2 i \xi }+2 \cosh \left(\frac{\pi }{\nu }\right) e^{\frac{4 \pi  j}{\nu }+i \xi }+1\right)},
\end{equation}
\begin{equation}
    g^{(-,-)}(\xi) = \frac{\exp{\left(-i\xi +\frac{2i\varphi _j}{\nu }-4it \epsilon _j\right)}}{\cosh \left(\frac{\pi }{\nu }\right)+\cos \left(\xi -\frac{4 i \pi  j}{\nu }\right)}.
\end{equation}
After taking the $N \rightarrow \infty$ limit, the normalization $\mathcal{N}$ of the matrix elements introduces a multiplicative factor of $\exp\left(-\frac{i}{2} \xi_0^{(1)} \delta \right)$ on the g-functions. The value of $\delta$ is $-1$ for the matrix element of $\hat{s}^{-}$ and $-2$ for that of $(\hat{s}^{-})^2$. This  enables us to obtain the normalized versions of \eqref{eq:matrelemsminus} and \eqref{eq:matrelemsminussqr} in the thermodynamic limit:
\begin{align}
&\frac{\braket{\Psi_{\scalebox{0.75}{$\infty$}}^{\scalebox{0.55}{$(N_+\!\!-\!1)$}}(t)|\hat{s}_j^{-}|\Psi_{\scalebox{0.75}{$\infty$}}^{\scalebox{0.55}{$(N_+)$}}(t)}}{(\mathcal{N}_{-1})^{-1}} = \frac{e^{-2 i t \varepsilon _j+i \varphi _j}}{\cosh\left(\frac{2 \pi \left (j-\mu_{(1)}\right)}{\nu }\right)},\\ 
&\frac{\braket{\Psi_{\scalebox{0.75}{$\infty$}}^{\scalebox{0.55}{$(N_+\!\!-\!2)$}}(t)|\scalebox{0.85}{$(\hat{s}_j^{-})^2$}|\Psi_{\scalebox{0.75}{$\infty$}}^{\scalebox{0.55}{$(N_+)$}}(t)}}{(\mathcal{N}_{-2})^{-1}}  = \frac{e^{-4 i t \varepsilon _j+2 i \varphi _j}}{2\cosh^2\left(\frac{2 \pi  \left(j-\mu_{(1)}\right)}{\nu }\right)}. 
\end{align}
where the chemical potential $\mu_{(1)}$ is given by Eq.~\eqref{eq:s1chemicalpotential}. Both results above exactly match  the mean-field predictions. 
\section{Higher moments of \texorpdfstring{$\hat{s}_j^z$}{szj}}\label{sec_App:higher_moments}
The general expression for the higher moments of $\hat{s}_j^{z}$ in the thermodynamic limit is
\begin{subequations}
\begin{align}
\label{eq:generalszavg}
\braket{(\hat{s}^{z}_{j})^n}^{\!(s)} &= \frac{(-s)^n +\sum_{k=1}^{2s}(-s+k)^n\begin{pmatrix}
2s \\ k    
\end{pmatrix}\omega^k}{1+\sum_{k=1}^{2s}\begin{pmatrix}
2s \\ k    
\end{pmatrix}\omega^k}, \\   \label{eq:generalszavgparam} \omega &= \exp\left(-\frac{4s\pi(j-\mu_{(s)})}{\nu}\right).   
\end{align}    
\end{subequations}

It is non-trivial to simplify the above expression for general $n$. For example, the closed expressions of $\braket{(\hat{s}^{z}_{j})^n}$ for $n=3, 4$ are as follows:
\begin{subequations}
\begin{equation}
\begin{split}
\braket{(\hat{s}^{z}_{j})^3}^{\!(s)} &= -\frac{s\left(s^2\cosh{\left(\frac{4s\pi (j-\mu_{(s)})}{\nu}\right)}-s(s-3)-1\right)}{2\cosh{\left(\frac{2s\pi (j-\mu_{(s)})}{\nu}\right)^2}}\times\\&\tanh{\left(\frac{2s\pi (j-\mu_{(s)})}{\nu}\right)},
\end{split}
\end{equation}
\begin{equation}
\begin{split}
\braket{(\hat{s}^{z}_{j})^4}^{\!(s)} &= s^4+\frac{s(2s-1)}{4\cosh{\left(\frac{2s\pi (j-\mu_{(s)})}{\nu}\right)}^2}\times\\&\left(\frac{(2s-3)(s-1)}{\cosh\left(\frac{2s\pi (j-\mu_{(s)})}{\nu}\right)^2}-4s(s-1)-2\right).    
\end{split}
\end{equation}
\end{subequations}

After setting $\nu=sN/\eta$, taking the $s\rightarrow\infty$ limit reveals that $ \braket{(\hat{s}^{z}_{j})^n}^{\!(s)}=  \left(\braket{\hat{s}^{z}_{j}}^{\!(s)}\right)^n$ for $n=2,3$ and $4$. However, we show that this property holds for any non-negative integer $n$. The denominator of \eqref{eq:generalszavg} simplifies to $(1+\omega)^{2s}$, and the numerator expands as follows
\begin{equation}\label{eq:numerator_expanded}
(-s)^n\left[1+\sum_{r=0}^{n}\begin{pmatrix}
n \\ r    
\end{pmatrix}\frac{(-1)^r}{s^r}\sum_{k=1}^{2s}\begin{pmatrix}
2s \\ k   
\end{pmatrix}k^r\omega^k\right].    
\end{equation}
Assuming $\omega = e^{\lambda}$, the entire summation is written in the following form:
\begin{equation}
(-s)^n\left\{(1+e^{\lambda})^{2s}+\sum_{r=1}^{n}\begin{pmatrix}
n \\ r    
\end{pmatrix}\frac{(-1)^r}{s^r}\frac{\partial^r}{\partial\lambda^{r}}\left[(1+e^{\lambda})^{2s}\right]\right\}.
\end{equation}
In the $s\rightarrow\infty$ limit, we then have
\begin{equation}
\frac{1}{s^r}\frac{\partial^r}{\partial\lambda^{r}}\left((1+e^{\lambda})^{2s}\right) \approx (2e^\lambda)^r (1+e^\lambda)^{2s-r} + \mathcal{O}(s^{-1}),  
\end{equation}
which allows the summation in the numerator \eqref{eq:numerator_expanded} to be simplified to
\begin{equation}
s^n(1+e^\lambda)^{2s}\left(\frac{e^{\lambda}-1}{e^{\lambda}+1}\right)^{n},
\end{equation}
up to corrections that vanish in the large $s$ limit. Identifying $\lambda$ from \eqref{eq:generalszavgparam} and plugging in the simplifications then gives us
\begin{equation}
\begin{split}
\braket{(\hat{s}^{z}_{j})^n}^{\!(s)} &\approx (-s)^n\tanh\left(\frac{2s\pi(j-\mu_{(s)})}{\nu}\right)^{n} \\
&= \left(\braket{\hat{s}^{z}_{j}}^{\!(s)}\right)^n.
\end{split}
\end{equation}

\section{Realizations in cavity QED}\label{app:Cavity}
One possible platform for implementing the model in Eq.~\eqref{eq:mainBCS} is a system of higher-spin atoms in an optical cavity, such as that described in Ref.~\cite{davis_photon-mediated_2019}. In that work, ${}^{87}\mathrm{Rb}$ atoms are operated on the $\ket{5\mathrm{S}_{1/2},F=1} \leftrightarrow \ket{5\mathrm{P}_{3/2}}$ transition, which enables access to three hyperfine spin states labeled by $m_F = \{-1, 0, 1\}$ within the $\ket{5\mathrm{S}_{1/2},F=1}$ manifold.
Effective spin operators for the $i$-th atom are defined as 
\begin{equation}
f_i^\alpha = \psi^\dag_{i,m_F} (F^\alpha)_{m_F, m_F'} \psi_{i,m_F'},   
\end{equation}
where $\psi_{i,m_F}$ and $\psi^{\dag}_{i,m_F}$ are the  bosonic  annihilation and creation operator, respectively, for an atom $i$ in spin state $m_F$, and $F^\alpha$ denote the spin-$1$ matrices for $\alpha=\{x,y,z\}$ \cite{stamper-kurn_spinor_2013}. In this section, we adopt the notation used in Ref.~\cite{davis_photon-mediated_2019} for ease of comparison. The effective Hamiltonian derived in Ref.~\cite{davis_photon-mediated_2019} is given by
\begin{equation}
H = \sum_{i,j} \chi_{ij} (f_i^x f_j^x + f_i^y f_j^y) + \sum_i h_i f_i^z,
\label{App_eq:H_spinor}
\end{equation}
where
\begin{subequations}
\label{eq:relations_cavity}
\begin{align}
\chi_{ij} &= \chi_{ij}^+ + \chi_{ij}^-, \\
h_i &= \chi_{ii}^+ - \chi_{ii}^-, \label{eq_App:hi} \\
\chi_{ij}^\pm &= \bar{n}(\delta_c) \Omega_i \Omega_j {\cal{A}}(\delta_\pm) \frac{1}{\kappa}, \label{eq_App:chi} \\
\bar{n}(\delta_c) &=\frac{|A|^2}{\delta_c^2 + \left( \frac{\kappa}{2} \right)^2}, \\
\Omega_i &= \frac{g^2_{m_F=-1}(x_i)-g^2_{m_F=+1}(x_i)}{2\Delta_c}, \\
{\cal A}(\delta_\pm) &= \frac{\delta_\pm \kappa}{16\left[ \delta_\pm^2 + \left( \frac{\kappa}{2} \right)^2 \right]}, \\
\delta_\pm &= \delta_c \mp \omega_z.
\end{align}
\end{subequations}
Here, $\omega_z$ denotes the splitting between adjacent $m_F$ levels, $\delta_c$ is the drive laser--cavity detuning, $\kappa$ is the cavity decay rate, and $\bar{n}$ is the intracavity photon number in the absence of atoms. The parameter $A$ represents the drive laser field amplitude, and $\Delta_c$ is the detuning of the cavity mode relative to the $\ket{5{\rm S}_\frac{1}{2},F=1} - \ket{5{\rm P}_\frac{3}{2}}$  atomic transition. The quantity $g_{m_F=-1}(x_i)$ denotes the single-photon Rabi frequency, and $\Omega_i$ is the induced light shift at the position $x_i$ of the atom along the cavity axis. To clarify the correspondence with the notation used throughout this paper: $f_i^{(x,y,z)} \equiv \hat{s}_i^{(x,y,z)}$, $h_i \equiv \varepsilon_i$, and $\chi_{ij}$ plays the role of the time-dependent coupling $g(t)$; see below for details.

A complete analysis of the experiment in Ref.~\cite{davis_photon-mediated_2019}, including effects such as the transverse and longitudinal distribution of atoms and inhomogeneities in the couplings, lies beyond the scope of this work. In particular, we note that both $h_i$ and $\chi_{ij}$ originate from the same underlying quantities and, according to equations~\eqref{eq:relations_cavity}, their spatial dependence (on $i$ and $j$) is governed by $\Omega_i = \Omega(x_i)$. We do not explore possible strategies to mitigate this issue---namely, how to achieve $\chi_{ij} = \text{const}$ while allowing for site-dependent $h_i$, for example by making $\omega_z$ position dependent.

For the purposes of our analysis, we simply note that both the spread and the mean value of $\Omega_i$ are of comparable magnitude [cf. Fig.~2b in \cite{davis_photon-mediated_2019} and Eq.~\eqref{eq:Omega_i}]. In what follows, to estimate the range of experimentally accessible regimes, we assume constant couplings that are representative of the experiment, namely
\begin{subequations}
\begin{align}
    \kappa & = 2\pi \times 200 \, {\rm kHz}, \\
    \Omega_i &= \Omega \approx 2\pi \times 20 \, {\rm Hz} \label{eq:Omega_i}\\
    A &= 2\pi \times 95 \; {\rm MHz}.
\end{align}
\label{App_eq:parameters}
\end{subequations}
We therefore drop the indices in Eqs.~\eqref{eq_App:hi} and \eqref{eq_App:chi} and write simply $h$ and $\chi$. Next, while $\omega_z$ is also tunable, we adopt the experimental value $\omega_z = 2\pi \times 3 \; {\rm MHz}$ for the estimates below. Similarly, we use the experimental value $\Delta_c = 2\pi \times 10 \; {\rm GHz}$ when applying the relations in equations~\eqref{eq:relations_cavity}. Furthermore, the dynamics of spins governed by the Hamiltonian in Eq.~\eqref{App_eq:H_spinor} is subject to both thermal broadening and collective dissipation with two associated jump operators
\begin{equation}
L_\pm = \sqrt{\gamma_\pm} {\cal F}_\pm,
\label{App_eq:L}
\end{equation}
with
\begin{equation}
\gamma_\pm = \bar{n}(\delta_c) \frac{\Omega^2}{16}\frac{\kappa}{\delta_\pm^2 + \left( \frac{\kappa}{2} \right)^2},
\end{equation}
where we assume homogeneous light field intensity $\Omega_i =\Omega$. The operators ${\cal F}_\pm$ are weighted sums of the single-atom spin operators $f_i^\pm$. which account for spatial inhomogeneities in the couplings to the cavity field.

\subsubsection*{Experimental estimates}
\begin{figure}
\centering
\includegraphics[width=\linewidth]{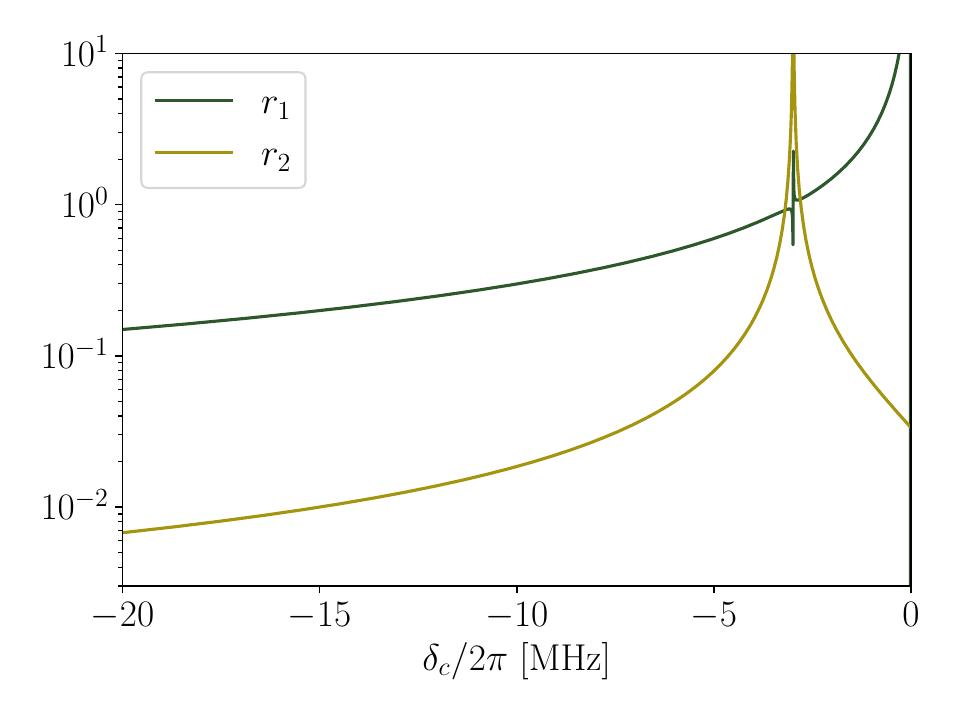}
\caption{Ratios $r_1$ and $r_2$ defined by equations~\eqref{eq:ratios_cavity} as  functions of the detuning $\delta_c$.
The resonance at $\delta^{\rm res}_c / 2\pi \approx -3.5~\text{MHz}$ corresponds to a discontinuity arising from a sign change in $\chi_{ij}$.}
\label{fig:ratios_cavity}
\end{figure}

To assess the feasibility of implementing the model in Eq.~\eqref{eq:mainBCS}, we recall that the time required to reach the steady state is expected to scale as (see Sec.~\ref{sec:Experimental_Realizations}):
\begin{equation}
\tilde{t}_{\rm ss} = \Delta \varepsilon \nu t_{\rm ss} \propto N^\alpha,
\label{eq:scaled_time_cavity}
\end{equation}
with a certain exponent $\alpha$. For example,  Ref.~\cite{zabalo_nonlocality_2022} argued that $\alpha=2$. Based on the discussion above, we identify $\Delta \varepsilon \sim h$ and $g = \chi$. We also require that, during the relevant time window, the decay \eqref{App_eq:L} remains negligible compared to the unitary dynamics. Noting that the transverse field $h$ and the spin-spin coupling $\chi$ have opposite signs for $\delta_c<0$ [except within a narrow region close to the $\delta^{\rm res}_c \equiv -\sqrt{(\kappa/2)^2 + \omega_z^2}$ resonance---as follows from the definitions in equations~\eqref{eq:relations_cavity}], and that for $\delta_c<0$ the decay is dominated by the $\gamma_-$ channel, we define the following two ratios
\begin{subequations}
\begin{align}
    r_1 & = -\frac{h}{\chi} \\
    r_2 & =  \frac{\gamma_-}{{\rm max}(|h|,|\chi|)}
\end{align}
\label{eq:ratios_cavity}
\end{subequations}
which are plotted in Fig.~\ref{fig:ratios_cavity}. We note that 
\begin{equation}
|r_1| = \left| \frac{h}{\chi} \right| \sim \Delta \varepsilon \nu t = \tilde{t},
\end{equation}
so that the scaled time $\tilde{t}$ associated with tuning $\delta_c$ can be directly inferred from the figure. Returning back to the estimate in Eq.~\eqref{eq:scaled_time_cavity} for the steady state time, and using the reported number of atoms $N \approx 10^5$, we obtain $\tilde{t}_{\rm ss} \sim 10^{10}$, which seems way out of reach of a realistic experiment.  Even after relaxing certain constraints---for example, by focusing on transient rather than steady state behavior, or by reducing the number of atoms $N$---it remains challenging to access a physically interesting regime. This motivates us not to study investigate experimental subtleties such as the role of the resonance $\delta^{\rm res}_c$ or strategies to achieve constant $\chi$ and tunable $h_i$.

Despite these findings, an alternative route towards realizing the physics explored in this work is to define coarse-grained spin operators ${\cal F}^\alpha_n = \sum_{i \in n} f^\alpha_i$ with $\alpha=x,y,z$, where $n$ labels a specific spatial region of the cavity. This leads to a effective Hamiltonian of the form
\begin{equation}
H = \sum_{m, n} \chi_{mn} \left( {\cal F}^x_m {\cal F}^x_n + {\cal F}^y_m {\cal F}^y_n  \right) + \sum_n h_n {\cal F}^\alpha_n,
\end{equation}
see Eq.~(S40) in Ref.~\cite{davis_photon-mediated_2019}. We leave this interesting opening for future investigations.

\section{Additional Figures}\label{sec_App:Additional_Figures}
This appendix contains additional figures verifying that the asymptotic solution presented in the main text becomes exact when the corrective term is included.

\begin{figure}[h]
\centering
\includegraphics[width=0.95\linewidth]{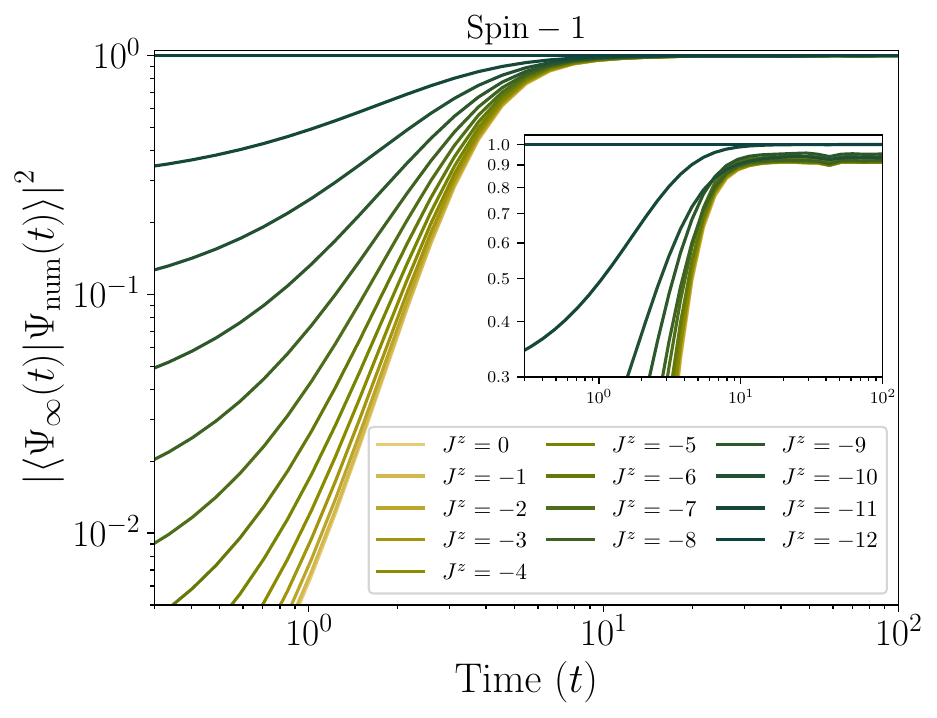}
\caption{Overlap between the asymptotic wavefunction $\ket{\Psi_\infty}$ of the spin-$1$ model and the numerically simulated wavefunction $\ket{\Psi_{\text{num}}}$, shown for various magnetization sectors $J^z$ with $N = 12$, $\nu = N$, and $\varepsilon_i = i/N$. As $t \rightarrow \infty$, the overlap approaches unity, validating the asymptotic solution. The inset displays the same overlap computed without applying  proper corrective terms.}
\label{fig_App:overlap_N=10}
\includegraphics[width=0.95\linewidth]{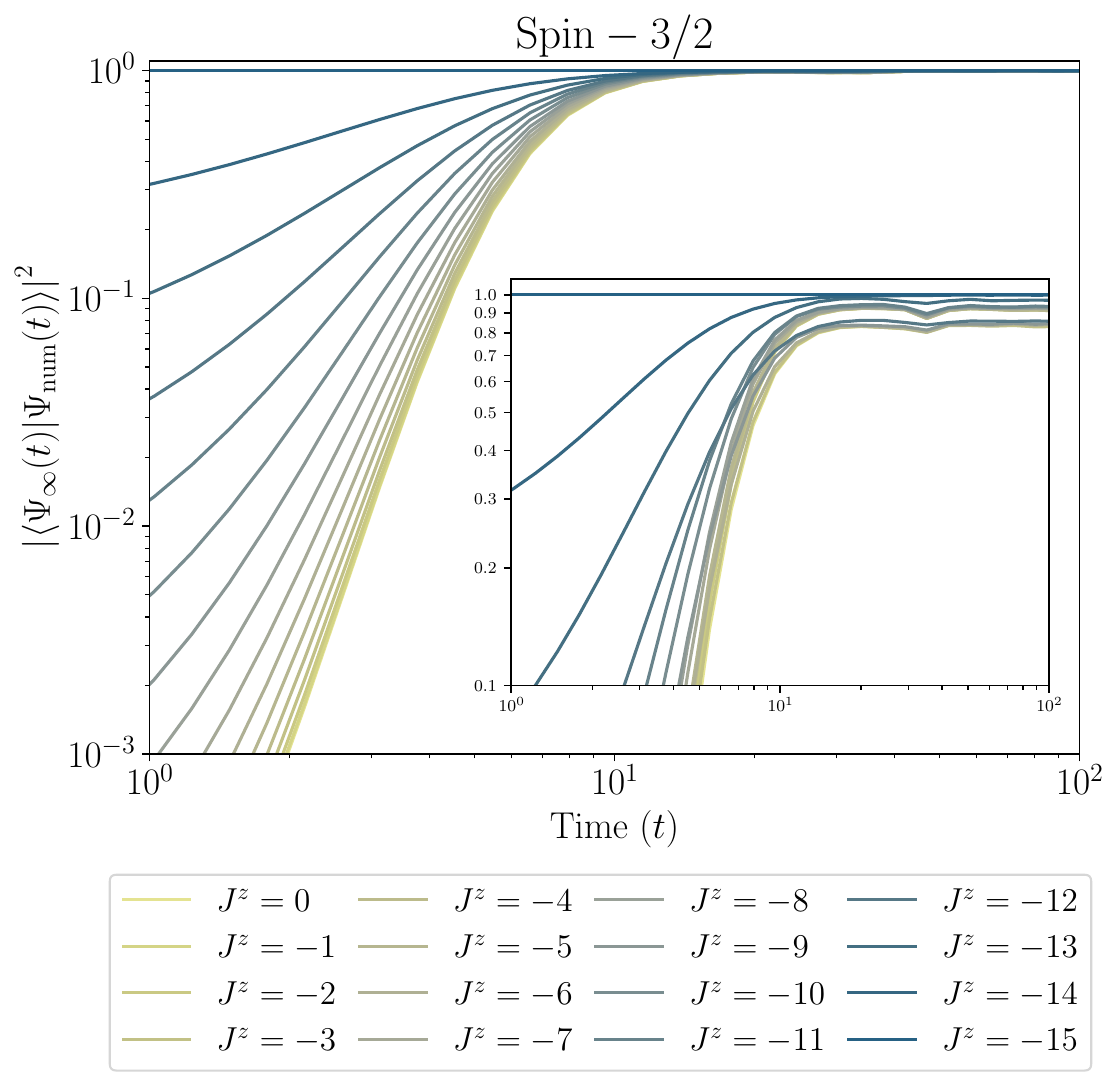}
\caption{Overlap between the asymptotic wavefunction $\ket{\Psi_\infty}$ of the spin-$3/2$ model and the numerically simulated wavefunction $\ket{\Psi_{\text{num}}}$, shown for various magnetization sectors $J^z$ with $N = 10$, $\nu = N$, and $\varepsilon_i = i/N$. As $t \rightarrow \infty$, the overlap approaches unity, confirming the accuracy of the asymptotic solution. The inset shows the overlap computed without applying   the proper corrective terms.}
\label{fig_App:overlap_N=10spin=3/2}
\end{figure}

\begin{figure}[h]
\centering
\includegraphics[width=0.95\linewidth]{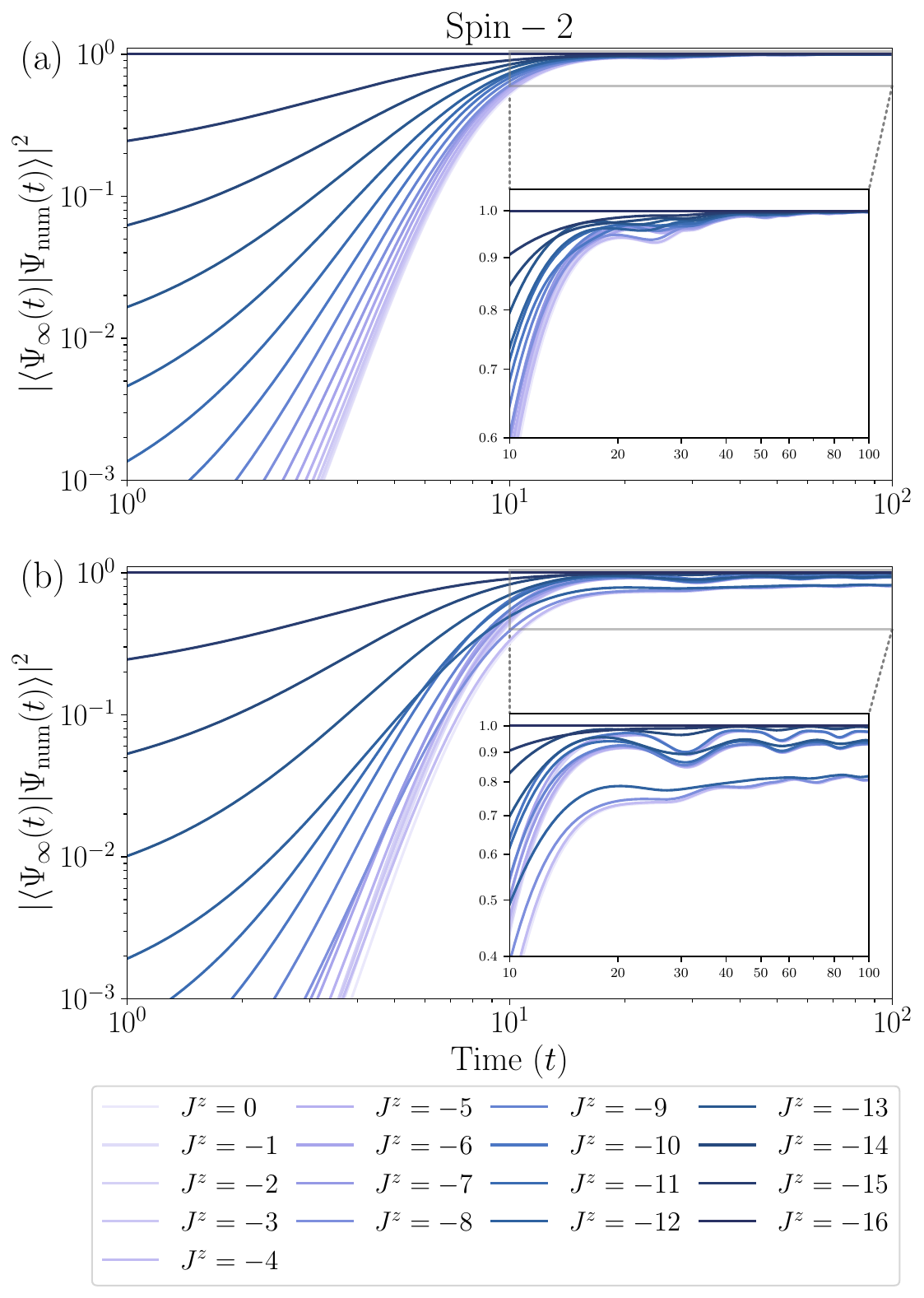}
\caption{(a) Overlap between the asymptotic wavefunction of the spin-$2$ model, $\ket{\Psi^{(2)}_\infty}$ with the weights given by Eq.~\eqref{App_eq:Gamma_Spin2_Guess}, and the numerically simulated wavefunction, $\ket{\Psi_{\text{num}}}$, for various magnetization sectors $J^z$ with $N=8$, $\nu = N$, and $\varepsilon_i = i/N$. As $t\rightarrow\infty$, the overlap approaches unity. The inset provides a zoomed in view of the asymptotic region where the overlap is close to unity. (b) same as (a) but using the weights in Eq.~\eqref{App_eq:Gamma_Spin2_Incorrect}. In this case, the overlap does \textit{not} converge to unity as $t\rightarrow\infty$. The inset highlights the asymptotic region.  }
\label{fig_App:N8Spin2}
\end{figure}

\clearpage
\bibliography{reflist}
\end{document}